\documentclass[aps,prc,tightenlines,floats,floatfix,superscriptaddress,showpacs]{revtex4}
\usepackage{amsmath}
\usepackage{bm}
\usepackage{dcolumn}
\usepackage{psfig}
\usepackage{graphicx}

\renewcommand{\Re}{{\rm{Re}}}
\renewcommand{\Im}{{\rm{Im}}}

\begin{document}

\title{Recoil Polarization Measurements for Neutral Pion Electroproduction
at $Q^2 = 1$~(GeV/$c$)$^2$ Near the Delta Resonance}
\author{J.~J.~Kelly}
\affiliation{Department of Physics, University of Maryland, College Park, Maryland 20742, USA}

\author{O.~Gayou}
\affiliation{Massachusetts Institute of Technology, Cambridge, Massachusetts 02139, USA}

\author{R.~E.~Roch\'e}
\affiliation{Florida State University, Tallahassee, Florida 32306, USA}

\author{Z.~Chai}
\affiliation{Massachusetts Institute of Technology, Cambridge, Massachusetts 02139, USA}

\author{M.~K.~Jones}
\affiliation{Thomas Jefferson National Accelerator Facility, Newport News, Virginia 23606, USA}

\author{A.~J.~Sarty}
\affiliation{Saint Mary's University, Halifax, Nova Scotia, Canada B3H 3C3}

\author{S.~Frullani}
\affiliation{Istituto Nazionale di Fisica Nucleare, Sezione Sanit\`a 
and Istituto Superiore di Sanit\`a, Physics Laboratory, 00161 Roma, Italy}

\author{K.~Aniol}
\affiliation{California State University Los Angeles, Los Angeles, California 90032, USA}

\author{E.~J.~Beise}
\affiliation{Department of Physics, University of Maryland, College Park, Maryland 20742, USA}

\author{F.~Benmokhtar}
\affiliation{Rutgers, The State University of New Jersey, Piscataway, New Jersey 08854, USA}

\author{W.~Bertozzi}
\affiliation{Massachusetts Institute of Technology, Cambridge, Massachusetts 02139, USA}

\author{W.~U.~Boeglin}
\affiliation{Florida International University, Miami, Florida 33199, USA}

\author{T.~Botto}
\affiliation{University of Athens, Athens, Greece}

\author{E.~J.~Brash}
\affiliation{University of Regina, Regina, Saskatchewan, Canada S4S 0A2}

\author{H.~Breuer}
\affiliation{Department of Physics, University of Maryland, College Park, Maryland 20742, USA}

\author{E.~Brown}
\affiliation{University of Georgia, Athens, Georgia 30602, USA}

\author{E.~Burtin}
\affiliation{CEA Saclay, F-91191 Gif-sur-Yvette, France}

\author{J.~R.~Calarco}
\affiliation{University of New Hampshire, Durham, New Hampshire 03824, USA}

\author{C.~Cavata}
\affiliation{CEA Saclay, F-91191 Gif-sur-Yvette, France}

\author{C.~C.~Chang}
\affiliation{Department of Physics, University of Maryland, College Park, Maryland 20742, USA}

\author{N.~S.~Chant}
\affiliation{Department of Physics, University of Maryland, College Park, Maryland 20742, USA}

\author{J.-P.~Chen}
\affiliation{Thomas Jefferson National Accelerator Facility, Newport News, Virginia 23606, USA}

\author{M.~Coman}
\affiliation{Florida International University, Miami, Florida 33199, USA}

\author{D.~Crovelli}
\affiliation{Rutgers, The State University of New Jersey, Piscataway, New Jersey 08854, USA}

\author{R.~De Leo}
\affiliation{Istituto Nazionale di Fisica Nucleare, Sezione Sanit\`a 
and Istituto Superiore di Sanit\`a, Physics Laboratory, 00161 Roma, Italy}

\author{S.~Dieterich}
\affiliation{Rutgers, The State University of New Jersey, Piscataway, New Jersey 08854, USA}

\author{S.~Escoffier}
\affiliation{CEA Saclay, F-91191 Gif-sur-Yvette, France}

\author{K.~G.~Fissum}
\affiliation{University of Lund, Box 118, SE-221 00 Lund, Sweden}

\author{V.~Garde}
\affiliation{Universit\'e Blaise Pascal Clermont Ferrand et CNRS/IN2P3 LPC 63, 177 Aubi\`ere Cedex, France}

\author{F.~Garibaldi}
\affiliation{Istituto Nazionale di Fisica Nucleare, Sezione Sanit\`a 
and Istituto Superiore di Sanit\`a, Physics Laboratory, 00161 Roma, Italy}

\author{S.~Georgakopoulos}
\affiliation{University of Athens, Athens, Greece}

\author{S.~Gilad}
\affiliation{Massachusetts Institute of Technology, Cambridge, Massachusetts 02139, USA}

\author{R.~Gilman}
\affiliation{Rutgers, The State University of New Jersey, Piscataway, New Jersey 08854, USA}

\author{C.~Glashausser}
\affiliation{Rutgers, The State University of New Jersey, Piscataway, New Jersey 08854, USA}

\author{J.-O.~Hansen}
\affiliation{Thomas Jefferson National Accelerator Facility, Newport News, Virginia 23606, USA}

\author{D.~W.~Higinbotham}
\affiliation{Massachusetts Institute of Technology, Cambridge, Massachusetts 02139, USA}

\author{A.~Hotta}
\affiliation{University of Massachusetts, Amherst, Massachusetts 01003, USA}

\author{G.~M.~Huber}
\affiliation{University of Regina, Regina, Saskatchewan, Canada S4S 0A2}

\author{H.~Ibrahim}
\affiliation{Old Dominion University, Norfolk, Virginia 23529, USA}

\author{M.~Iodice}
\affiliation{Istituto Nazionale di Fisica Nucleare, Sezione Sanit\`a 
and Istituto Superiore di Sanit\`a, Physics Laboratory, 00161 Roma, Italy}

\author{C.~W.~de~Jager}
\affiliation{Thomas Jefferson National Accelerator Facility, Newport News, Virginia 23606, USA}

\author{X.~Jiang}
\affiliation{Rutgers, The State University of New Jersey, Piscataway, New Jersey 08854, USA}

\author{A.~Klimenko}
\affiliation{Old Dominion University, Norfolk, Virginia 23529, USA}

\author{A.~Kozlov}
\affiliation{University of Regina, Regina, Saskatchewan, Canada S4S 0A2}

\author{G.~Kumbartzki}
\affiliation{Rutgers, The State University of New Jersey, Piscataway, New Jersey 08854, USA}

\author{M.~Kuss}
\affiliation{Thomas Jefferson National Accelerator Facility, Newport News, Virginia 23606, USA}

\author{L.~Lagamba}
\affiliation{Istituto Nazionale di Fisica Nucleare, Sezione Sanit\`a 
and Istituto Superiore di Sanit\`a, Physics Laboratory, 00161 Roma, Italy}

\author{G.~Laveissi\`ere}
\affiliation{Universit\'e Blaise Pascal Clermont Ferrand et CNRS/IN2P3 LPC 63, 177 Aubi\`ere Cedex, France}

\author{J.~J.~LeRose}
\affiliation{Thomas Jefferson National Accelerator Facility, Newport News, Virginia 23606, USA}

\author{R.~A.~Lindgren}
\affiliation{University of Virginia, Charlottesville, Virginia 22901, USA}

\author{N.~Liyanage}
\affiliation{Thomas Jefferson National Accelerator Facility, Newport News, Virginia 23606, USA}

\author{G.~J.~Lolos}
\affiliation{University of Regina, Regina, Saskatchewan, Canada S4S 0A2}

\author{R.~W.~Lourie}
\affiliation{Renaissance Technologies Corporation, Setauket, New York 11733, USA}

\author{D.~J.~Margaziotis}
\affiliation{California State University Los Angeles, Los Angeles, California 90032, USA}

\author{F.~Marie}
\affiliation{CEA Saclay, F-91191 Gif-sur-Yvette, France}

\author{P.~Markowitz}
\affiliation{Florida International University, Miami, Florida 33199, USA}

\author{S.~McAleer}
\affiliation{Florida State University, Tallahassee, Florida 32306, USA}

\author{D.~Meekins}
\affiliation{Florida State University, Tallahassee, Florida 32306, USA}

\author{R.~Michaels}
\affiliation{Thomas Jefferson National Accelerator Facility, Newport News, Virginia 23606, USA}

\author{B.~D.~Milbrath}
\affiliation{Eastern Kentucky University, Richmond, Kentucky 40475, USA}

\author{J.~Mitchell}
\affiliation{Thomas Jefferson National Accelerator Facility, Newport News, Virginia 23606, USA}

\author{J.~Nappa}
\affiliation{Rutgers, The State University of New Jersey, Piscataway, New Jersey 08854, USA}

\author{D.~Neyret}
\affiliation{CEA Saclay, F-91191 Gif-sur-Yvette, France}

\author{C.~F.~Perdrisat}
\affiliation{College of William and Mary, Williamsburg, Virginia 23187, USA}

\author{M.~Potokar}
\affiliation{Institut Jo\^zef Stefan, University of Ljubljana, SI-1001 Ljubljana, Slovenia}

\author{V.~A.~Punjabi}
\affiliation{Norfolk State University, Norfolk, Virginia 23504, USA}

\author{T.~Pussieux}
\affiliation{CEA Saclay, F-91191 Gif-sur-Yvette, France}

\author{R.~D.~Ransome}
\affiliation{Rutgers, The State University of New Jersey, Piscataway, New Jersey 08854, USA}

\author{P.~G.~Roos}
\affiliation{Department of Physics, University of Maryland, College Park, Maryland 20742, USA}

\author{M.~Rvachev}
\affiliation{Massachusetts Institute of Technology, Cambridge, Massachusetts 02139, USA}

\author{A.~Saha}
\affiliation{Thomas Jefferson National Accelerator Facility, Newport News, Virginia 23606, USA}

\author{S.~\v{S}irca}
\affiliation{Massachusetts Institute of Technology, Cambridge, Massachusetts 02139, USA}

\author{R.~Suleiman}
\affiliation{Massachusetts Institute of Technology, Cambridge, Massachusetts 02139, USA}

\author{S.~Strauch}
\affiliation{Rutgers, The State University of New Jersey, Piscataway, New Jersey 08854, USA}

\author{J.~A.~Templon}
\affiliation{University of Georgia, Athens, Georgia 30602, USA}

\author{L.~Todor}
\affiliation{Old Dominion University, Norfolk, Virginia 23529, USA}

\author{P.~E.~Ulmer}
\affiliation{Old Dominion University, Norfolk, Virginia 23529, USA}

\author{G.~M.~Urciuoli}
\affiliation{Istituto Nazionale di Fisica Nucleare, Sezione Sanit\`a 
and Istituto Superiore di Sanit\`a, Physics Laboratory, 00161 Roma, Italy}

\author{L.~B.~Weinstein}
\affiliation{Old Dominion University, Norfolk, Virginia 23529, USA}

\author{K.~Wijesooriya}
\affiliation{University of Illinois at Urbana-Champaign, Urbana, Illinois 61801, USA}

\author{B.~Wojtsekhowski}
\affiliation{Thomas Jefferson National Accelerator Facility, Newport News, Virginia 23606, USA}

\author{X.~Zheng}
\affiliation{Massachusetts Institute of Technology, Cambridge, Massachusetts 02139, USA}

\author{L.~Zhu}
\affiliation{Massachusetts Institute of Technology, Cambridge, Massachusetts 02139, USA}

\collaboration{The Jefferson Laboratory E91011 and Hall A Collaborations}
\noaffiliation

\date{August 31, 2005}

\begin{abstract}
We measured angular distributions of differential cross section, beam
analyzing power, and recoil polarization for neutral pion electroproduction 
at $Q^2 = 1.0$ (GeV/$c$)$^2$ in 10 bins of $1.17 \leq W \leq 1.35$ GeV
across the $\Delta$ resonance.
A total of 16 independent response functions were extracted, of
which 12 were observed for the first time.
Comparisons with recent model calculations show that response functions
governed by real parts of interference products are determined
relatively well near the physical mass, $W=M_\Delta \approx 1.232$ GeV, 
but the variation among models is large for response functions
governed by imaginary parts and for both increases rapidly with $W>M_\Delta$.
We performed a multipole analysis that adjusts suitable subsets of 
$\ell_\pi \leq 2$ amplitudes with higher partial waves constrained by 
baseline models.   This analysis provides both real and imaginary parts.
The fitted multipole amplitudes are nearly model-independent --- there
is very little sensitivity to the choice of baseline model or truncation
scheme.
By contrast, truncation errors in the traditional Legendre analysis of 
$N \rightarrow \Delta$ quadrupole ratios are not negligible.
Parabolic fits to the $W$ dependence around $M_\Delta$ for the 
multiple analysis gives values for 
$\Re(S_{1+}/M_{1+}) = (-6.61 \pm 0.18)\%$ and 
$\Re(E_{1+}/M_{1+}) = (-2.87 \pm 0.19)\%$ for the $p\pi^0$ channel at
$W=1.232$ GeV and $Q^2 = 1.0$ (GeV/$c$)$^2$ that are distinctly larger
than those from the Legendre analysis of the same data.
Similarly, the multipole analysis gives 
$\Re(S_{0+}/M_{1+}) = (+7.1 \pm 0.8)\%$ at $W=1.232$ GeV, 
consistent with recent models, while the traditional Legendre analysis 
gives the opposite sign because its truncation errors are quite severe.  
Finally, using a unitary isobar model (UIM), we find that excitation of the 
Roper resonance is dominantly longitudinal with 
$_p S_{1/2} = (0.05 \pm 0.01)$ GeV$^{-1/2}$ at $Q^2 = 1.0$ (GeV/$c$)$^2$.
The $\Re S_{0+}$ and $\Re E_{0+}$ multipoles favor pseudovector $\pi NN$ 
coupling over pseudoscalar coupling or a recently proposed mixed-coupling
scheme, but the UIM does not reproduce the imaginary parts of $0+$ multipoles,
especially $\Im S_{0+}$, well.
\end{abstract}

\maketitle

\section{Introduction}
\label{sec:intro}

The electroexcitation of a nucleon resonance is described by two transverse 
form factors, $A_{1/2}(Q^2)$ and $A_{3/2}(Q^2)$, and one longitudinal or 
scalar form factor, $S_{1/2}(Q^2)$, where the subscript denotes the  
projection of the resonance spin upon the virtual photon momentum and 
the dependence upon photon virtuality, $Q^2$, describes the spatial 
structure of the transition. 
Measurement of all three transition form factors could provide stimulating 
tests of QCD-inspired models of baryon structure \cite{Capstick00}.
For example, one might be able to determine the admixtures of small 
components with mixed symmetry or orbital excitation into wave functions
dominated by $SU(6)$ symmetry.
Alternatively, it has been speculated that the Roper resonance at 1440 MeV 
could be a hybrid baryon based upon gluonic excitation \cite{Li91}.
The electromagnetic transition form factors would be distinctly different
for a hybrid baryon or for a radial single-quark excitation \cite{Li92}.
However, the electroexcitation of a resonance, $R$, by a virtual photon,
$\gamma_v$, in a reaction of the form 
$\gamma_v N \rightarrow R \rightarrow N x$,
where $x$ could be one or even several mesons, inevitably is accompanied 
by nonresonant contributions and is complicated by final-state interactions.
Furthermore, the nucleon resonances are broad and overlapping.
Therefore, measurements of differential cross sections alone are not 
sufficient to permit clean, model-independent determination of the 
multipole amplitudes for meson electroproduction or, ultimately, 
the transition form factors.

There has been long-standing interest in deformed components of the 
$N$ and $\Delta$ wave functions \cite{Glashow79}.
The dominant amplitude for pion electroproduction at the $\Delta$ resonance 
is the $M_{1+}$ amplitude, but smaller $S_{1+}$ and $E_{1+}$ amplitudes 
arise from configuration mixing within the quark core \cite{Isgur82},
often described as quadrupole deformation,
or from meson and gluon exchange currents between quarks \cite{Meyer01}, 
or coupling to the pion cloud outside the quark core 
\cite{Kamalov99,Fiolhais96}.
Thus, the quadrupole deformation is related to the ratios of isospin-3/2
electroproduction multipole amplitudes 
$R_{EM}^{(3/2)} = \Re (E^{(3/2)}_{1+}/M^{(3/2)}_{1+})$ and 
$R_{SM}^{(3/2)} = \Re (S^{(3/2)}_{1+}/M^{(3/2)}_{1+})$ 
evaluated at $W=M_\Delta$ where $M_\Delta=1.232$ GeV is the physical mass.
These quantities are often labelled as EMR and SMR instead and we will
use both notations interchangeably.
It has been argued that the intrinsic $N$ and $\Delta$ quadrupole moments
are small \cite{Buchmann01} and that the observed EMR and SMR  
are dominated by nonvalence degrees of freedom.

Most previous determinations of EMR and SMR analyzed differential
cross section data using Legendre expansions truncated according to 
the assumption of $M_{1+}$ dominance.
The data prior to 1990 generally suffered from poor statistics, limited
acceptance, and relatively large systematic uncertainties, leaving
even the sign of EMR at low $Q^2$ undetermined \cite{Papanicolas01}.
More recent experiments using polarized photons \cite{Blanpied97,Beck00}
now find $R_{EM} \approx -2.5\%$ at $Q^2 = 0$ with relatively little 
model dependence \cite{Arndt02}.
Similarly, recoil polarization for parallel kinematics gave
$R_{SM} \approx -6.4\%$ for $Q^2 = 0.12$ (GeV/$c$)$^2$ \cite{Pospischil01}.
Furthermore, using more precise measurements of the azimuthal dependence of 
the differential cross section give $R_{SM} \approx -6.1\%$ at 
$Q^2 \rightarrow 0$ \cite{Mertz01,Kunz03,Sparveris05} and
have mapped the $Q^2$ dependence up to 4 (GeV/$c$)$^2$ 
\cite{Joo02,Frolov99,Laveissiere04}, 
but these analyses still rely upon $M_{1+}$ dominance.
However, there are several indications that this truncation may be inadequate.
First, at $Q^2 \approx 0.13$ (GeV/$c$)$^2$ there is a strong disagreement
between the the SMR value obtained at Bonn \cite{Kalleicher97} detecting
a forward pion and those obtained at MIT-Bates \cite{Mertz01} and
Mainz \cite{Pospischil01} detecting a forward proton that might be
explained by an unexpectedly large $S_{0+}$ amplitude \cite{Schmieden01}.
Second, the dynamical models fail to reproduce the induced polarization for 
parallel kinematics at $Q^2 \lesssim 0.2$ (GeV/$c$)$^2$ where the pion cloud
is important \cite{Warren98,Bartsch02}.
Similar problems have also been observed at $Q^2 = 0.4$ and 0.65 (GeV/$c$)$^2$
\cite{Joo03}.
Third, it appears to be difficult to obtain a consistent description of the
real and imaginary parts of the longitudinal-transverse interference
for $Q^2 \lesssim 0.25$ (GeV/$c$)$^2$ and $W \lesssim 1.23$ GeV where one 
might expect $M_{1+}$ dominance to suffice \cite{Kunz03,Sparveris03}, 
but a multipole analysis is not possible without more complete angular
coverage and sensitivity to phases.

In principle, this model dependence can be reduced considerably by 
measurement of recoil and/or target polarization observables that are 
sensitive to the relative phase between resonant and nonresonant 
mechanisms \cite{Raskin89,Lourie90,Lourie91,Schmieden98}.
Polarization often enhances the sensitivity to small amplitudes by 
interference with a large amplitude.
Helicity-dependent recoil polarization is sensitive to real parts
and helicity-independent recoil polarization is sensitive to imaginary 
parts of products of multipole amplitudes.
Furthermore, by measuring the azimuthal dependencies one can determine 
response functions sensitive to the linear polarization of the virtual photon.
Thus, it may be possible to perform a nearly model-independent multipole 
analysis if complete angular distributions are measured for
a comprehensive set of recoil-polarization observables. 
These multipole amplitudes can then be interpreted with the aid of
dynamical or coupled-channels models that enforce unitary and relate 
the complex multipole amplitude to the real transition form factor. 

Note that coarsely binned measurements of asymmetries with respect
to longitudinal target polarization have been made recently by
Biselli {\it et al}. \cite{Biselli03} for the $p\pi^0$ channel and by
De Vita {\it et al}. \cite{DeVita02} for the $n\pi^+$ channel in the
resonance region with $0.35 < Q^2 <1.5$ (GeV/$c$)$^2$.
However, those data have not been resolved into response functions
and are not suitable for detailed multipole analysis. 

In this paper we report the first extensive measurements of angular 
distributions for recoil polarization in pion electroproduction.
We have measured recoil polarization response functions for the 
$p(\vec{e},e^\prime \vec{p})\pi^0$ reaction at $Q^2 \approx 1$ 
(GeV/$c$)$^2$ near the $\Delta$ resonance, 
obtaining angular distributions for a total of 16 independent response 
functions in 10 steps of $W$ covering $1.17 \leq W \leq 1.35$ GeV;
the angular coverage and statistical precision are best in the 
central region, $1.21 \leq W \leq 1.29$ GeV.
The data for $W=1.23$ GeV were reported in Ref. \cite{Kelly05c}.
Twelve of these response functions are observed here for the first time.  
We compare a traditional truncated Legendre analysis with a more general
multipole analysis of these data.
Although the two analyses are qualitatively consistent,
the multipole analysis shows that truncation errors in the traditional
Legendre analysis of the quadrupole ratios are not negligible.
Furthermore, we find that excitation of the Roper resonance is
dominantly longitudinal.

Section \ref{sec:rsfns} defines the response functions,
Sec. \ref{sec:experiment} describes the experiment, and
Sections \ref{sec:xsec} and \ref{sec:polarization} describe the 
cross section and  polarization analyses, respectively.
We compare the results with selected models in  
Sec. \ref{sec:comparisons} and present Legendre and 
multipole analyses in Sections \ref{sec:legendre} and 
\ref{sec:mpamps}.
Further discussion of the relationship between the two
analyses is given in Sec. \ref{sec:discussion}.
An interpretation of the multipoles amplitudes using a
unitary isobar model is then given in Sec. \ref{sec:UIM}.
Finally, Sec. \ref{sec:conclusions} summarizes our 
conclusions.

\section{Response Functions}
\label{sec:rsfns}

\subsection{Definitions}
\label{sec:rsfns-def}

\begin{figure}
\centering
\includegraphics[width=4.5in]{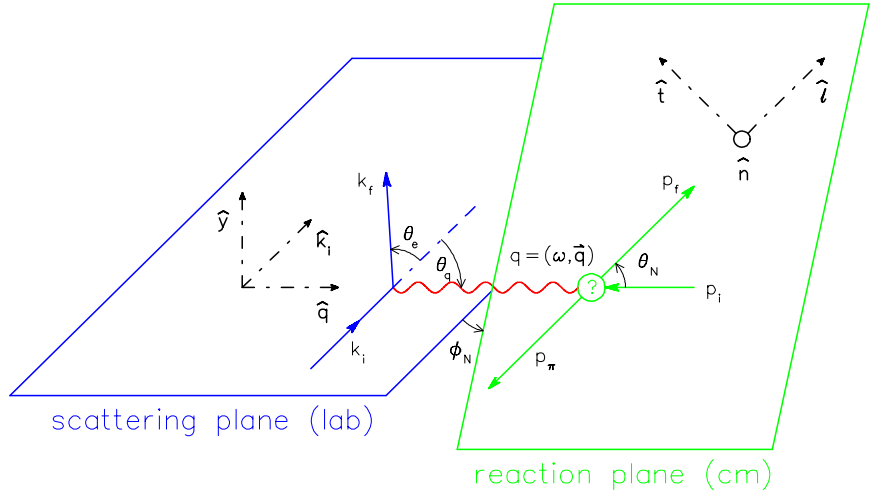}
\caption{(Color online)
Kinematics of the $p(\vec{e},e^\prime \vec{p})\pi^0$ reaction
and the polarization basis vectors.}
\label{fig:pikin}
\end{figure}

The differential cross section for recoil polarization in the pion 
electroproduction reaction
$p(\vec{e},e^\prime \vec{N})\pi$ can be expressed in the form
\begin{equation}
 \frac{d^5 \sigma}{dk_f d\Omega_e d\Omega^\ast} = \Gamma_\gamma \bar{\sigma} 
\left[1 + h A + \bm{\Pi}\cdot \bm{S} \right] 
\end{equation}
where $\bar{\sigma}$ is the unpolarized cross section, 
$h$ is the electron helicity, 
$\bm{S}$ is the spin direction for the recoil nucleon,
$A$ is the beam analyzing power,
$\bm{\Pi}=\bm{P} + h \bm{P}^\prime$ is the recoil polarization,
\begin{equation}
  \Gamma_\gamma= \frac{\alpha}{2 \pi^2} \frac{k_f}{k_i}
         \frac{k_\gamma}{Q^2}\frac{1}{1-\epsilon}
\end{equation}
is the virtual photon flux for initial (final) electron momenta $k_i$ ($k_f$),
$\epsilon = \left( 1+2\frac{{\bf q}^2}{Q^2}\tan^2 \frac{\theta_e}{2}
\right)^{-1}$
is the transverse polarization of the virtual photon,
$\theta_e$ is the electron scattering angle, and
$k_\gamma = (W^2 - m_p^2)/2m_p$
is the laboratory energy a real photon would need to excite the same
transition.
Note that electron kinematics and solid angle $d\Omega_e$ are
normally expressed in the lab frame while hadron kinematics and
nucleon solid angle $d\Omega^\ast$ are expressed in the $\pi N$ cm frame.
Figure \ref{fig:pikin} illustrates the kinematics of this reaction 
and the definitions for polarization vectors.

It is convenient to express polarization vectors in a basis where
$\hat{\ell}$ is along the nucleon momentum in the cm frame,
$\hat{n} \propto \hat{q}\wedge\hat{\ell}$ is normal to the reaction plane,
and $\hat{t}=\hat{n}\wedge\hat{\ell}$.
The azimuthal dependence can then be extracted and the observables can be 
decomposed into kinematical factors, $\nu_\alpha$, which depend only upon 
electron kinematics and response functions, $R_\alpha(x,W,Q^2)$, which carry 
the hadronic information, such that
\begin{widetext}
\begin{subequations}
\label{eq:obs}
\begin{eqnarray}
\bar{\sigma}    &=& \nu_0 \left[ \nu_L R_L + \nu_T R_T 
       + \nu_{LT}  R_{LT} \sin \theta \cos \phi 
       + \nu_{TT} R_{TT} \sin^2 \theta \cos 2\phi \right]  
\label{eq:sigma_rsfns}\\
 A \bar{\sigma}  &=&  \nu_0 
\left[ \nu_{LT}^\prime  R_{LT}^\prime \sin \theta \sin\phi  \right] \\
 P_{t} \bar{\sigma} &=&  \nu_0 \left[ \nu_{LT} R_{LT}^{t} \sin\phi
 + \nu_{TT} R_{TT}^{t} \sin\theta \sin 2\phi \right]     \\
 P_n \bar{\sigma} &=& \nu_0 \left[
   (\nu_L  R_L^n + \nu_T R_T^n) \sin\theta + \nu_{LT} R_{LT}^n \cos\phi
        +\nu_{TT} R_{TT}^n \sin\theta\cos 2\phi  \right]  \\
 P_{\ell} \bar{\sigma} &=&  \nu_0 \left[ 
           \nu_{LT} R_{LT}^{\ell} \sin\theta \sin\phi
         + \nu_{TT} R_{TT}^{\ell} \sin^2 \theta \sin 2\phi \right]     \\
 P^\prime_{t} \bar{\sigma} &=&  \nu_0 \left[
             \nu_{LT}^\prime R_{LT}^{\prime {t}} \cos \phi + 
             \nu_{TT}^\prime R_{TT}^{\prime {t}}  \sin\theta \right] \\
 P^\prime_n \bar{\sigma} &=& \nu_0 \left[
 \nu_{LT}^\prime R_{LT}^{\prime n} \sin\phi \right] \\
 P^\prime_{\ell} \bar{\sigma} &=&  \nu_0 \left[
             \nu_{LT}^\prime R_{LT}^{\prime {\ell}} \sin\theta \cos \phi + 
             \nu_{TT}^\prime R_{TT}^{\prime {\ell}}   \right]
\end{eqnarray}
\end{subequations}
\end{widetext}
where the response functions depend upon $W$, $Q^2$, and $x=\cos{\theta}$ 
where $\theta$ is the pion angle relative to $\vec{q}$ in the $\pi N$ 
center of mass frame. 
The azimuthal angle $\phi$ also refers to the pion momentum.
The overall factor $\nu_0$ permits phase-space factors to be 
extracted from the response functions, 
thereby simplifying multipole expansions.
This factor is usually taken as the ratio between the cm momentum in the
final state and the cm momentum a real photon needs for the same
transition, such that $\nu_0 = k/q_0$ where 
\begin{subequations}
\begin{eqnarray}
k^2 &=& \frac{ (W^2 + m_\pi^2 - m_p^2 )^2}{4W^2} - m_\pi^2 \\
q_0 &=& \frac{ W^2 - m_p^2 }{ 2W }
\end{eqnarray}
\end{subequations}

Regrettably, no single convention for the signs and normalizations
of the response functions has gained wide acceptance.  
We chose
\begin{subequations}
\begin{eqnarray}
\nu_T &=& 1 \\
\nu_{TT} &=& \epsilon \\
\nu_{TT}^\prime &=& \sqrt{1-\epsilon^2} \\
\nu_L &=& \epsilon_S \\
\nu_{LT} &=& \sqrt{2\epsilon_S(1+\epsilon)} \\
\nu_{LT}^\prime &=& \sqrt{2\epsilon_S(1-\epsilon)} 
\end{eqnarray}
\end{subequations}
where
\begin{equation}
\label{eq:epsS}
\epsilon_S = \frac{Q^2}{q^2} \epsilon
\end{equation}
and use the azimuthal angle for the pion in Eq. (\ref{eq:obs}).
Note that although $\epsilon$ is invariant, $\epsilon_S$ is not
and that Eq. (\ref{eq:epsS}) is evaluated in the $\pi N$ cm frame.
When there is insufficient information to perform a Rosenbluth separation,
we employ $\epsilon$-dependent combinations
\begin{subequations}
\label{eq:RL+T}
\begin{eqnarray}
\nu_T R_{L+T} &=& \nu_L R_L + \nu_T R_T \\
\nu_T R_{L+T}^n &=& \nu_L R_L^n + \nu_T R_T^n 
\end{eqnarray}
\end{subequations}

\subsection{Legendre Expansions}
\label{sec:legexp}

Legendre expansions often provide the most efficient representation of
the angular dependence of a response function.
Each of the response functions can be represented by a Legendre expansion of
the form
\begin{equation}
\label{eq:RLegendre}
R_\eta(x,W,Q^2) = \sum_{m=0}^\infty A^{\eta}_{m}(W,Q^2) P_m(x)
\end{equation}
where $\eta$ represents all of the labels needed to identify a particular
response function.
Notice that by extracting the leading dependencies upon $\sin{\theta}$,
the response functions defined by Eq. (\ref{eq:obs}) reduce to polynomials
in $x$ that are expected to be of low order for modest $W$ and $Q^2$.
This convention should also improve the accuracy of acceptance averaging
within bins of $x$. 
The Legendre coefficients $A^{\eta}_{m}(W,Q^2)$ can be obtained by fitting
response-function angular distributions for each $(W,Q^2)$.
Alternatively, often the most efficient method for extracting the 
$x$ dependence of response functions for a particular $(W,Q^2)$ bin 
is to perform a two-dimensional fit of the $(x,\phi)$ dependencies of the 
appropriate observable, cross section or polarization, with the aid
of the Legendre expansion.
This type of analysis can be used to extrapolate response functions to
parallel or antiparallel kinematics where interesting symmetry relations
have been developed but where the experimental acceptance vanishes
\cite{Kelly99d,Schmieden00}.

Each of the Legendre coefficients can be further expanded in terms
of products of pairs of multipole amplitudes, 
but these expansions quickly become unwieldy as the number of 
participating partial waves increases.
Analyses of previous data for differential cross section have
usually found that fits using truncated Legendre expansions 
based upon $M_{1+}$ dominance describe the data well in 
the immediate vicinity of the $\Delta$ resonance. 
Tables \ref{table:RtoLegendre1}-\ref{table:RtoLegendre3}
display truncated expansions that are limited to $s$ and $p$ waves and
that require $M_{1+}$ to appear in every term.
To streamline the notation, each coefficient is obtained from either
the real or imaginary part of $M_{1+}^\ast B_{\eta,m}$ where
the phase type is indicated in the second column and $B_{\eta,m}$ 
is listed in the fourth column of these tables.
The considerable redundancy among the $B_{\eta,m}$ functions
can be used to perform a model-independent evaluation of the
accuracy of $M_{1+}$ dominance.
The last column gives the response-function labels that will appear in 
multipanel figures.

\begin{table}
\caption{Truncated Legendre expansion of the differential cross section.  
Legendre coefficients based upon the assumption of $M_{1+}$ dominance
are obtained as either the real or imaginary parts of 
$M_{1+}^\ast B_{\eta,m}$.
The final column lists the labels used in figures.
\label{table:RtoLegendre1}}
\begin{ruledtabular}
\begin{tabular}{|l|l|l|l|l|}
$R_\eta$ & type & $m$ & $B_{\eta,m}$ & label \\ \hline
$R_L$ & Re & 0 & 0 & \\ \hline
$R_T$ & Re & 0 & $2 M_{1+}$ & L+T(0) \\ 
      &    & 1 & $2 E_{0+}$ & \\
      &    & 2 & $-M_{1+} +6 E_{1+} -2 M_{1-}$ & \\ \hline
$R_{TT}$ & Re & 0 & $ -\frac{3}{2} M_{1+} - 3 E_{1+} -3 M_{1-}$ & TT(0) \\ \hline
$R_{LT}$ & Re & 0 & $S_{0+}$ & LT(0) \\
         &    & 1 & $6 S_{1+}$ & \\ \hline
$R^\prime_{LT}$ & Im & 0 & $S_{0+}$ & LTh(0) \\
                &    & 1 & $6 S_{1+}$ & \\ \hline
\end{tabular}
\end{ruledtabular}
\end{table}

\begin{table}
\caption{Truncated Legendre expansion of the helicity-dependent recoil polarization.
Legendre coefficients based upon the assumption of $M_{1+}$ dominance
are obtained as either the real or imaginary parts of 
$M_{1+}^\ast B_{\eta,m}$.
\label{table:RtoLegendre2}}
\begin{ruledtabular}
\begin{tabular}{|l|l|l|l|l|}
$R_\eta$ & type & $m$ & $B_{\eta,m}$ & label \\ \hline
$R^{\prime n}_{LT}$ & Re & 0 & $  S_{1-}$ & LTh(n) \\
                    &    & 1 & $  S_{0+}$ & \\
                    &    & 2 & $4 S_{1+}$ & \\ \hline
$R^{\prime \ell}_{LT}$ & Re & 0 & $ -2 (S_{1-} + S_{1+})$ & LTh(l) \\
                       &    & 1 & $ -3 S_{0+}$ & \\
                       & Re & 2 & $-12 S_{1+}$ & \\ \hline
$R^{\prime t}_{LT}$ & Re & 0 & $-S_{0+}$ & LTh(t) \\   
                    &    & 1 & $S_{1-} - \frac{16}{5} S_{1+}$ & \\
                    &    & 2 & $2 S_{0+}$ & \\ 
                    &    & 3 & $\frac{36}{5} S_{1+}$ & \\ \hline

$R^{\prime \ell}_{TT}$ & Re & 1 & $-M_{1+} + \frac{6}{5} E_{1+} + 2 M_{1-}$ & TTh(l) \\
                       &    & 2 & $2 E_{0+}$ & \\ 
                       &    & 3 & $-\frac{36}{5} E_{1+}$ & \\ \hline
$R^{\prime t}_{TT}$ & Re & 0 & $ -2 M_{1+} + M_{1-}$ & TTh(t) \\
                    &    & 1 & $ -3 E_{0+}$ & \\
                    &    & 2 & $-12 E_{1+}$ & \\  \hline
\end{tabular}
\end{ruledtabular}
\end{table}

\begin{table}
\caption{Truncated Legendre expansion of the helicity-independent recoil polarization.
Legendre coefficients based upon the assumption of $M_{1+}$ dominance
are obtained as either the real or imaginary parts of 
$M_{1+}^\ast B_{\eta,m}$.
\label{table:RtoLegendre3}}
\begin{ruledtabular}
\begin{tabular}{|l|l|l|l|l|}
$R_\eta$ & type & $m$ & $B_{\eta,m}$ & label \\ \hline
$R^{n}_L$ & Im & 0 & 0 & \\ \hline
$R^{n}_T$ & Im & 0 & $-E_{0+}$ & L+T(n) \\
          &    & 1 & $3M_{1-}$ & \\ \hline
$R^{n}_{LT}$ & Im & 0 & $-S_{1-}$ & LT(n) \\
             &    & 1 & $-S_{0+}$ & \\
             &    & 2 & $-4 S_{1+}$ & \\ \hline
$R^{\ell}_{LT}$ & Im & 0 & $ -2 (S_{1-} + S_{1+})$ & LT(l) \\
                &    & 1 & $ -3 S_{0+}$ & \\
                &    & 2 & $-12 S_{1+}$ & \\ \hline
$R^{t}_{LT}$ & Im & 0 & $-S_{0+}$ & LT(t) \\
             &    & 1 & $S_{1-} - \frac{16}{5} S_{1+}$ & \\
             &    & 2 & $2 S_{0+}$ & \\ 
             &    & 3 & $\frac{36}{5} S_{1+}$ & \\ \hline

$R^{n}_{TT}$ & Im & 0 & $  3 E_{0+}$ & TT(n) \\
             &    & 1 & $ 12 E_{1+} - 3 M_{1-}$ & \\ \hline
$R^{\ell}_{TT}$ & Im & 0 & $ 3 E_{0+}$ & TT(l) \\
                &    & 1 & $18 E_{1+}$ &  \\ \hline
$R^{t}_{TT}$ & Im & 0 & $  3 M_{1-}$ & TT(t)  \\
             &    & 1 & $ -3 E_{0+}$ & \\
             &    & 2 & $ -12 E_{1+}$ & \\  \hline
\end{tabular}
\end{ruledtabular}
\end{table}

\section{Experiment}
\label{sec:experiment}

The experiment was performed in Hall A of the Continuous Electron
Beam Accelerator Facility (CEBAF) at Jefferson Laboratory in 
three periods between May and August 2000
using standard equipment described in detail in Ref. \cite{HallA-nim}.
Details of the design and performance of the focal-plane polarimeter
(FPP) can be found in Ref.\ \cite{Punjabi05}.
Further information specific to this experiment may be found in
Refs.\ \cite{Roche-thesis,Chai-thesis,Escoffier01}.
Here we summarize the salient features.

The beam energy of $4531 \pm 1$ MeV was measured by two independent
methods and the results were in agreement.
The arc method uses wire scanners to measure the beam deflection angle 
between the accelerator and the experimental beam lines.
The $eP$ method measures electron and proton angles for elastic
scattering from hydrogen using silicon strip detectors placed
symmetrically in the vertical plane of a dedicated instrument.

The beam current was monitored by a pair of resonant RF cavities 
that were periodically calibrated with respect to an Unser monitor.
The Unser monitor is a Parametric Current Transformer that provides
absolute current measurements and is calibrated with respect to a 
precision current source, but it suffers from a variable offset.
The RF cavities are much more stable and can be used as continuous
monitors.
The calibration procedure alternates between beam off and current 
ramping between about 10 and 100 $\mu$A in about 5 steps,
using the beam off periods to determine the drift of the Unser 
offset and the current steps to determine the RF gains.
This procedure was repeated several times during the experiment, 
with negligible differences between calibrations.
The accuracy of these monitors is better than 0.5\% \cite{HallA-nim}. 

The beam position and direction were measured by Beam Position Monitors
(BPMs) located 7.524 and 1.286 m upstream of the target.
Each BPM is a 4-wire antenna array tuned to the fundamental RF frequency
of the beam and provides the relative position to within 100 $\mu$m.
The absolute positions of the BPMs are calibrated with respect
to wire scanners that are surveyed regularly.

\begin{figure}
\centering
\includegraphics[width=3in,angle=-90]{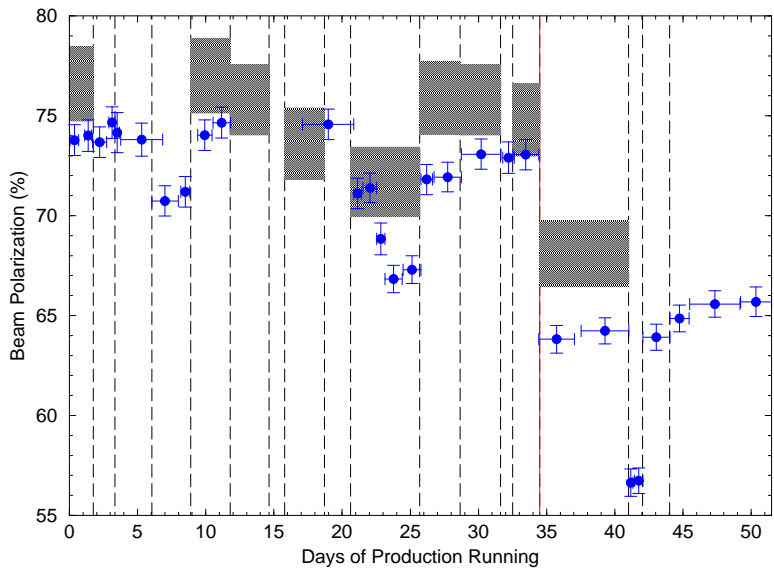}
\caption{(Color online) 
Beam polarization measurements: gray boxes indicate Moller 
measurements and points Compton measurements, with systematic uncertainties
indicate for both.
Vertical dashed lines mark movements of the laser spot.
The red vertical line makes the change of polarized source.}
\label{fig:beampol}
\end{figure}

The polarized electron beam was produced by circularly polarized
light from a 780 nm diode laser, pulsed at 1497 MHz, illuminating a 
strained gallium arsenide wafer.
A Pockels cell was used to reverse the laser polarization at 30 Hz.
The beam polarization was measured nearly continuously using a
Compton polarimeter, with systematic uncertainties estimated to be
about 1\% \cite{Escoffier01}.
In addition, periodic measurements were also made with a Moller
polarimeter, with systematic uncertainties of about 2.4\%.
Statistical uncertainties were negligible for both methods.
Fig.\ \ref{fig:beampol} summarizes the polarization measurements, 
where the vertical dashed lines indicate movement of the laser spot 
needed to maintain high current.
The polarization averaged about 72\% for the first two running periods,
but for the third it was necessary to use a different source with
lower polarization, about 65\%.
Fig.\ \ref{fig:beampol} shows that the time dependence of the beam 
polarization is minimal between spot changes.
Moller results were used during the few occasions when the Compton 
measurements were unavailable.
A comparison between the five beam polarimeters at Jefferson Lab
\cite{Grames04} indicates that Compton measurements in Hall A are
consistent with Mott measurements at the injector and that 
the ratio between Hall A Moller and Compton analyzing powers is 
approximately $1.034 \pm 0.028$, which is consistent with the observation 
in Fig.\ \ref{fig:beampol} that the Moller results are generally slightly 
larger than the Compton results.
However, because the statistical precision of this ratio remains poor
and relatively few runs rely on the Moller polarimeter, a correction
for the relative normalization of the two polarimeters was not applied.

The production target was a cryogenic loop for liquid hydrogen 
operating at 19 K and 0.17 MPa with a density of 0.723 g/cm$^3$.
The beam passes through a cylindrical aluminum cell  
that is 15 cm long and 6.35 cm in diameter with sidewall thickness of 
178 $\mu$m and entrance and exit windows approximately 
71 and 102 $\mu$m thick, respectively.
To avoid damage to the target cell and to minimize luminosity variations
due to local heating or boiling, the beam was rastered over a 
$4\times 4$ mm$^2$ pattern.
Measurements of rates versus current demonstrate that the 
average target density decreases by less than 2\% for rastered currents
in the range 20-60 $\mu$A relevant to the cross section measurements
\cite{Chai-thesis}.
Furthermore, the electron singles rate provided a continuous luminosity
monitor because the electron spectrometer remained fixed throughout
the experiment.

Additional data were taken with about 5 $\mu$A of unrastered beam
on a thin carbon foil and on dummy targets consisting of aluminum foils 
separated by either 4, 10, or 15 cm that mimic the windows of a cryogenic cell.
These data were used to determine the mispointing of the spectrometers 
and the data for the 15 cm dummy cell were used to determine the 
contribution of the cell walls for the production data.

Particles were detected by a pair of high resolution QQDQ spectrometers 
of identical design, with electrons observed in the left and protons 
in the right spectrometer.
These spectrometers have a central bend angle of $45^\circ$ and nominal
acceptances of $\pm 4.5\%$ in momentum, $\pm 60$ mr in vertical angle,
and $\pm 30$ mr in horizontal angle. 
The cross section data taken during the first of three running periods
were taken with 80 mm tungsten collimators at distances of 1.109 m 
for the left or 1.100 m for the right with nominal apertures of 6 msr.
Polarization data taken in the next two running periods were acquired with 
open apertures.
The entrance windows were 178 $\mu$m Kapton while the exit windows
were 100 $\mu$m titanium.

The position of the electron spectrometer was determined by survey.
The initial position of the proton spectrometer was also determined
by survey for each run period and subsequent angles were measured
using accurately placed floor marks observed through a closed-circuit
television camera mounted on a linear translation stage without parallax.
These raw angle measurements were then corrected for roll and pitch 
as measured by bi-axial inclinometers.
However, because the spectrometers do not rotate about a fixed pivot,
it is necessary to correct for possible mispointing.  
In principle, mispointing can be determined by a system that uses 
a linear variable differential transformer (LVDT) to measure the gap 
between an arm parallel to the spectrometer mid-plane and the outer 
surface of the scattering chamber, but this method proved to be unreliable.
The LVDT mispointing data were compared with mispointing measurements
deduced from scattering data for the carbon target.
The latter were found to be more stable and were used in the 
subsequent analysis.
The spectrometer offsets deduced for the electron arm, which was
stationary for each run period, were consistent with constant values
that were less than 2 mm and consistent with survey.
The offsets for the hadron arm varied between about -4 and +1.5 mm
with the motion of the spectrometer, but were reproducible without motion
\cite{Roche-thesis}. 

Both arms use two vertical drift chambers (VDC) for tracking that are 
inclined by $\pm 45^\circ$ with respect to the central trajectory and are
separated by 50 cm.
Each chamber contains two planes of sense wires inclined at $\pm 45^\circ$
with respect to the midplane ($uv$ configuration).
Valid tracks typically produce signals on 3-5 sense wires.
These detectors typically provide position resolutions of
$\sigma_{x,y}= 100$ $\mu$m and angular resolutions of
$\sigma_{\theta,\phi}$ = 0.5 mrad.
Further details about the VDCs are provided in Ref. \cite{Fissum01}.

Both arms also use two trigger planes, S1 and S2, consisting of 6 
plastic scintillator paddles that are 5 mm thick, overlap by about 0.5 cm, 
and are viewed by photomultiplier tubes on both ends.
Five types of trigger were produced by the trigger supervisor module
using signals from the two trigger planes in each spectrometer.
A T1 (T3) trigger in the electron (proton) arm requires coincidence
between paddle $i$ of S1 and paddle $j$ of S2 with $|i-j| \leq 1$, 
with PMT signals above threshold in both ends of both paddles.
A T2 (T4) trigger in the electron (proton) arm misses one or more of 
the PMT signals or fails the directivity requirement.
The T5 primary coincidence trigger requires T1 and T3 within 100 ns
of each other, where T3 and T4 include adjustable delays depending
upon hadron momentum.
The time resolution for the coincidence trigger is typically about
$\sigma=0.3$ ns.
Events corresponding to any of these triggers can be recorded, but
triggers other than T5 were generally prescaled to limit computer deadtime.

The electron arm also contained a CO$_2$ gas Cerenkov and 
lead-glass pre-shower and shower detectors for particle identification, 
but these were not used for this experiment because electrons were 
overwhelmingly dominant for these conditions.
A simple cut on flight time between scintillator planes was sufficient.
The gas Cerenkov detector normally present in the proton arm was 
removed for this experiment.
For the first running period a single large scintillator paddle, labeled S0, 
was included in the proton trigger.
After the first running period the S0 scintillator was
replaced with an aerogel Cerenkov detector with $n=1.025$ and
the aerogel signal was used in the proton trigger to suppress 
pion background with momentum greater than $0.63$ GeV/$c$.
We could then make polarization measurements at forward angles with 
higher beam currents without excessive deadtime.

Recoil polarization measurements were made using the focal plane
polarimeter (FPP) installed in the proton arm.  
More complete details of the construction and operation of the FPP
can be found in Refs. \cite{HallA-nim,Punjabi05}.  
The FPP consists of two front straw chambers, 
a carbon analyzer of adjustable thickness, and two rear straw chambers. 
The front chambers were separated by 114 cm and each contains 6 
planes in a $vvvuuu$ configuration.
The rear chambers were separated by 38 cm.
Chamber 3 has a $uuvvxx$ and chamber 4 a $uuuvvv$ configuration,
where $x$ denotes sensitivity to the dispersive direction.
The analyzer consisted of five graphite plates with thicknesses of
1.9, 3.8, 7.6, 15.2, and 22.9 cm separated by about 1.6 cm that can
be deployed in any combination to optimize the analyzing power for
specified proton momentum.

Spectrometer settings were chosen to obtain 
$(W,Q^2) = $ (1.232 GeV, 1.0 (GeV/$c$)$^2)$ 
for a nominal beam energy of 4.535 GeV.
Thus, the electron spectrometer remained fixed at $14.095^\circ$ with a 
central momentum of 3.660 GeV/$c$.
The nominal settings for the proton spectrometer, with accumulated charge,
are summarized in Table \ref{table:kin}.
It is convenient to define $\theta_{pq}$ as the nominal center-of-mass angle 
between the nucleon momentum and the momentum transfer in the lab frame 
with positive (negative) sign corresponding to forward (backward) angles 
with respect to the beam direction.
The experiment was divided into three periods.
The first period was used to scan the angular distribution and to obtain
differential cross sections for all settings except $\theta_{pq}=180^\circ$ 
using the 6 msr collimator in the hadron arm and relatively low currents to
limit deadtime.
The second and third periods used the highest possible currents to obtain 
adequate statistics for recoil polarization and used the aerogel detector
for forward angles to suppress deadtime due to accidental $\pi^+$ 
coincidences. 

\begin{table}
\caption{Settings for the proton spectrometer.
Here $\theta_p$ is the laboratory angle with respect to the beam,
$p_p$ is the central momentum, and $\theta_{pq}$ is the cm angle with
respect to $\vec{q}$.
The fifth column lists the $^{12}$C thickness used for the FPP and
the final column lists the nominal precession angle for the central
momentum.
\label{table:kin}}
\begin{ruledtabular}
\begin{tabular}{|r|r|r|r|r|r|}
$\theta_{pq}$ & $\theta_p$ & $p_p$ & charge & $^{12}$C thickness & $\chi_0$ \\ 
degree & degree & GeV/$c$ & Coulomb & cm & degree \\ \hline
    0 & 42.31 & 1.378 & 25.9 & 49.5 & 143.3 \\
   25 & 38.12 & 1.350 &  7.9 & 49.5 & 141.3 \\
  -25 & 46.33 & 1.350 &  4.6 & 49.5 &       \\
   50 & 34.29 & 1.270 & 18.9 & 34.3 & 135.7 \\
  -50 & 50.18 & 1.270 & 12.9 & 34.3 &       \\
   90 & 29.81 & 1.066 & 15.5 & 22.9 & 122.1 \\
  -90 & 54.79 & 1.066 & 20.1 & 22.9 &       \\
  135 & 30.81 & 0.819 & 14.6 & 11.4 & 107.1 \\
 -135 & 53.64 & 0.819 & 27.6 & 11.4 &       \\
  155 & 34.71 & 0.742 & 13.9 & 11.4 & 102.8 \\
 -155 & 49.72 & 0.742 & 13.6 & 11.4 &       \\
  180 & 42.28 & 0.703 &  5.0 & 7.6  & 100.8
\end{tabular}
\end{ruledtabular}
\end{table}

\begin{figure}
\centering
\includegraphics[width=3in]{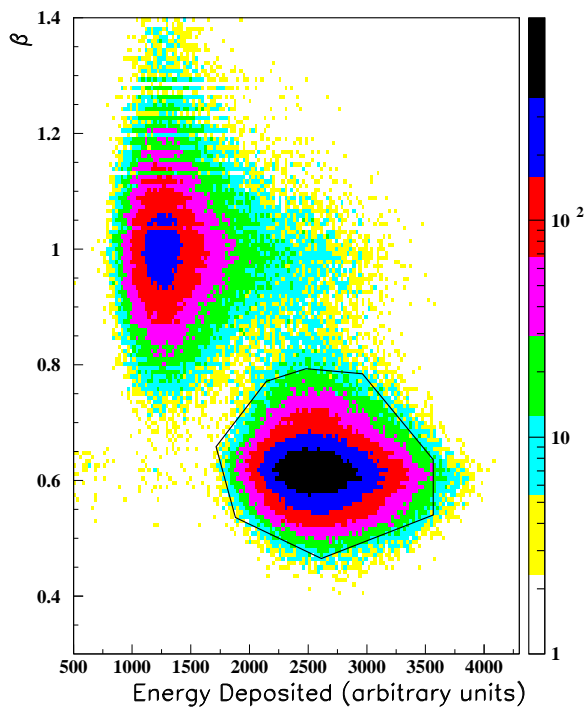}
\caption{(Color online) 
Proton identification using the correlation between velocity
and energy deposition in S1.  The polygon selects protons.
Notice that the intensity scale is logarithmic.}
\label{fig:PID}
\end{figure}

\begin{figure}
\centering
\includegraphics[width=3in]{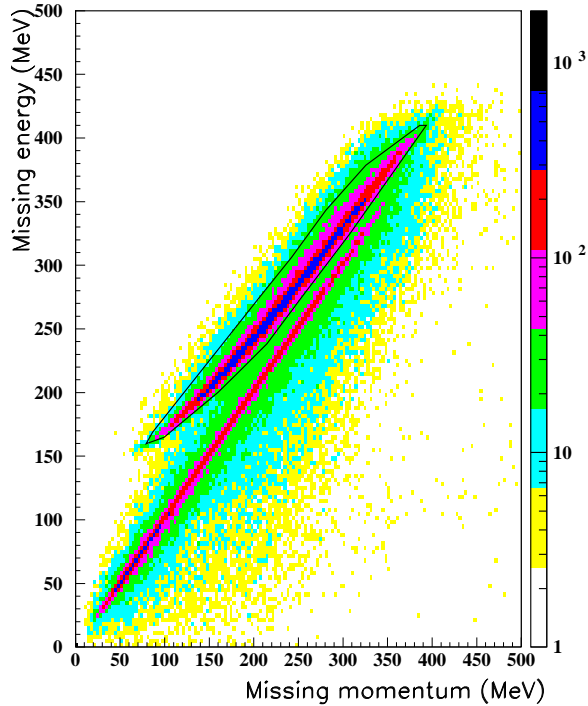}
\caption{(Color online) 
Pion production events for $\theta_{pq}=-50^\circ$
are selected by the polygon in $(E_m,p_m)$.
The diagonal population arises from the elastic radiative tail.
The intensity scale is logarithmic.}
\label{fig:elastic-tail}
\end{figure}

Electrons were identified using time of flight between the S1 and S2 
scintillator planes.
Protons were identified using the velocity measured by time of
flight between the S1 and S2 scintillator planes and the geometric mean 
between pulse heights for the two sides of S1.
Fig.\ \ref{fig:PID} shows a two-dimensional distribution for 
$\theta_{pq}=-155^\circ$ where the pion background is relatively large.
The polygonal cut selects protons with little loss and with little 
contamination.
A cut was also made on the reconstructed vertex position.
This cut was typically $\pm 2$ cm;
the resolution is limited mostly by the forward angle of the
electron spectrometer.
Finally, a cut on the missing energy and momentum, $(E_m,p_m)$, for the
$p\pi^0$ final state was used in the polarization analysis to suppress 
background from elastic scattering.
This background was visible only for the $\theta_{pq}=-50^\circ$ and 
$-90^\circ$ settings; the $p\pi^0$ event selection is illustrated in 
Fig. \ref{fig:elastic-tail} for $\theta_{pq}=-50^\circ$.
However, this cut was not needed for the cross-section analysis because
the more restrictive acceptance cuts described in Sec. \ref{sec:acceptance}
eliminated the elastic background quite effectively.

\section{Cross section analysis}
\label{sec:xsec}

\subsection{Acceptance}
\label{sec:acceptance}

\begin{figure*}
\centering
\includegraphics[width=3.0in]{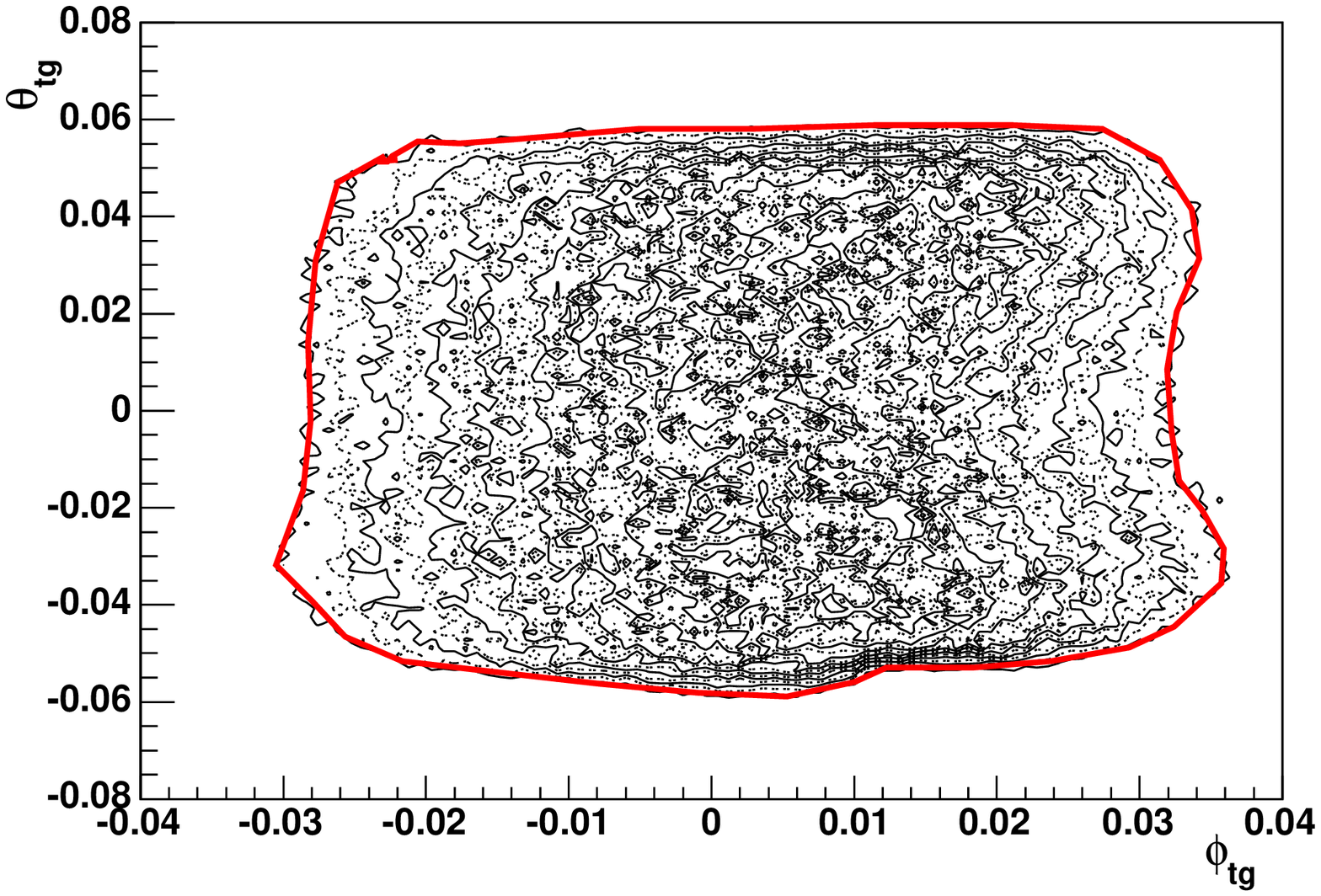}
\includegraphics[width=3.0in]{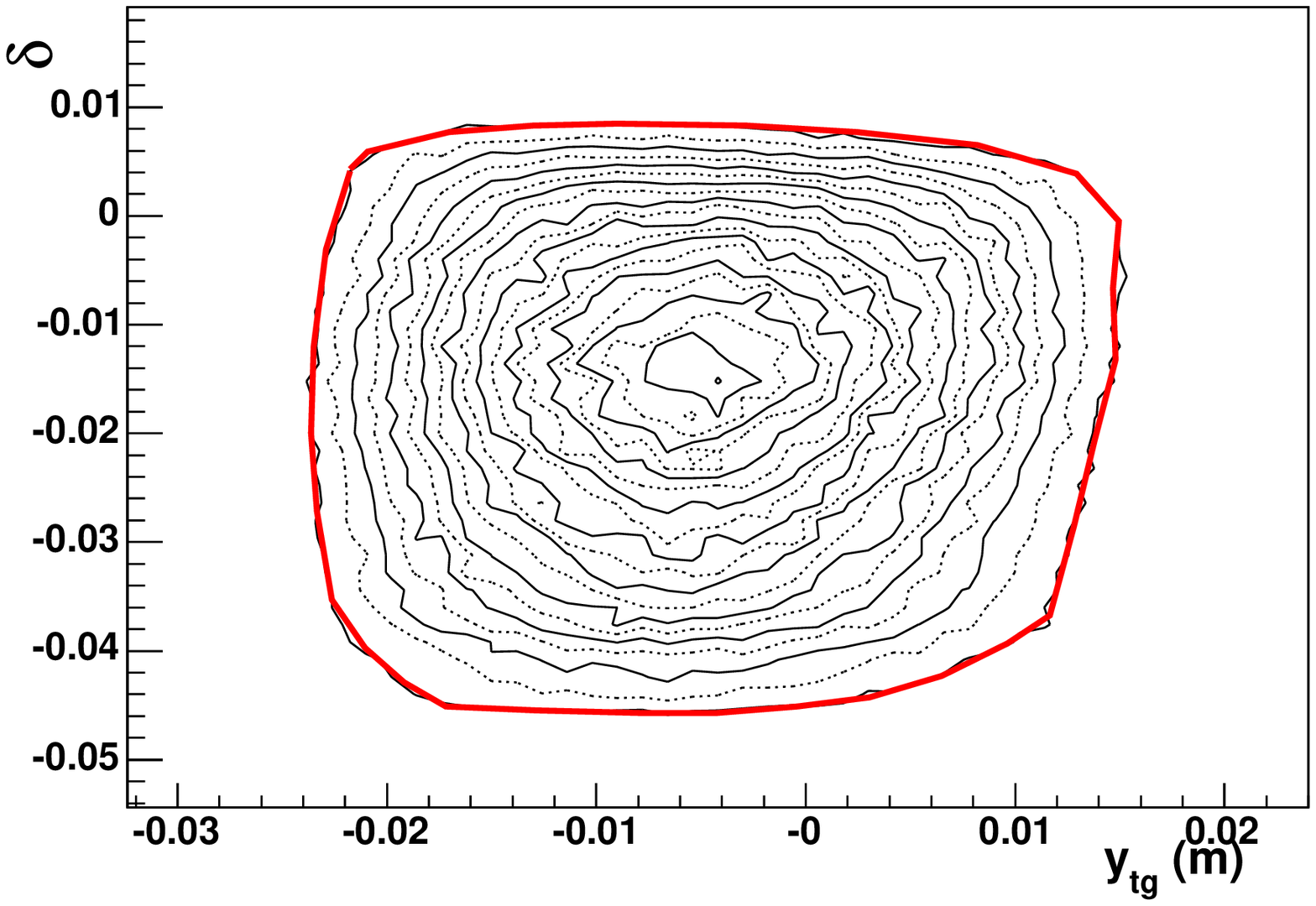}
\caption{(Color online)
Polygonal boundaries for two pairs of reconstructed electron 
variables for coincidence events.}
\label{fig:E-polygons}
\end{figure*}

\begin{figure}
\centering
\includegraphics[width=3.0in]{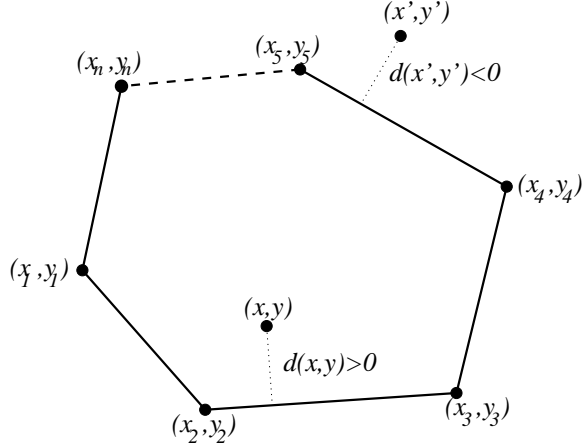}
\caption{Definition of signed distances from the boundary of
a two-dimensional projection.}
\label{fig:polygon}
\end{figure}

\begin{figure}
\centering
\includegraphics[width=3.0in]{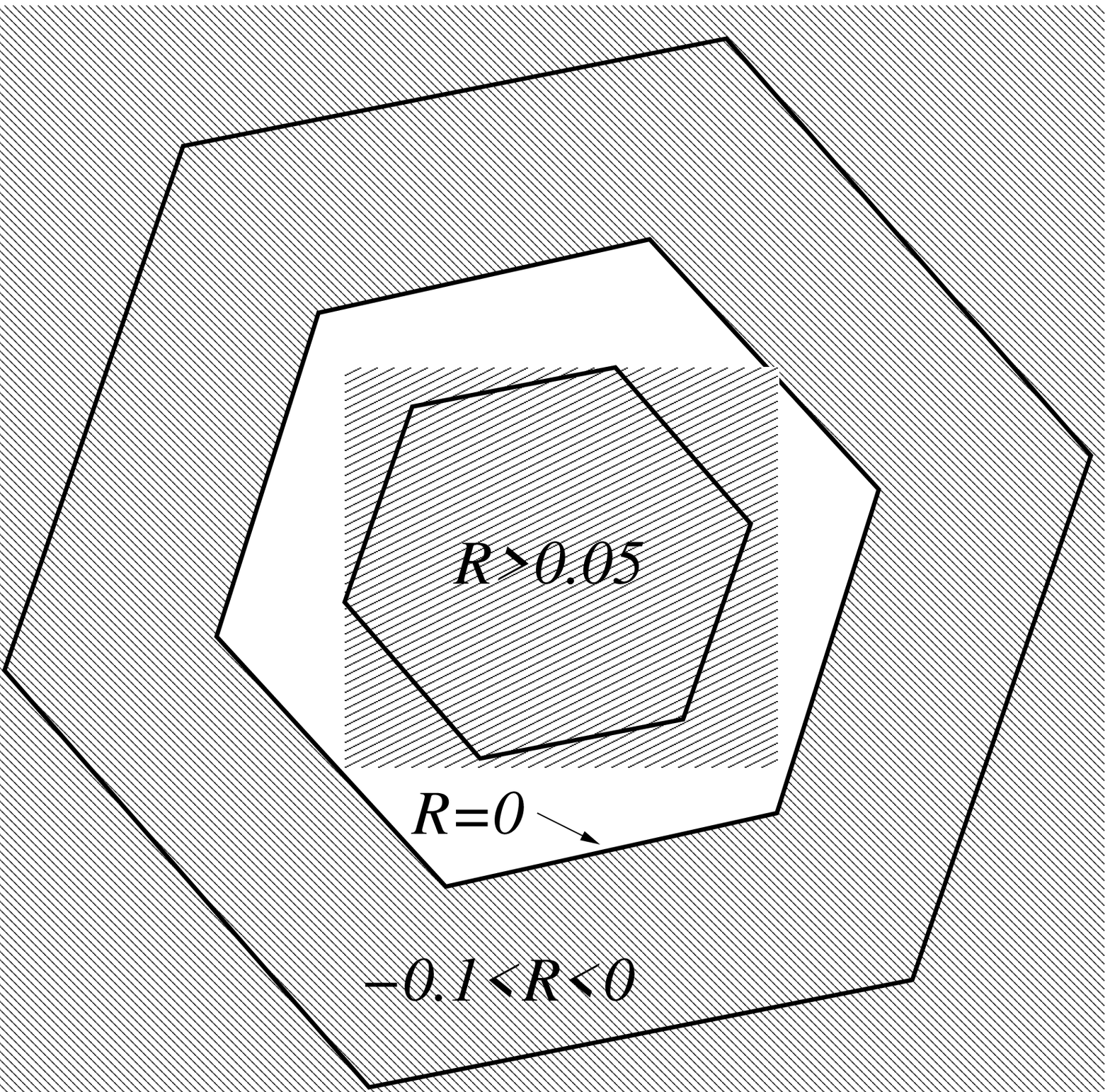}
\caption{
Regions selected by an $R$-cut in two dimensions.}
\label{fig:areas}
\end{figure}

In each spectrometer a track is defined by the four reconstructed
target variables $(\delta,y,\theta,\phi)_\text{tgt}$; 
these variables are defined according to {\tt TRANSPORT} 
\cite{TRANSPORT} conventions.
Rather than attempt to visualize the four-dimensional volumes populated 
by event coordinates for each spectrometer, 
one normally inspects two-dimensional projections, 
such as those shown in Fig.\ \ref{fig:E-polygons} for the electron 
spectrometer.
An event that is safely near the center of the distribution in one
projection may be found near the edge of another where either the 
experimental acceptance or the calibration of the magnetic optics 
may be suboptimal.
Therefore, it is useful to construct a measure of the distance between
event coordinates and the boundary of a multidimensional acceptance
volume that is based upon two-dimensional projections that are more
amenable to visualization.
We employ a variation of the $R$-function method that was originally
developed by V. Rvachev \cite{Rvachev82,Rvachev95} and applied to 
$(e,e^\prime p)$ reactions by M. Rvachev \cite{Rvachev03,Rvachev05}.
We begin by drawing regular polygons that encompass 
most of the pion electroproduction events within each of the 6 
two-dimensional projections formed by pairs of variables.
The two coordinates $(x,y)$ are then rescaled according to
\begin{equation}
x^\prime = \frac{x}{x_{max}-x_{min}} \hspace{1cm} 
y^\prime = \frac{y}{y_{max}-y_{min}}
\end{equation}
where the denominators represent the extremes of the polygonal boundaries
and where {\tt TRANSPORT} conventions ensure that the polygon
is centered near the origin.
For each event with coordinates $(x^\prime,y^\prime)$ in projection $i$,
we define $d_i$ to be plus or minus the minimum distance to a side of
the polygon for that projection, with a positive (negative) sign used 
for points that are inside (outside) the enclosed area.
Figure \ref{fig:polygon} illustrates these signed distances. 
Differences between the units employed for the various transport
variables are then compensated by defining a normalized distance
$\xi_i=d_i/\sqrt{A_i}$ where $A_i$ is the area enclosed by polygon $i$
in terms of normalized coordinates.  
Thus, $\xi_i$ represents the normalized distance of an event from the 
acceptance boundary in projection $i$ with positive signs inside, 
negative signs outside, and zero on the boundary.
The inner region of Fig. \ref{fig:areas} with $\xi_i>0.05$ represents
events that are safely inside the polygonal boundary represented by $\xi_i=0$.
Finally, for each spectrometer we define $R=\pm \min{\{\xi_i\}}$ with a 
positive sign when all $\xi_i$ are positive and a negative sign whenever
an event falls outside the border for any of the projections. 

This method can be applied to each spectrometer independently or
to coincidence data for which kinematic correlations between tracks
in the two spectrometers suppresses accidentals.
For the coincidence method we define $R=\min(R_e,R_p)$ as the smaller
of the $R$-functions for the two spectrometers.
However, the disadvantage of the coincidence method is that the
polygons must be constructed for each kinematical setting independently.
Nevertheless, we chose the coincidence method for this experiment
because there were often substantial accidental populations.
Further details concerning the algorithms for the present implementation
of the $R$-function method can be found in Ref. \cite{Chai-thesis}.

\subsection{Simulation}
\label{sec:simulation}

The differential cross sections for each kinematical bin were
obtained from comparison between experimental and simulated yields.
The simulations used {\tt MCEEP} \cite{mceep} to apply radiative
corrections to a theoretical model, resulting in a six-fold
differential cross section, and to integrate the theoretical
cross section over the experimental acceptance.
{\tt MCEEP} samples the phase space uniformly, over a volume larger
than the experimental acceptance, and evaluates the yield
\begin{equation}
Y_i = {\cal L} \int{ {\cal K}_i } d^6\sigma \otimes {\cal R} 
\end{equation}
where ${\cal L}$ is the luminosity,
${\cal K}_i$ represents the acceptance function for bin $i$, 
$d^6\sigma$ represents the model cross section for each event,
$\otimes$ represents convolution, and 
${\cal R}$ represents resolution functions for quantities measured
in the focal-plane.
Here ${\cal K}_i=1$ if $R>R_\text{cut}$ or 0 otherwise based upon 
target variables that are reconstructed from focal-plane coordinates 
convoluted with resolution functions.

The model cross section was based upon tabulated multipole amplitudes
for MAID2000 \cite{MAID}.
The kinematics for each event were used to interpolate 
the multipole amplitudes with respect to $(W,Q^2)$ and then to compute
the 5-fold differential cross section in the laboratory frame.

The radiative corrections include bremsstrahlung in the target before 
and after scattering, internal soft-photon processes according to 
the Schwinger prescription, and radiation of hard photons using
the Borie-Drechsel \cite{Borie71} prescription with the peaking approximation.
Multiple scattering within the target and windows is included also.
These corrections do not account for polarization effects.
Further details can be found in Refs. \cite{Chai-thesis,mceep}.

\begin{figure}
\centering
\includegraphics[height=5.0in]{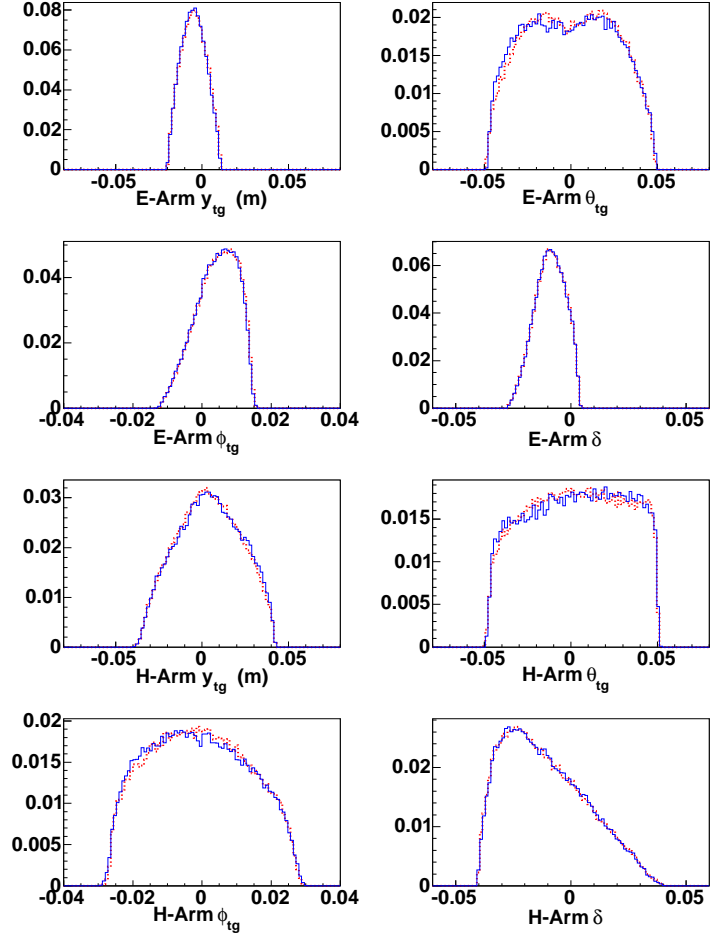}
\caption{(Color online)
Comparison between observed (solid blue) and simulated 
(dotted red) distributions of target variables for 
$\theta_\text{pq}=0$ with $R_\text{cut}>0.05$.
E-arm refers to the electron and H-arm to the hadron spectrometer.}
\label{fig:tgt}
\end{figure}

\begin{figure}
\centering
\includegraphics[height=5.0in]{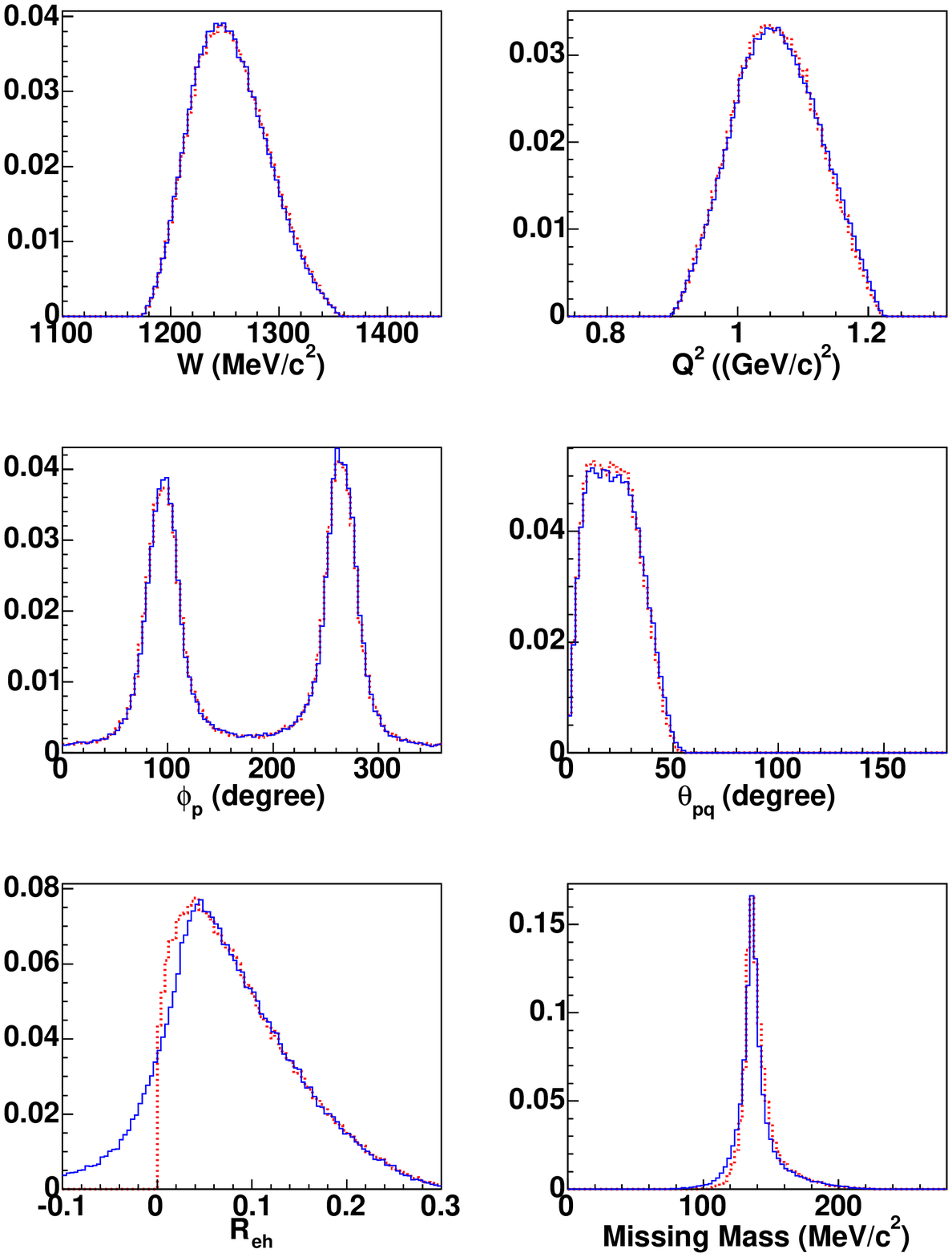}
\caption{(Color online) 
Comparison between observed (solid blue) and simulated 
(dotted red) distributions of kinematical variables for 
$\theta_\text{pq}=0$ with $R_\text{cut}>0.05$.  
The lower left panel shows the uncut distribution of $R_\text{eh}$.}
\label{fig:kin}
\end{figure}

Figure \ref{fig:tgt} compares simulated and experimental distributions
for target variables at $\theta_{cm}=0$ using an acceptance cut $R>0.05$ 
for both, and applying a common normalization factor to the simulation.
Although a slight discrepancy can be observed for $\theta_\text{tg}$
in the electron spectrometer,
the simulation reproduces the experimental distributions very well.
Similarly, Fig. \ref{fig:kin} shows that the simulation also reproduces
the distributions for reaction kinematics very well.
Note that the experimental acceptance function, $R_{eh}$, is shown
without applying the cut. 
The optimal choice for $R_\text{cut}$ is somewhere below the center of 
the plateau in the ratio between experimental and simulated yields.
The systematic uncertainty due to the acceptance function, estimated 
from the flatness of the plateau, is typically less than 1\%.

The missing mass for this reaction is quite sensitive to laboratory angles;
for example, at $\theta_{pq}=-155^\circ$ the sensitivity to electron
angle is 13 MeV/degree for our kinematics.
Thus, comparing the simulation with data for $\theta_{pq}=-155^\circ$
we adjusted the electron angle by $-0.05^\circ$, 
which is well within the survey uncertainty, 
and then find good agreement with the missing mass peaks
for all other settings as well.
Furthermore, the width of the missing mass peak is underestimated by
the simulation unless resolution functions are applied to the track 
coordinates.
We found that Gaussian resolutions with $\sigma=0.5$ mm applied to the hit 
positions in each VDC plane provide good agreement with those widths.

\subsection{Background}
\label{sec:xsec-bkgd}

The acceptance cut $R>0.05$ suppresses the background from the elastic 
radiative tail quite strongly.
The residual contribution is less than $0.4\%$ at $\theta_{pq} = -90^\circ$
and much smaller at $\theta_{pq} = -50^\circ$.
Hence, no corrections for this background were needed for the 
cross section or the unpolarized response functions.
The background due to accidental coincidences was subtracted using
time windows on both sides of the coincidence peak, 
applying the same particle identification and acceptance tests, and 
normalizing by width.

\subsection{Cross section calculation}
\label{sec:xsec-calc}

The virtual photoproduction cross section $\bar{\sigma}$ for a particular
kinematical bin was determined by scaling the model cross section
$\bar{\sigma}_{\rm model}$ 
for that bin, evaluated for bin-centered kinematics, and applying 
various deadtime and efficiency corrections according to
\begin{equation}
\bar{\sigma} = \frac{Y}{Y_\text{MC}} \bar{\sigma}_\text{model} 
\frac{f_\text{CDT} f_\text{EDT} f_\text{abs}}{ \epsilon_\text{trigger} 
\epsilon_\text{track} }
\end{equation}
where 
$Y$ and $Y_\text{MC}$ are the observed and simulated yields,
$f_\text{CDT}$ corrects for computer deadtime, 
$f_\text{EDT}$ corrects for electronics deadtime,
$\epsilon_\text{trigger}$ corrects for trigger efficiency, 
$\epsilon_\text{track} $ corrects for wire chamber and tracking efficiency, 
and $f_\text{abs}$ corrects for proton absorption in materials
between the scattering and detection.
No correction was made for variation of luminosity with current because
no systematic variation was observed in the luminosity monitor for
the currents employed in the cross section measurements (20-60 $\mu$A). 

The computer deadtime was determined by comparing the coincidence scaler
with the number of coincidence events recorded.
In addition, the trigger supervisor has an internal deadtime of approximately
100 ns, such that the electronics deadtime for a 1 MHz rate in the
trigger scintillators is about 10\%.
The electronics deadtime was measured by sending pulser signals to one 
scintillator paddle in each arm and comparing the number of pulser signals
recorded with the number counted by a scaler, 
correcting for computer deadtime.
The dependence of the electronic deadtime upon strobe rate was then
parametrized.
The systematic uncertainty in the correction for electronic deadtime
was estimated to be about 1\% at the highest rates \cite{Jones-EDT}.

The event reconstruction software rejects events with more than one
track in either spectrometer.
For the electron spectrometer 10-12\% of the events contained multiple
tracks while for the proton spectrometer 1-12\% contained multiple tracks
depending upon the momentum and angle settings.
We assume that the fraction of multiple-track events that contain a 
particle that would have satisfied the particle-identification criteria and
other tests is the same as that for single-track events 
and apply corrections for each arm independently.
In addition, we required valid tracks to contain 3-8 hits in each VDC plane. 
For the two settings with significant population by elastic scattering,
the elastic scattering events were excluded from the calculation of 
tracking efficiency to minimize position-dependent effects upon 
trigger efficiency and to improve factorization of the tracking
efficiencies for the two arms.

The triggers in each arm require coincidence between two scintillator
planes and test the track direction.
Thus, the trigger efficiency compares the total number of valid triggers 
with the total number of events with a least one hit in a scintillator.
For the electron arm, we require events in both the numerator and the
denominator to satisfy the Cerenkov test for electrons, and to contain
only one track.
For the proton arm we also use one-track events but use the S0
scintillator instead of the Cerenkov detector.
The net trigger efficiency of approximately 96.7\%, 
with a systematic uncertainty of about 1\%, 
is then the product of the efficiencies for the two arms 

Finally, we used a compilation of proton reaction cross sections to
estimate the probability for proton absorption between scattering and
detection.
The net correction factor varied between 1.008 and 1.017 depending upon
momentum.

\subsection{Cross section results}
\label{sec:xsec-results}

We assume that the ratio between observed and simulated yields over the
acceptance for a kinematic bin is very nearly equal to the ratio between
actual and model differential cross sections for the central kinematics
of that bin.
The accuracy of this assumption depends, of course, upon the bin size 
and the curvature of the differential cross section with respect to the
binned variables.
Events were accepted for $Q^2 = 1.0 \pm 0.2$ (GeV/$c$)$^2$.
We used 10 bins in $W$ between 1.17 and 1.35 GeV of width $\pm 0.01$ GeV,  
20 bins of $x$ between -0.95 and +0.95 in steps of 0.1, 
and 12 bins of $\phi$ of $30^\circ$ width. 
After dropping bins with negligible acceptance, approximately 1140 data 
were obtained for both differential cross section and beam analyzing power.
These data are reported for central kinematics.

\begin{figure*}
\centering
\includegraphics[width=4.0in,angle=90]{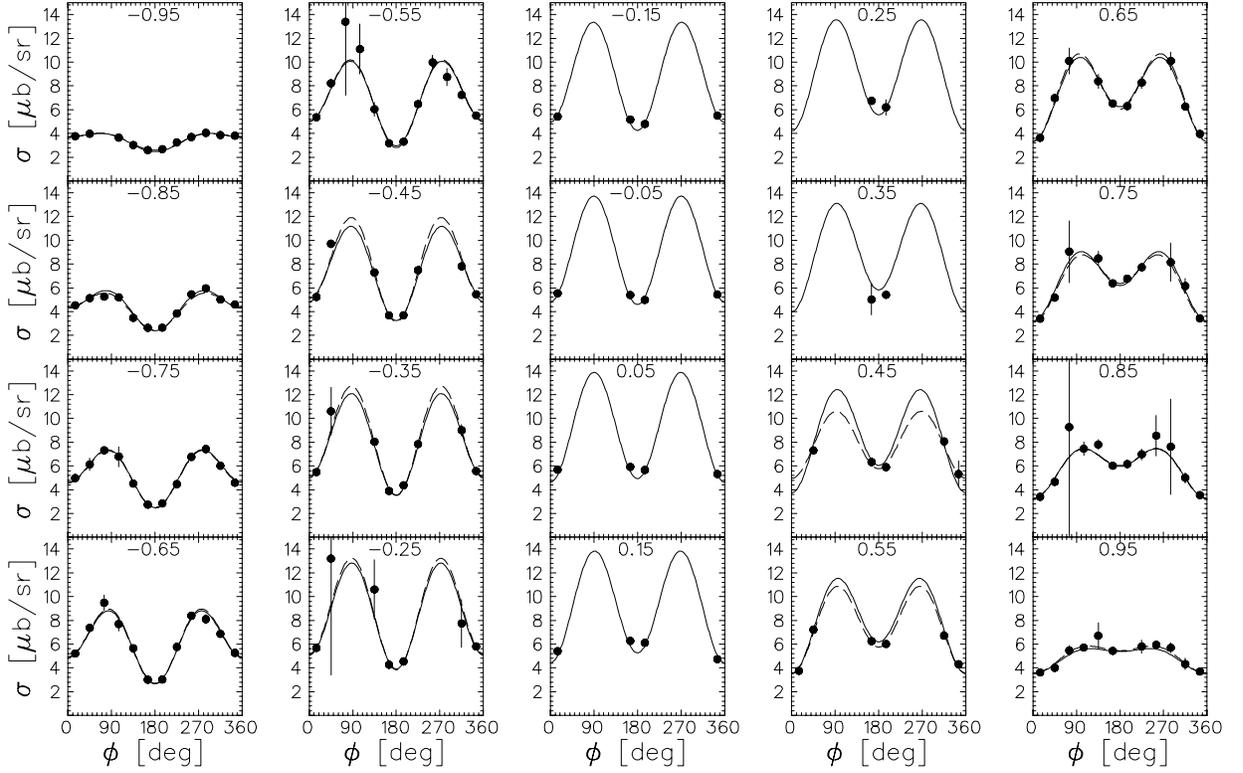}
\caption{Fits to the $\phi$ dependence of unpolarized cross section data with
$(W,Q^2)=(1.23,1.0)$.  Each panel is labeled by the central $x$.
Solid curves fit Legendre coefficients to the entire data set shown while 
dashed curves fit response functions within each panel independently.}
\label{fig:sigma_phi_1230}
\end{figure*}

Figure \ref{fig:sigma_phi_1230} shows the $\phi$ dependence of the
differential cross section for each $x$ bin with $(W,Q^2)=(1.23,1.0)$.
The dashed curves fit $R_{L+T}$, $R_{LT}$, and $R_{TT}$ for each $(x,W,Q^2)$
independently to the $\phi$ dependence of Eq. (\ref{eq:sigma_rsfns}).
Unfortunately, this procedure did not permit model-independent separation 
of $R_{TT}$ from $R_{L+T}$ for $x \approx 0$ because correlations were too
large given the present $\phi$ acceptance.
The solid curves fit Legendre coefficients to the $(x,\phi)$ dependence, 
thereby imposing a smooth $x$ dependence that is not required by the 
extraction of unpolarized response functions.
Nevertheless, both methods fit the data well and agree within the
uncertainties estimated from covariances.
Similar figures can be made for each $(W,Q^2)$ bin, but are too numerous
to display here.

\begin{figure*}
\centering
\includegraphics[width=2.0in]{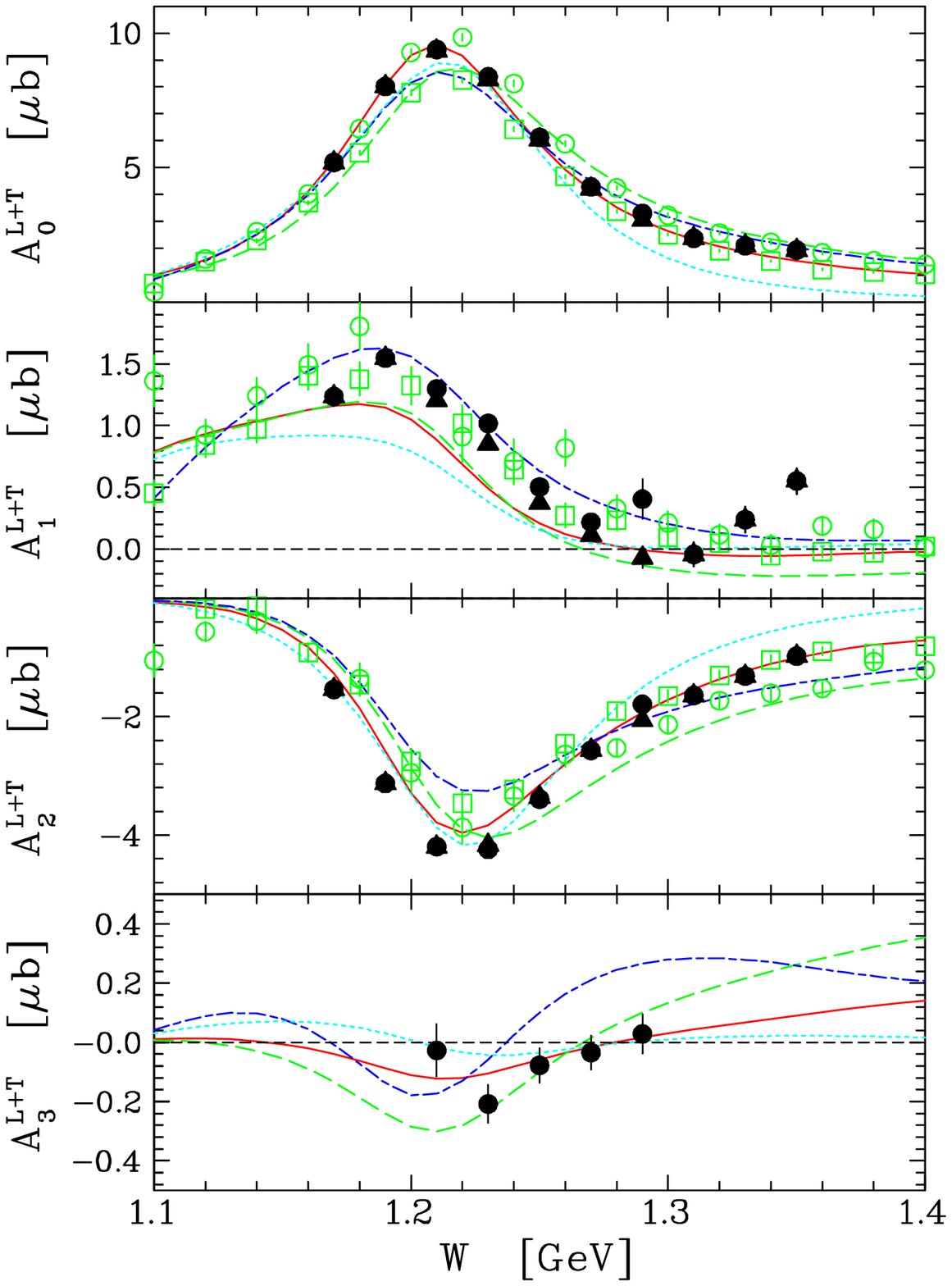}
\includegraphics[width=2.0in]{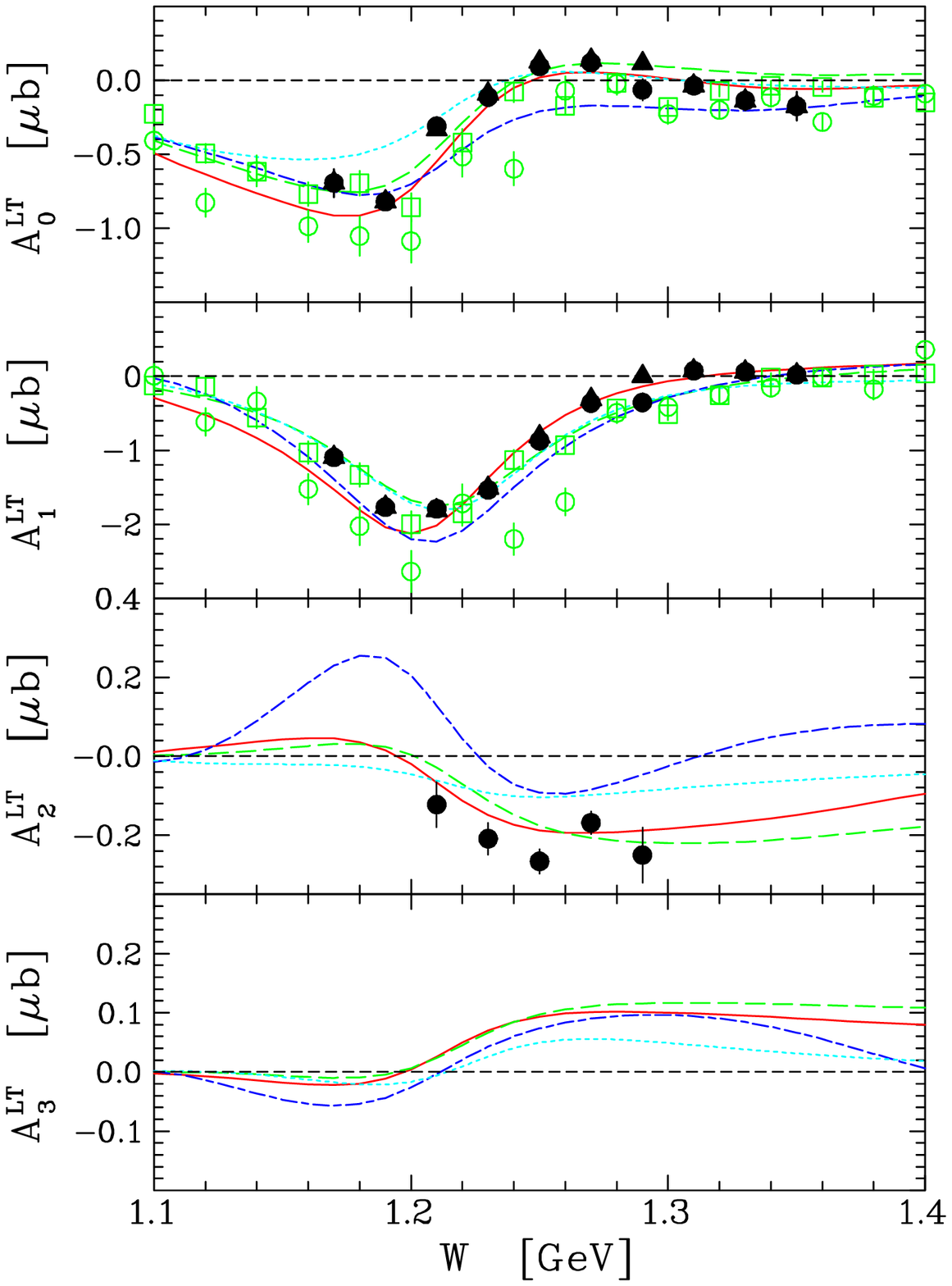}
\includegraphics[width=2.0in]{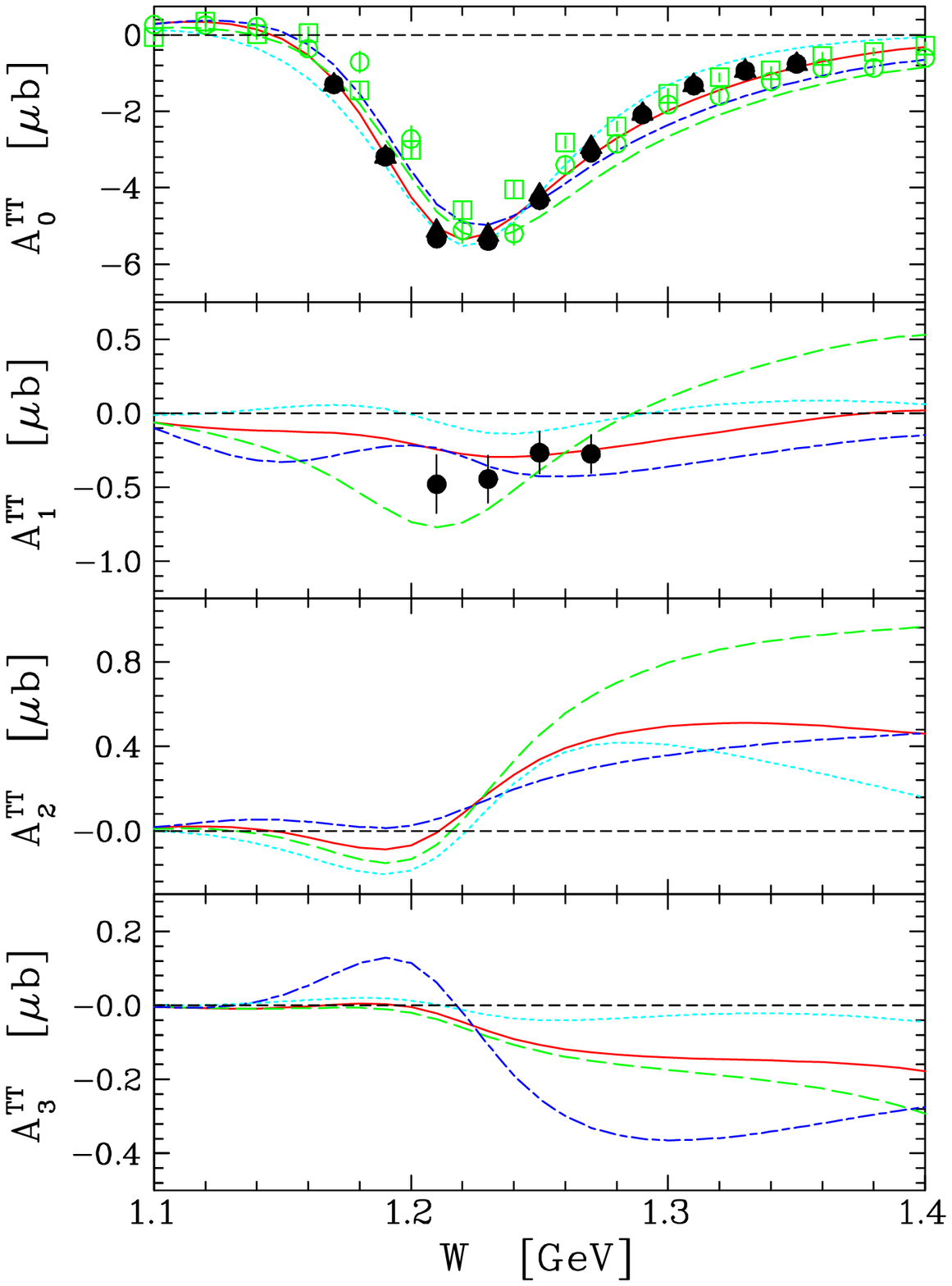}
\caption{(Color online)
Legendre coefficients fit to differential cross sections at 
$Q^2 = 1.0$ (GeV/$c$)$^2$.
These quantities are defined in Eq. (\protect{\ref{eq:RLegendre}}) 
where superscripts identify the response function and subscripts
the degree of the Legendre polynomial.
The filled triangles use the $sp$ truncation while in the central $W$ range
the filled circles include an extra term for each response function.
CLAS results scaled to $Q^2 = 1.0$ (GeV/$c$)$^2$ using a dipole form factor
are shown as open green circles for $Q^2 = 0.9$ (GeV/$c$)$^2$ and 
open green squares for $Q^2 = 1.15$ (GeV/$c$)$^2$; 
only data for a beam energy of 2.445 GeV are shown. 
These results are compared with MAID2003 (red solid), DMT (green dashed), 
SAID (blue dash-dotted), and SL (cyan dotted) calculations.}
\label{fig:compare_clas}
\end{figure*}

The Legendre coefficients fit to the unpolarized cross section for
$Q^2 = 1.0$ (GeV/$c$)$^2$ are compared in Fig. \ref{fig:compare_clas}
with expansion coefficients for calculations based upon the 
MAID2003, DMT, SAID, and SL models obtained by inversion of 
Eq. (\ref{eq:RLegendre}).
(More details about model calculations are given in Sec. 
\ref{sec:comparisons}).
Although these calculations suggest that the $sp$ truncation 
is probably adequate in the immediate vicinity of the $\Delta$
resonance, it appears that additional terms may be necessary
elsewhere.
Therefore, in addition to fits based upon $sp$ truncation, we
show fits with one additional free parameter for each response 
function within the the central $W$ range where the angular coverage
and statistical precision are best.
The models reproduce the even $L+T$ coefficients relatively well, 
although the $W$ dependence of the SAID calculation for $A^{L+T}_2$
is somewhat too flat.
The models also reproduce the low-order coefficients for $R_{LT}$ and
$R_{TT}$ relatively well.
For $R_{LT}$ the additional coefficient is determined relatively
well near the middle of the $W$ range and is consistent in both sign
and magnitude with most model calculations.
The resulting curvature in $R_{LT}$ is small but definitely visible.
Similarly, the data are consistent with the small negative linear 
coefficient predicted for $R_{TT}$ but cannot determine higher-order
coefficients.
The additional term for $R_{L+T}$ appears to be rather weak.
The extra terms have very small effects upon the fitted value for
lower coefficients of the same parity, but negligible effect upon those
of opposite parity.
Note that $A^{L+T}_1$ is appreciably stronger than MAID, DMT, or SL
predictions and exhibits an upturn for $W \gtrsim 1.3$ GeV that is absent
from those models and that this result is not affected by the inclusion 
of terms beyond $sp$ truncation.

Figure \ref{fig:compare_clas} also shows similar results obtained
by Joo {\it et al.}\ \cite{Joo02,CLAS98} at Jefferson Laboratory using CLAS.
Here we show their results for the higher beam energy, 2.445 GeV, that
has better statistical precision.
However, the two experiments used different binnings with respect to $Q^2$.
For the purposes of this comparison, we assume that the Legendre
coefficients are proportional to the square of a dipole form factor and 
rescale the CLAS data for $Q^2 = 0.9$ and 1.15 (GeV/$c$)$^2$ to a 
common value of 1.0 (GeV/$c$)$^2$.
Note, however, that for a given $W$, $\epsilon$ is higher for 
$Q^2 = 0.9$ than for 1.15 (GeV/$c$)$^2$ and that $\epsilon$ for 
our experiment is higher than for either of the CLAS data sets.
We observe good qualitative agreement between these data sets,
but there are significant differences in detail.
For example, our $A_2^{L+T}$ is systematically stronger for low $W$
than in CLAS data.
Nor does the form-factor scaling prescription bring the two 
CLAS data sets for $LT$ coefficients in agreement with each 
other, but the higher $Q^2$ data also appear to show more scatter.
On the other hand, the curvature we see in the $x$ dependence of
$R_{LT}$ clearly requires at least one term beyond $sp$ truncation;
this is shown in more detail in Sec. \ref{sec:legendre}.
Perhaps the omission of $A_2^{LT}$ from the CLAS analysis is 
partly responsible for discrepancies in the lower coefficients.

\section{Polarization analysis}
\label{sec:polarization}

\subsection{Polarization Analysis using Likelihood Method}
\label{sec:PALM-P}

Let $\vec{T}=(T_t,T_n,T_\ell)=\vec{P}+h P_e\vec{P}^\prime$ represent
the proton recoil polarization at the target in the $\pi N$ center of 
mass system, 
where $h$ denotes the sign of the electron helicity and $P_e$ is
the magnitude of the beam polarization,
and let $\vec{F}=(F_x,F_y,F_z)$ represent the polarization
at the focal-plane polarimeter with $\hat{z}$ along the nucleon
momentum and $\hat{y}$ leftward with respect to the vertical plane 
containing the nucleon momentum.
These vectors are related by a spin transport matrix $S$, representing
a sequence of transformations from the target $cm$ frame to the local 
FPP coordinate system, such that $F=S T$.
The spin transport matrix is evaluated for each event.
Details of the individual transformations are given in 
Appendix \ref{appendix:spin}

The polarization components at the target were extracted from the azimuthal
distribution for scattering by the FPP analyzer using the method of
maximum likelihood \cite{Besset79}.
The likelihood function takes the form
\begin{equation}
\label{eq:P_likelihood}
L = \prod_{\rm events} \frac{1}{2\pi} \left( 1 + \xi - 
\varepsilon_x \sin{\phi_\text{fpp}} + \varepsilon_y \cos{\phi_\text{fpp}}
\right)
\end{equation}
of a product of the scattering probabilities for each event that
satisfies the selection criteria for a given kinematical bin.   
The azimuthal scattering angle $\phi_\text{fpp}$ is measured 
counterclockwise from the final $\hat{x}$ axis for each event and
$\xi$ represents the false (instrumental) asymmetry, 
discussed in Sec. \ref{sec:fa}. 
The $\varepsilon$ coefficients are given by
\begin{equation}
\label{eq:epsilon}
\varepsilon_\alpha =  A_y(\theta_\text{fpp})
\sum_{\beta} S_{\alpha\beta} T_\beta
\end{equation}
where 
$A_y(\theta_\text{fpp})$ is the analyzing power for polar scattering angle
$\theta_\text{fpp}$, 
$\alpha\in\{x,y,z\}$ identifies the polarization components at the FPP,
$\beta\in\{t,n,\ell\}$ identifies components of $\vec{T}$ at the target
in the $\pi N$ cm frame, 
and $S_{\alpha\beta}$ are elements of the spin-transport matrix
given by Eq. (\ref{eq:spin}).
Although the scattering probability for each event is independent of the
longitudinal polarization, the variation of spin transport within
the experimental acceptance offers access to all three components
of polarization at the target.

An iterative method for extracting the target polarization that maximize
the likelihood is presented in Appendix \ref{sec:maximization}.
If the asymmetries $(\varepsilon_x,\varepsilon_y,\xi)$ are small, the
problem reduces to the linear system
\begin{equation}
V = \Lambda \cdot R
\end{equation}
where
\begin{equation}
R=(P_t,P_n,P_\ell,P^\prime_t,P^\prime_n,P^\prime_\ell) 
\end{equation}
is the result vector,
\begin{equation}
V_\alpha = \sum_i \frac{\lambda_{i\alpha}}{1+\xi_i}
\end{equation}
is an element of the measurement vector, and
\begin{equation}
\Lambda_{\alpha,\beta} = \sum_{i} 
\frac{\lambda_{i\alpha}}{1+\xi_i}\frac{\lambda_{i\beta}}{1+\xi_i}
\end{equation}
is an element of the design matrix where the Greek indices
$\{\alpha,\beta=1,6\}$ correspond to elements of the result
vector and 
the Latin index $i$ enumerates events that satisfy the selection
criteria for a particular kinematical bin.
Elements of the result vector represent acceptance-averaged
components of recoil polarization that are taken to be 
constant within each kinematical bin.
Conversely, the elements of the measurement vector and design
matrix accumulate contributions 
\begin{eqnarray*}
\xi &=& a_0 \sin{\phi_{\rm fpp}} + b_0 \cos{\phi_{\rm fpp}} +
c_0 \sin{2\phi_{\rm fpp}} + d_0 \cos{2\phi_{\rm fpp}} \\
\lambda_1 &=& A(\theta_{\rm fpp}) (S_{yt}\cos{\phi_{\rm fpp}} - S_{xt}\sin{\phi_{\rm fpp}})  \\ 
\lambda_2 &=& A(\theta_{\rm fpp}) (S_{yn}\cos{\phi_{\rm fpp}} - S_{xn}\sin{\phi_{\rm fpp}}) \\
\lambda_3 &=& A(\theta_{\rm fpp}) (S_{y\ell}\cos{\phi_{\rm fpp}} - S_{x\ell}\sin{\phi_{\rm fpp}}) \\
\lambda_4 &=& h P_e \lambda_1 \\
\lambda_5 &=& h P_e \lambda_2 \\
\lambda_6 &=& h P_e \lambda_3
\end{eqnarray*}
that are evaluated independently for each event, where the
event indices have been suppressed.

\begin{figure*}
\centering
\includegraphics[width=2.5in]{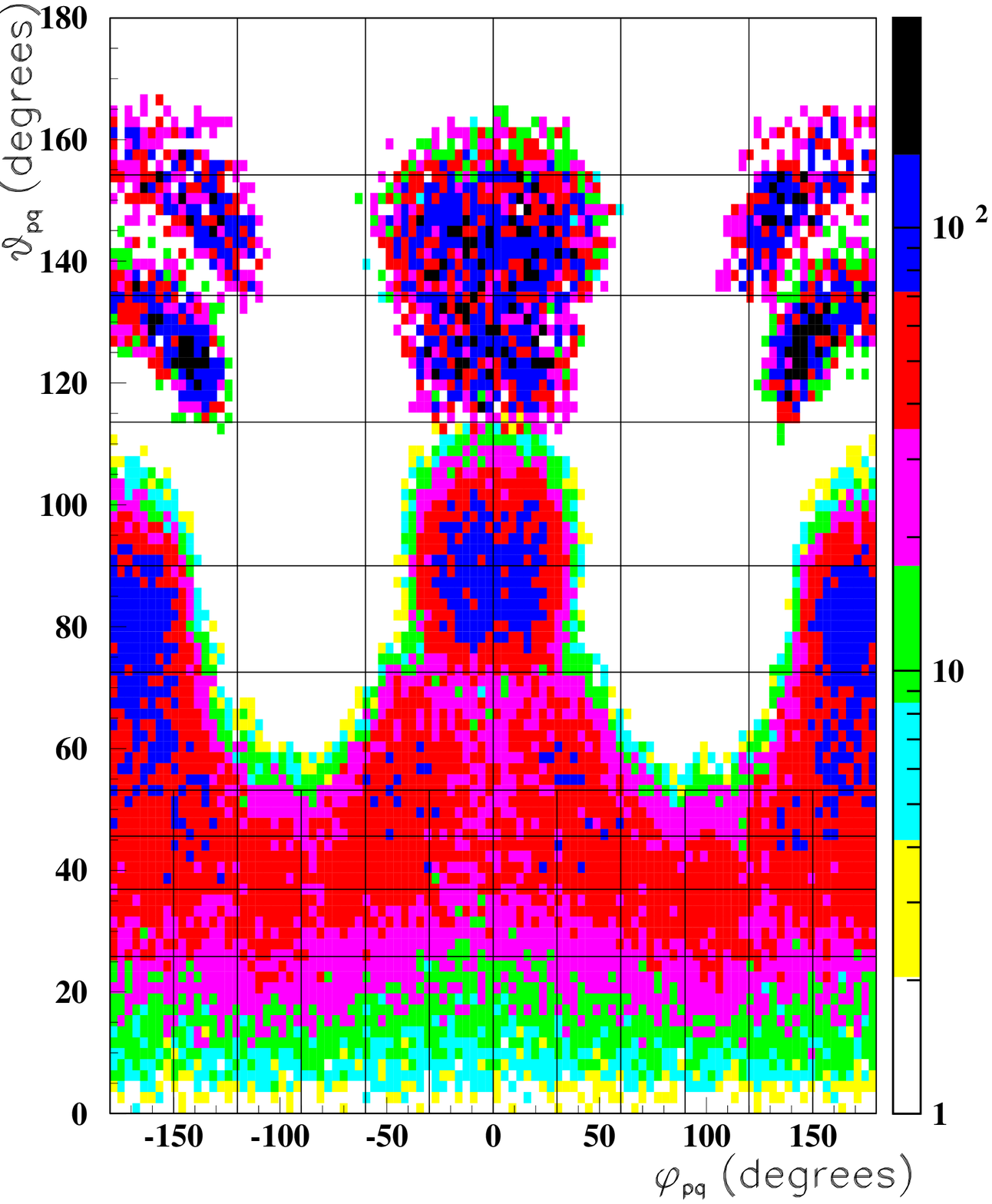}
\includegraphics[width=2.5in]{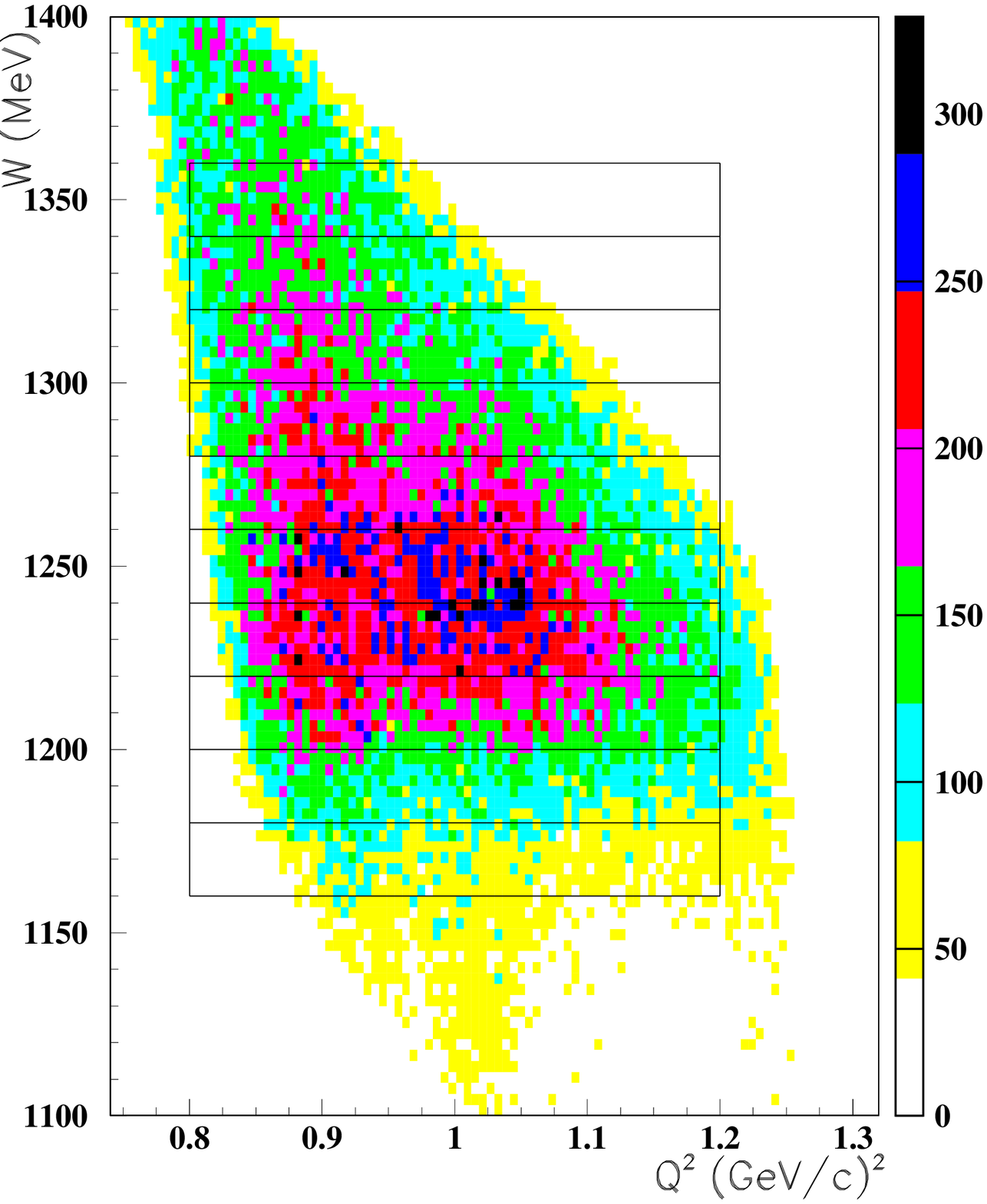}
\caption{(Color online) 
Kinematical acceptance and the binning for the polarization analysis.  
Left: angular acceptance for ${\theta_N,\phi_N}$ in the cm frame.
Right: $(W,Q^2)$ acceptance.}
\label{fig:acceptance}
\end{figure*}

\subsection{Extraction of polarized response functions using Likelihood Method}
\label{sec:PALM-R}

Binning with respect to $\phi$ can be avoided by using
Eq. (\ref{eq:obs}) to express the likelihood function 
\begin{equation}
\label{eq:R_likelihood}
L = \prod_{\rm events} \frac{1}{2\pi} \left(
1 + \xi + \eta \cdot R
\right)
\end{equation}
in terms of response functions 
\begin{equation}
R = (R^t_{LT},R^t_{TT},R^n_{L+T},R^n_{LT},R^n_{TT},
R^\ell_{LT},R^\ell_{TT},R^{\prime t}_{LT},R^{\prime t}_{TT},
R^{\prime n}_{LT},R^{\prime \ell}_{LT},R^{\prime \ell}_{TT})
\end{equation}
with coefficients 
\begin{eqnarray*}
 \bar{\sigma} \, \eta_{1}  &=& \lambda_1 \nu_0\nu_{LT} \sin{\phi} \\ 
 \bar{\sigma} \, \eta_{2}  &=& \lambda_1 \nu_0\nu_{TT} \sin{\theta} \sin{2\phi} \\
 \bar{\sigma} \, \eta_{3}  &=& \lambda_2 \nu_0\nu_T \sin{\theta} \\
 \bar{\sigma} \, \eta_{4}  &=& \lambda_2 \nu_0\nu_{LT} \cos{\phi} \\ 
 \bar{\sigma} \, \eta_{5}  &=& \lambda_2 \nu_0\nu_{TT} \sin{\theta} \cos{2\phi} \\ 
 \bar{\sigma} \, \eta_{6}  &=& \lambda_3 \nu_0\nu_{LT} \sin{\theta} \sin{\phi} \\ 
 \bar{\sigma} \, \eta_{7}  &=& \lambda_3 \nu_0\nu_{TT} \sin^2{\theta} \sin{2\phi} \\ 
 \bar{\sigma} \, \eta_{8}  &=& \lambda_4 \nu_0\nu^\prime_{LT} \cos{\phi} \\ 
 \bar{\sigma} \, \eta_{9}  &=& \lambda_4 \nu_0\nu^\prime_{TT} \sin{\theta} \\ 
 \bar{\sigma} \, \eta_{10} &=& \lambda_5 \nu_0\nu^\prime_{LT} \sin{\phi} \\ 
 \bar{\sigma} \, \eta_{11} &=& \lambda_6 \nu_0\nu^\prime_{LT} \sin{\theta} \cos{\phi} \\ 
 \bar{\sigma} \, \eta_{12} &=& \lambda_6 \nu_0\nu^\prime_{TT} 
\end{eqnarray*}
that incorporate the azimuthal dependencies event by event.
The coefficients for response functions depend upon the 
unpolarized differential cross section for each event, 
which varies within the kinematical bin.
This cross section was obtained by scaling the model cross section (MAID2000)
calculated at the event kinematics by the ratio between the $sp$ Legendre 
fit to the experimental cross section and the model cross section for the
central kinematics of the bin.
This Legendre fit is discussed in Sec.\ \ref{sec:xsec-results}.
The Legendre parametrization can sometimes produce nonpositive cross sections
for some events with kinematics at the edges of the acceptance; for those
events we simply use the MAID2000 cross section and recognize that these
extreme kinematics contribute very little to acceptance-averaged quantities
anyway.

\subsection{Track reconstruction and selection}

The chambers were aligned with respect to each other and the VDCs using
{\it straight-through} events obtained by removing the carbon analyzer.
The track reconstruction algorithms are described in Ref. \cite{Roche-thesis}.
For our purposes it is sufficient to note that the hit multiplicity within the
straw chambers is sufficient to define tracks before and after the carbon
analyzer.
Thus, we can impose a requirement that the scattering vertex lie within 
the carbon plates used for a particular measurement.
We also require that the polar scattering angle be in the range
$5^\circ \leq \theta_\text{fpp} \leq 20^\circ$, where the lower limit
enhances the analyzing power by suppressing unpolarized Coulomb 
scattering and the upper limit keeps instrumental asymmetries small.
Finally, to minimize false asymmetries due to the finite size of the rear
chambers, we impose a {\it cone test} that demands that the entire cone  
subtended by the polar scattering angle for each track intercepts both 
rear chambers.
The rear chambers are actually large enough that only a few percent
of the events in the accepted $\theta_\text{fpp}$ range
fail the cone test.

\subsection{Calibration}

We fitted an extension of the McNaughton parametrization \cite{McNaughton85} 
of the $p+^{12}$C analyzing power using earlier Hall A data supplemented
by new measurements of elastic scattering by the proton for momenta of 
0.818, 1.066, 1.188, and 1.378 GeV/$c$ in order to provide analyzing power 
data closer to some of the present kinematical settings.
Our measurements of $G_{Ep}/G_{Mp}$ at $Q^2 = 1.0$ and 1.4 (GeV/$c$)$^2$ 
are in good agreement with those of Ref. \cite{MKJones00}. 

\subsection{False asymmetry}
\label{sec:fa}

The one-photon exchange approximation predicts that the helicity-independent
recoil polarization for elastic electron-proton scattering vanishes.
Assuming that the two-photon contribution is negligible, we used
this reaction to measure the false instrumental asymmetries arising
from misalignment, detector or tracking inefficiencies, 
variations of pathlengths in the analyzer, 
and other mechanisms.
We express the detection probability in the form
\begin{eqnarray}
f_h(\theta_\text{fpp},\phi_\text{fpp}) &=& 
f_0 (\theta_\text{fpp},\phi_\text{fpp}) \frac{1}{2\pi}
\left( 1 - h \varepsilon_x \sin{\phi_\text{fpp}} 
+ h \varepsilon_y \cos{\phi_\text{fpp}}  \right. \nonumber \\
&+& \left.
a_0 \sin{\phi_\text{fpp}}  + b_0 \cos{\phi_\text{fpp}} + 
c_0 \sin{2\phi_\text{fpp}} + d_0 \cos{2\phi_\text{fpp}} \right)
\end{eqnarray}
where the coefficients $(a_0,b_0,c_0,d_0)$ parametrize the false asymmetry
while the coefficients $(\varepsilon_x,\varepsilon_y)$ depend upon the
helicity-dependent recoil polarization and the FPP analyzing power.
Thus, the false asymmetry coefficients are obtained by Fourier analysis of 
\begin{equation}
\frac{f_+ + f_-}{2f_0} =  \frac{1}{2\pi}(1+\xi) =
\frac{1}{2\pi}
\left( 1 + a_0 \sin{\phi_\text{fpp}} + b_0 \cos{\phi_\text{fpp}} + 
c_0 \sin{2\phi_\text{fpp}} + d_0 \cos{2\phi_\text{fpp}} \right)
\end{equation}

\begin{figure}
\centering
\includegraphics[angle=-90,width=4.0in]{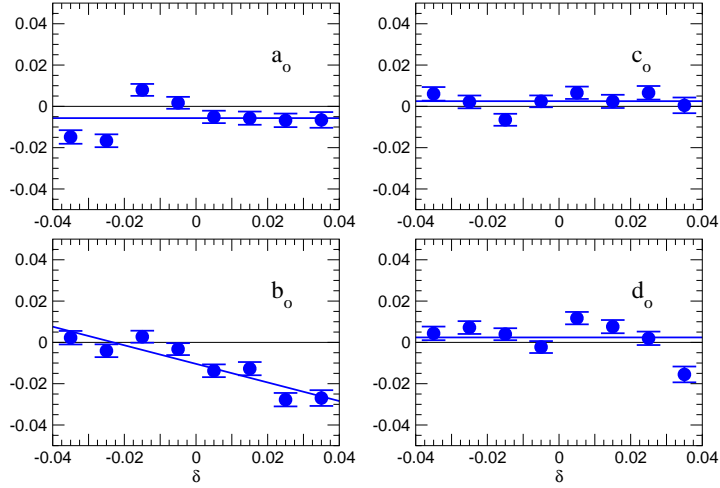}
\caption{(Color online)
The dependence of measured Fourier coefficients for false asymmetry
upon spectrometer relative momentum are shown with average values 
($a_0$, $c_0$, $d_0$) or a linear fit ($b_0$) as blue lines.}
\label{fig:fa}
\end{figure}

Data for elastic scattering were taken at five proton momenta between
0.785 and 0.851 GeV/$c$ and at 1.066 and 1.188 GeV/$c$.  
The dependence of false asymmetries upon $\delta$ are shown in
Fig. \ref{fig:fa} for $5^\circ < \theta_\text{fpp} < 20^\circ$.
Only the coefficient of $\cos{\phi_\text{fpp}}$ shows a significant 
dependence upon $\delta$ that can be attributed, in part, 
to its correlation with the average pathlength of scattered particles
in the carbon analyzer.
This dependence was parametrized by a linear function.
The other three coefficients are essentially independent of $\delta$ with
average values less than $1\%$.
No significant dependence upon proton central momentum is apparent over 
this range.

\subsection{Background subtraction}
\label{sec:pol-bkgd}

The polarization response functions were corrected for two types of 
background: the elastic radiative tail and accidental coincidences.
Corrections for the unresolved contribution of the elastic radiative 
tail were made using the likelihood function because
those contributions varied strongly with both $\theta$ and $\phi$.
Thus, we generalize Eq. (\ref{eq:epsilon}) to
\begin{equation}
\label{eq:epsilon_bg}
\varepsilon_\alpha =  A_y(\theta_{\rm fpp})
\sum_{\beta} (f_1 S^{(1)}_{\alpha\beta} T^{(1)}_\beta + 
              f_2 S^{(2)}_{\alpha\beta} T^{(2)}_\beta)
\end{equation}
where $f_1+f_2=1$ and where $T^{(1)}$ is the polarization for
$\Delta$ excitation and $T^{(2)}$ is the polarization for elastic scattering.
Note that the spin transformation matrix $S^{(2)}$ for elastic scattering 
differs from $S^{(1)}$ for pion production because the polarization vectors
for the two reactions are normally evaluated in different frames.
Thus, $S^{(2)}$ omits $R_W$ (see Appendix \ref{appendix:spin}) and assumes 
that the proton emerges parallel to $\vec{q}$.
The relative weights depend upon the $(W,Q^2,x,\phi)$ bin and were obtained 
by fitting the missing-mass distributions for each bin with appreciable
elastic contamination.
This contribution is actually very small and is only visible for the
$\theta_\text{cm} = -90^\circ$ setting.
Generalization of the likelihood formula, Eq. (\ref{eq:R_likelihood}),
is straightforward.
The elastic polarizations were computed from the parametrizations of
$G_{Ep}$ and $G_{Mp}$ found in Refs. \cite{Gayou02,Brash02}, 
but the results are insensitive to the small differences between models.

Accidental background was subtracted by analyzing both in-time and out-time
events in the same manner.
For polarization we obtain
\begin{subequations}
\begin{eqnarray}
P &=& \frac{P_p - r P_b}{1-r} \\
(\delta  P)^2 &=& \frac{ (\delta P_p)^2 + (r \delta P_b)^2}{(1-r)^2}
\end{eqnarray}
\end{subequations}
where $P_p \pm \delta P_p$ is the measurement for the in-time region,
$P_b \pm \delta P_b$ is the result for the out-time region, and
$r$ is the ratio between the widths of these regions.
Similarly, for response functions we obtain
\begin{subequations}
\begin{eqnarray}
R &=& R_p - r R_b \\
(\delta R)^2 &=& (\delta R_p)^2 + (r \delta R_b)^2
\end{eqnarray}
\end{subequations}
where $R_p \pm \delta R_p$ and $R_b \pm \delta R_b$ are obtained for
in-time and out-time regions, respectively.
The effect of background subtraction is generally difficult to discern
in standard figures and is always much less than the statistical uncertainty
in these measurements.

\subsection{Pseudodata tests}
\label{sec:pseudodata}

The analysis procedures were tested using pseudodata.
For each accepted event, response functions and polarizations at the 
target were computed based upon the MAID2000 model.
The observed polar scattering angle $\theta_{\rm fpp}$ was retained but
the azimuthal scattering angle $\phi_{\rm fpp}$ was sampled randomly.
This value of $\phi_{\rm fpp}$ was retained if the likelihood $L$ calculated 
according to Eq. (\ref{eq:P_likelihood}) was greater than the next random 
number thrown and rejected otherwise.
This procedure was iterated until a value of $\phi_{\rm fpp}$ was selected.  
Contributions to $V$ and $\Lambda$ were then accumulated
and the pseudodata were analyzed in the same manner as real data.

These tests demonstrate that model input for response functions is 
recovered within statistical uncertainties, but that there are sometimes 
inconsistencies in the $\phi$ dependence of polarization data.
This problem arises because relatively large bins in $\phi$ are needed
to obtain useful statistical precision, but some of the 
spin-transport matrix elements can exhibit broad distributions with
respect to other variables in part due to kinematic focusing in the 
lab frame.
Under those conditions the acceptance-averaged polarization can differ
appreciably from model values for the central kinematics of a bin.
These difficulties are much smaller for response functions because 
binning with respect to $\phi$ is not necessary; all $\phi$ values
contribute to the determination of a response functions and their
coefficients are evaluated properly for each event.
Explicit eventwise weighting with the leading factors of 
$\sin{\theta}$ also reduces the effects of acceptance averaging
on the response functions as defined in Eq. (\ref{eq:obs}). 
Therefore, we focus upon the response-function data and do not consider
polarization binned with respect to $\phi$ further.
A more detailed report on the pseudodata analysis is provided in
Ref. \cite{e91011_pseudodata}.

\subsection{Acceptance averaging}
\label{sec:kincorr}

Multipole amplitudes and Legendre coefficients are functions of $(W,Q^2)$, 
but the acceptance averaged $(\overline{W},\overline{Q^2})$ depend upon $x$.
Consequently, extraction of these quantities from angular distributions 
can be distorted by the $x$ variations of $(\overline{W},\overline{Q^2})$.
Such distortions can artificially enhance terms for large $\ell_\pi$.
Two methods for compensating for such distortions have been tested using both 
pseudodata and real data.
The additive method is based upon the first-order expansion
\begin{equation}
\label{eq:additive-centering}
R(W,Q^2,\bar{x},\bar{\epsilon}) = 
R(\overline{W},\overline{Q^2},\bar{x},\bar{\epsilon}) -
\frac{\partial R}{\partial W} (\overline{W}-W) -
\frac{\partial R}{\partial Q^2} (\overline{Q^2}-Q^2)
\end{equation}
where overlines indicate acceptance averaging and where the derivatives 
are evaluated at central kinematics using a model, such as MAID.
For this experiment $\overline{Q^2}-Q^2$ tends to be much more 
important than $\overline{W}-W$.
Additive kinematical corrections have the advantage that variations 
of both $W$ and $Q^2$ can be accommodated, but this procedure has the 
disadvantage that it relies upon a model and we have no model that 
provides a uniformly good fit to all of the response functions.
While that is not a problem for pseudodata, the use of an inaccurate
model to make kinematical corrections to real data could introduce
more serious errors than it corrects.
Therefore, a second procedure based upon form factors was tested. 
Assuming that all response functions share a common form factor,
and that kinematical corrections are dominated by the $x$-dependence 
of $\overline{Q^2}$, we postulate
\begin{equation}
\label{eq:multiplicative-centering}
R(W,Q^2,\bar{x}) = R(\overline{W},\overline{Q^2},\bar{x}) 
\left( G(Q^2)/G(\overline{Q^2}) \right)^2
\end{equation}
and approximate $G(Q^2)$ by the usual dipole form factor
$G_D(Q^2) = (1+Q^2/\Lambda^2)^{-2}$ where $\Lambda^2 = 0.71$ (GeV/$c$)$^2$.
This multiplicative procedure does not compensate for variations of
$\overline{W}$, but for this experiment these variations are much smaller 
than those for $\overline{Q^2}$.

\begin{figure*}
\centering
\includegraphics[angle=90,width=5.0in]{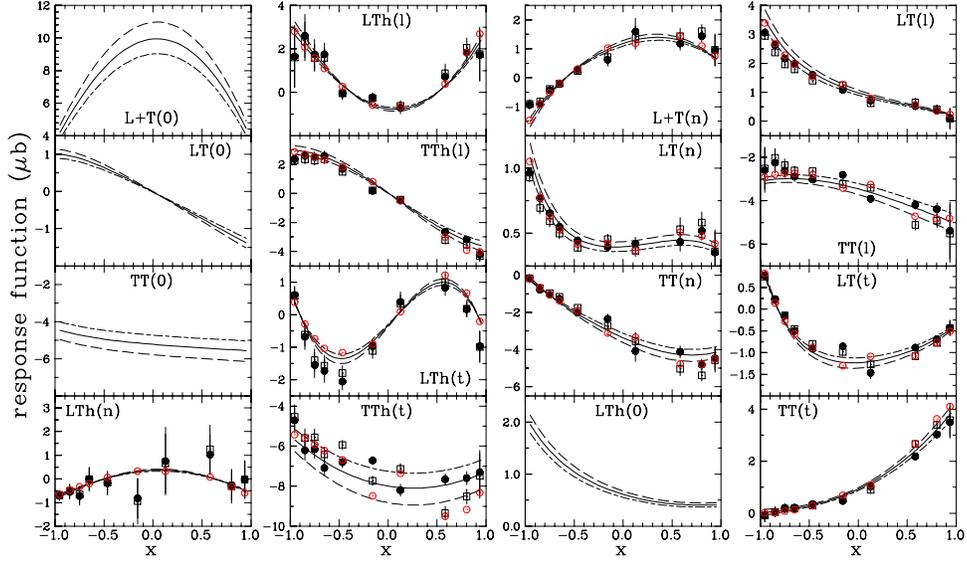}
\caption{(Color online)
Pseudodata data for response functions at
$W = 1.23 \pm 0.01$ GeV and $Q^2 = 1.0 \pm 0.2$ (GeV/$c$)$^2$
are compared with the input model (MAID2000) at the central 
kinematics (solid curves) and with neighboring values of $Q^2$ 
representative of the $x$ dependence of acceptance averaging; 
dashed curves show $Q^2 = 0.94$ and dashed-dotted curves show 
$Q^2 = 1.06$ (GeV/$c$)$^2$.
Acceptance-averaged calculations are shown as red open circles 
and pseudodata as open squares.
Filled circles show pseudodata with multiplicative kinematical 
corrections based upon the dipole form factor.}
\label{fig:multiplicative}
\end{figure*}

Figure \ref{fig:multiplicative} compares pseudodata with multiplicative
kinematical corrections with the model for central kinematics.
The open squares show raw response functions extracted from 
pseudodata while open red circles show acceptance-averaged
response functions from MAID2000.
The agreement between these data sets, modulo statistical 
fluctuations, demonstrates the internal consistency of the
simulation/analysis program.
However, the $x$ dependence of $\overline{Q^2}$ can produce 
significant systematic deviations from the input model 
(solid curves) evaluated at central kinematics, especially
for $R_{TT}^{\prime t}$.
Recognizing that $\overline{Q^2} \approx 0.94$ for $x>0.5$ or
1.06 (GeV/$c$)$^2$ for $x < -0.5$, we observe that the
pseudodata for $R_{TT}^{\prime t}$ do cluster around the
model curve for the appropriate $Q^2$.
Even the abrupt transition across $x=0$ is reproduced.
The solid circles show that ``centered'' pseudodata adjusted
according to Eq. (\ref{eq:multiplicative-centering}) cluster better 
around the model curves for central kinematics.
Therefore, distortion of multipole amplitudes by the $x$
dependence of $\overline{Q^2}$ should be minimized by fitting
centered data.
We find that the multiplicative corrections move the pseudodata 
in the directions indicated by ratios between acceptance averaged 
calculations and those for central kinematics.
The net effect is to reduce the scatter in the pseudodata and to
remove many systematic deviations attributable to the $x$
dependence of $\overline{Q^2}$.
On the other hand, we have the qualitative impression that the 
corrections are sometimes a little too large, although no attempt
has been made to quantify that impression.
One could reduce the size of the multiplicative correction by
increasing the dipole mass $\Lambda$, 
but the $N \rightarrow \Delta$ form factor is actually steeper 
(smaller $\Lambda$) than the standard dipole form factor.
Furthermore, changes to the correction by replacing the dipole
with a parameterization of the $N \rightarrow \Delta$ form factor
are quite small.

Therefore, we adopted the multiplicative correction based upon
the dipole form factor as the standard method for bin centering.
The figures in the remainder of this paper show recoil-polarization
response functions plotted at $\overline{x}$ with bin-centering 
corrections for $\overline{Q^2}$.
Legendre and multipole fits were made to the data in this form. 

\subsection{Systematic uncertainties}
\label{sec:syserr}

\subsubsection{Response functions}
\label{sec:syserr_rsfns}

There are several types of normalization uncertainty that affect the
response-function data.
These include uncertainties in the unpolarized differential cross section
used to normalize the likelihood, the FPP analyzing power, and for 
helicity-dependent responses the beam polarization.
Although these systematic uncertainties do vary to some degree with 
spectrometer settings, beam conditions, and time, those variations are 
small compared with the statistical uncertainties.
Therefore, we believe it sufficient to estimate average systematic
uncertainties for those quantities without tracking the propagation 
of particular settings through event sorting.
The typical systematic uncertainties in the differential cross section data 
are about $\pm 3\%$ point-to-point, so we assume that the uncertainty in
the parametrized cross section used in the likelihood analysis is also
about $\pm 3\%$. 
Similarly, the systematic relative uncertainty for Compton measurements 
of beam polarization is estimated to be about $1.4\%$ \cite{Escoffier01}.
Finally, the relative uncertainties in average analyzing power reported 
by Punjabi {\it et al.} \cite{Punjabi05} are in the $1-2\%$ range.  
Because we do not consider thickness or momentum variations, we adopt a 
fairly conservative estimate of $\delta A_y /A_y = 0.02$.
Therefore, the normalization uncertainties are approximately $\pm 3.6\%$
for helicity-independent or $\pm 3.9\%$ for helicity-dependent
response functions.

The evaluation of other types of systematic error requires replaying the
data subject to a perturbation of one of the analysis parameters.
Thus, the uncertainty due to subtraction of the elastic background was
obtained by comparing replays with and without that subtraction.
Because the contamination fractions binned in $\phi$ were difficult to 
determine, we assumed their relative uncertainties to be $100\%$ and
estimated the corresponding uncertainties in response functions as
\begin{equation}
\delta R_\alpha = |R_\alpha^{(2)} -  R_\alpha^{(1)}|
\end{equation}
where $R_\alpha^{(1)}$ and $R_\alpha^{(2)}$ represent response 
function $\alpha$ with and without elastic subtraction.
Similarly, the uncertainty in corrections for false asymmetry were estimated
as
\begin{equation}
\delta R_\alpha = 0.1 |R_\alpha^{(2)} -  R_\alpha^{(1)}|
\end{equation}
where the relative uncertainty in false asymmetry was estimated to 
be $\pm 10\%$ and is multiplied by the difference in response functions
obtained with and without false asymmetry in the likelihood function.

\begin{figure*}
\centering
\includegraphics[angle=90,width=5in]{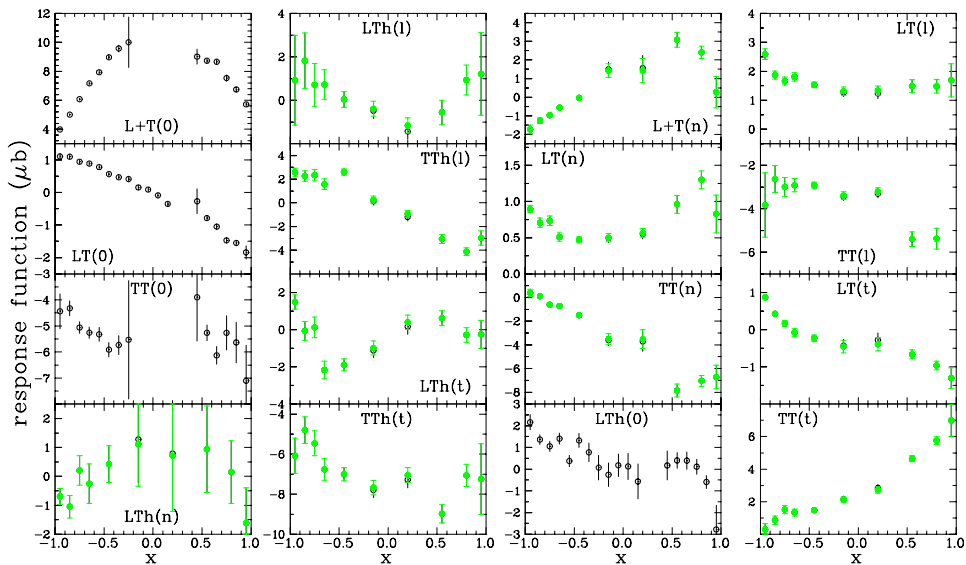}
\caption{(Color online)
Sensitivity of response functions for $W=1.23$ GeV to uncertainties
in the spin rotation matrix.
Open black points were obtained using the COSY model while filled 
green points are based upon the Pentchev model.
For the Pentchev analysis inner error bars are statistical while outer
error bars include systematic errors due to uncertainties in precession
angles and optical matrix elements; however, the systematic errors are
generally too small to see.}
\label{fig:spinrot_1230}
\end{figure*}

A similar procedure was also applied for the spin transport matrix.
The sensitivity of response functions to uncertainties in the
spin rotation matrix is illustrated in Fig. \ref{fig:spinrot_1230} 
for $W=1.23$ GeV.  
The open black points were obtained using the COSY model while
the filled green points were obtained using a simpler geometrical model
by Pentchev in terms of 6 parameters consisting of two trajectory angles 
and four matrix elements coupling spin components \cite{Pentchev03,Punjabi05}.
Obviously, this geometrical model accurately reproduces the COSY model.
Therefore, we can estimate systematic errors in response functions
due to uncertainties in the spin rotation by comparing results obtained
from independent perturbations of each of the 6 parameters of the Pentchev 
model by its estimated uncertainty and combining in quadrature differences 
with respect to nominal parameters.
We use the same systematic uncertainties for those parameters as in
Refs. \cite{Pentchev03,Punjabi05}.
The green error bars in Fig. \ref{fig:spinrot_1230} include an inner
statistical portion shown with endcaps and a total error without endcaps.
However, rarely can one discern the systematic contribution to the 
total bar because the composite contribution of spin rotation errors
is almost always small relative to statistical uncertainties. 

The net systematic uncertainties in response functions, consisting of the 
quadrature sum all contributions discussed in this section, tends to be
dominated by the normalizations and is almost always small compared with
statistical uncertainties.
Many of the figures show statistical uncertainties as inner error bars
with endcaps and total uncertainites as outer error bars without endcaps.
The systematic contributions are occassionally visible for Legendre
coefficients or multipole amplitudes, but are rarely visible for
response functions.

\subsubsection{Legendre coefficients and multipole amplitudes}
\label{sec:syserr_fits}
 
Let $y$ represent a fitted quantity, such as a quadrupole ratio, 
Legendre coefficient, or multipole amplitude and let $\beta$ represent
a calibration parameter or a scale factor applied to one of the 
corrections, such as false asymmetry.
We estimate the systematic uncertainty $(\delta y)_\text{sys}$ by
adding in quadrature numerical derivatives
\begin{equation}
\label{eq:sys}
(\delta y)^2_\text{sys} = 
\sum_i \left(\frac{\partial y}{\partial \beta_i} \delta \beta_i \right)^2 =
\sum_i \left( y(\beta_i+\delta \beta_i) - y(\beta_i) \right)^2
\end{equation}
estimated by performing a series of fits in which each calibration 
parameter is perturbed in turn.
Therefore, Legendre and multipole fits to such data sets begin with the
results of the best fit for nominal calibration parameters and usually
require only a few iterations to determine small displacements of the 
minimum on the $\chi^2$ hypersurface.
We assume that the desired local minimum is related to the best fit 
by a small distortion of the $\chi^2$ hypersurface produced by small
changes in the data set due to perturbation of an analysis parameter.
By starting with the nominal best fit, we minimize the chance that the
fitting procedure might find a different local minimum.  
With enough care in the fitting procedure, we find that changes in 
fitted Legendre coefficients or multipole amplitudes due to variation of 
spin-rotation parameters or omission of false asymmetry or elastic 
subtraction are typically small.

In addition to systematic uncertainties considered in the previous section,
Legendre and multipole analyses also include an estimate of the uncertainty
due to the kinematic or bin-centering correction.
The customary dipole form factor should describe the $Q^2$ dependence
of nonresonant contributions fairly well, but the $N \rightarrow \Delta$
form factor is known to have a more rapid $Q^2$ dependence.
The best description is probably intermediate between these models.
We estimated the uncertainty in fitted Legendre coeffcients and multipole
amplitudes due to the choice of bin-centering form factor by comparing
fits for the dipole and $N \rightarrow \Delta$ form factors, assigning 
a systematic uncertainty equal to the difference between the two fits.
The dipole form factor is given by $G_D(Q^2) = (1 + Q^2/\Lambda^2)^{-2}$
where $\Lambda^2 = 0.71$ (GeV/$c$)$^2$.
For the $N \rightarrow \Delta$ form factor we use the Sato-Lee 
parametrization $G_{N\Delta} = (1 + a Q^2) \exp{(-b Q^2)} G_D$ with
$a = 0.154$ and $b = 0.166$ (GeV/$c$)$^{-2}$ \cite{Sato01}. 
However, the difference between these form factors over the range of
$\overline{Q^2}-Q^2$ for this experiment is too small to produce a 
visible difference in the projected data or fitted angular distributions.

The systematic uncertainties in fitted Legendre coefficients and multipole
amplitudes contain a total of 12 contributions added in quadrature, 
each requiring a fit to the relevant data set.
Variations of the cross section, FPP analyzing power, beam polarization,
bin centering, false asymmetry, and elastic subtraction  are all compared 
with the best fit for data obtained using COSY spin rotation.  
The six contributions to the spin rotation uncertainty are estimated 
using differences with respect to data based upon the nominal Pentchev model.
The net systematic errors in these quantities are generally small 
compared with statistical uncertainties.  
Figures showing fitted quantities with statistical and total errors bars
can be found in the separate reports on Legendre coeffcients 
\cite{e91011_legfit} and multipole amplitudes \cite{e91011_mpamps}.

\subsection{Summary of experimental data}
\label{sec:data}

Near the middle of our $(W,Q^2)$ acceptance,
we have obtained complete angular distributions for 16 response functions,
14 separated plus 2 Rosenbluth combinations for $\varepsilon \sim 0.95$.
The angular coverage and statistical precision are clearly best in the 
central $W$ range, $1.21 \leq W \leq 1.29$ GeV.
Data tables are on deposit with EPAPS 
\footnote{See EPAPS Document No. [number will be inserted by publisher ] 
for tables of data, Legendre coefficients, and multipole amplitudes. 
We also include the three internal reports 
\cite{e91011_pseudodata,e91011_legfit,e91011_mpamps}
cited in the present article.
A direct link to this document may be found in the online article's HTML 
reference section. The document may also be reached via the EPAPS homepage 
(http://www.aip.org/pubservs/epaps.html) or from ftp.aip.org in the directory 
/epaps/. See the EPAPS homepage for more information.}.
These tables give both raw data and bin-centered data with both nominal and 
acceptance-averaged kinematics.
Tables of Legendre coefficients and multipole amplitudes are included also.

\section{Results}
\label{sec:results}

\subsection{Comparison with Models}
\label{sec:comparisons}

In this section we compare our data for response functions with calculations
using four recent models.
We provide very brief summaries of the models and refer to original
sources for more detailed information.
A recent review of these and related models has also been provided by
Burkert and Lee \cite{Burkert04}.

The SAID model \cite{Arndt90a,Arndt02} parametrizes a photoexcitation 
multipole amplitude $A$ in the form
\begin{equation}
\label{eq:SAID}
A = (A_B +A_Q) (1 + i t_{\pi N}) + A_R t_{\pi N}
+ (C + i D)(\Im t_{\pi N} - t_{\pi N}^2)
\end{equation}
where $t_{\pi N}$ is a $t$-matrix fit to $\pi N$ elastic scattering
data that enforces the Fermi-Watson theorem \cite{Watson54} below the 
two-pion threshold, 
$A_R$ is parametrized as a polynomial in $E_\pi$ with the correct
threshold behavior for each partial wave, 
$A_B$ is a partial wave of the pseudoscalar Born amplitude, and 
$A_Q$ is parametrized using Legendre functions of the second kind.
Recent extensions also include energy-dependent polynomials $C$ and $D$.
Electroexcitation amplitudes also include form factors in $Q^2$. 
We are now using the WI03 version of SAID \cite{Arndt03b}. 
Multipole amplitudes were projected from helicity amplitudes 
using the formalism in Appendix \ref{app:amplitudes}.

The Mainz unitary isobar model \cite{Drechsel99}, known as MAID, 
parametrizes resonant contributions to multipole amplitudes using the
Breit-Wigner form
\begin{equation}
\label{eq:MAID-resonance}
A = \bar{A}(Q^2) C_{\pi N} f_{\gamma N}(W) \frac{\Gamma_\text{tot}W_R 
e^{i \psi}} {W_R^2 - W^2 - i W_R \Gamma_\text{tot}} f_{\pi N}(W)
\end{equation}
where $W_R$ is the resonance mass, 
$\Gamma_\text{tot}$ is its total width at resonance,
$C_{\pi N}$ is an isospin factor, and $\bar{A}$ is a form factor.
The $W$ dependence of the electroexcitation vertex and its pseudothreshold 
behavior is represented by $f_{\gamma N}$ while
$f_{\pi N}$ describes the $R \rightarrow \pi N$ decay in terms
of an energy-dependent partial width, $\Gamma_{\pi N}(W)$, and 
appropriate phase-space and penetrability factors.
Nonresonant amplitudes are computed using Born and vector-meson
diagrams with a mixed $\pi NN$ coupling that interpolates between 
pseudovector coupling at low cm momentum, $p_\pi$, and 
pseudoscalar coupling at high $p_\pi$.
Background amplitudes are unitarized with the $(1 + i t_{\pi N})$ factor,
as above, while resonant contributions are unitarized by adjusting the 
phase $\psi$ such that the total phase of the resonant contribution is 
given by the SAID partial-wave analysis for $\pi N$ elastic scattering.
Thus, $\psi$ depends upon both $W$ and $Q^2$ and varies with multipole.
The event generator used for data analysis employed MAID2000, but here
we will also show calculations using the updated MAID2003 version
\cite{MAID2003,Tiator04}. 

The Dubna-Mainz-Taipei (DMT) model \cite{Kamalov01,DMT} is based upon MAID 
but employs a more sophisticated analysis of $\pi N$ rescattering.
Whereas MAID employs a $K$-matrix approximation for the background
contribution to the $t$-matrix, DMT includes off-shell contributions 
in the form of a principal-value integral.
Both models use similar Breit-Wigner parametrizations for resonances, 
but the electroexcitation amplitudes for MAID should be interpreted as 
``dressed'' while for DMT the resonant amplitudes are considered ``bare'' 
because the $\pi N$ rescattering terms account for background contributions
to resonant multipoles.

The Sato-Lee (SL) model \cite{Sato01} is formulated in terms of 
energy-independent effective Hamiltonian and current operators. 
This dynamical model provides coupled equations for the $\pi N$ and $\gamma N$ 
reactions that automatically satisfy unitarity.
The potential governing pion rescattering is optimized to reproduce $\pi N$
elastic scattering data.
By means of a unitary transformation one can distinguish between the
electroexcitation amplitudes for the $N \rightarrow \Delta$ transition 
and the contributions of the pion cloud and rescattering mechanisms.
Although differing in detail, both the DMT and SL analyses conclude that 
the pion cloud is responsible for enhancing the $M_{1+}$ amplitude relative 
to the quark model and for most of the observed quadrupole strength.
Thus, these models suggest that the intrinsic deformation is rather small.
Note that the SL model omits higher resonances and is limited to 
$W \lesssim 1.4$ GeV 
while the DMT model reaches larger $W$ by including contributions of 
higher resonances based upon MAID2000.

\begin{figure*}
\centering
\includegraphics[angle=90,width=5in]{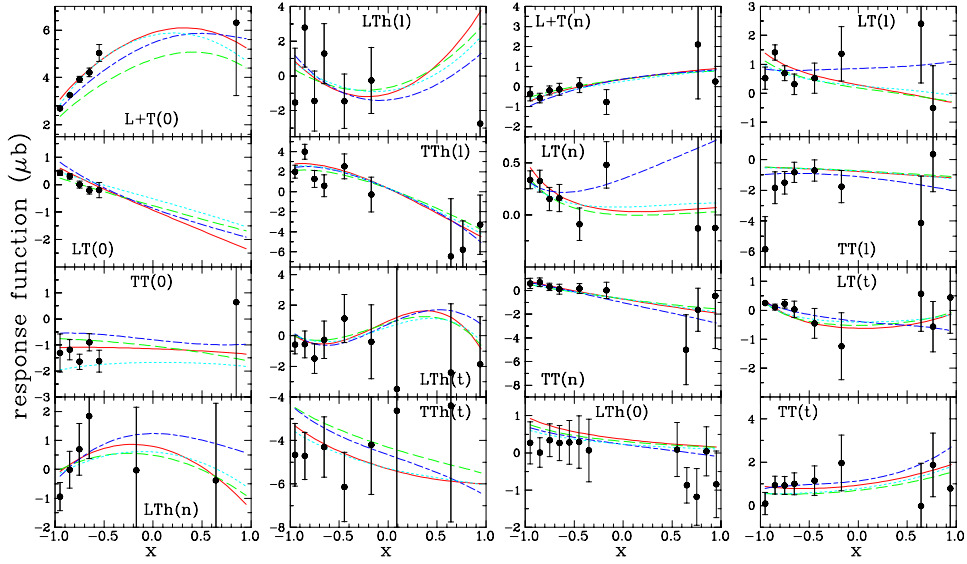}
\caption{(Color online) Data for response functions at $W=1.17$ GeV
are compared with selected models: 
MAID2003 (solid red), DMT (dashed green), 
SAID (dash-dotted blue), and SL (dotted cyan).
Inner error bars with endcaps are statistical while outer error bars 
without endcaps include systematic uncertainties; however, the
systematic contributions are often indistinguishable.}
\label{fig:models_rsfns_1170}
\end{figure*}

\begin{figure*}
\centering
\includegraphics[angle=90,width=5in]{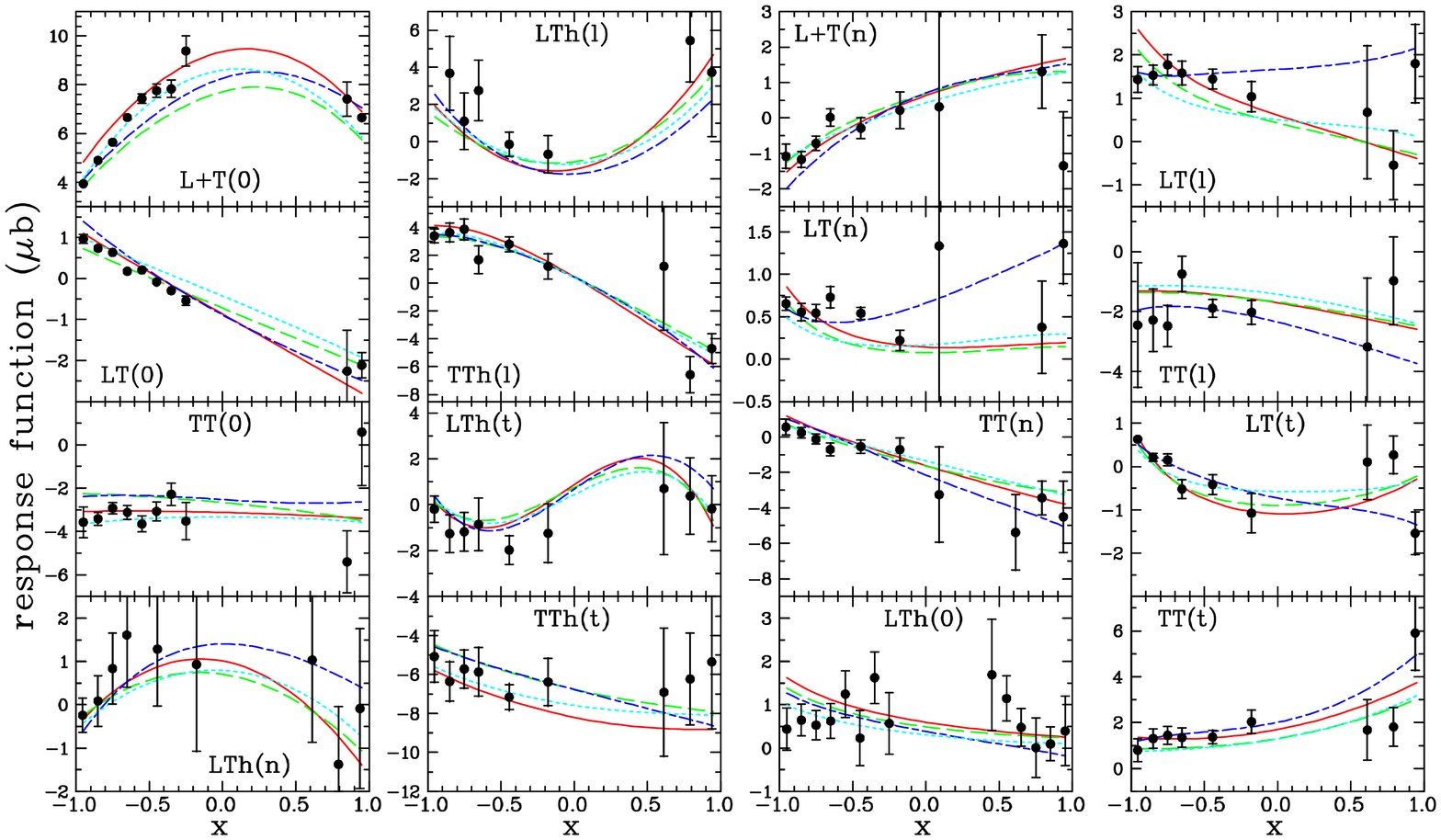}
\caption{(Color online) Data for response functions at $W=1.19$ GeV
are compared with selected models.
See Fig. \protect{\ref{fig:models_rsfns_1170}} for legend.}
\label{fig:models_rsfns_1190}
\end{figure*}

\begin{figure*}
\centering
\includegraphics[angle=90,width=5in]{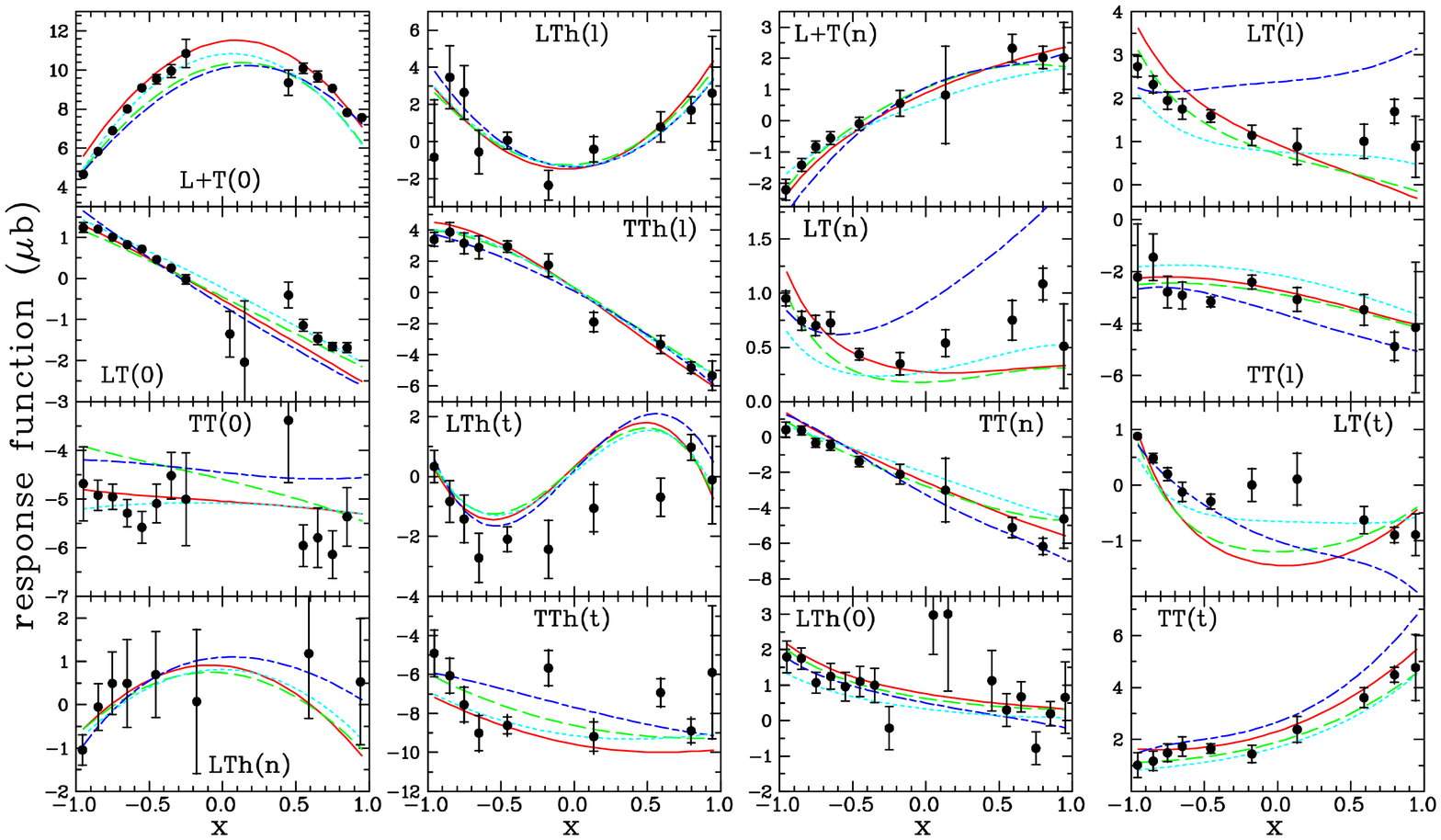}
\caption{(Color online) Data for response functions at $W=1.21$ GeV
are compared with selected models.
See Fig. \protect{\ref{fig:models_rsfns_1170}} for legend.}
\label{fig:models_rsfns_1210}
\end{figure*}

\begin{figure*}
\centering
\includegraphics[angle=90,width=5in]{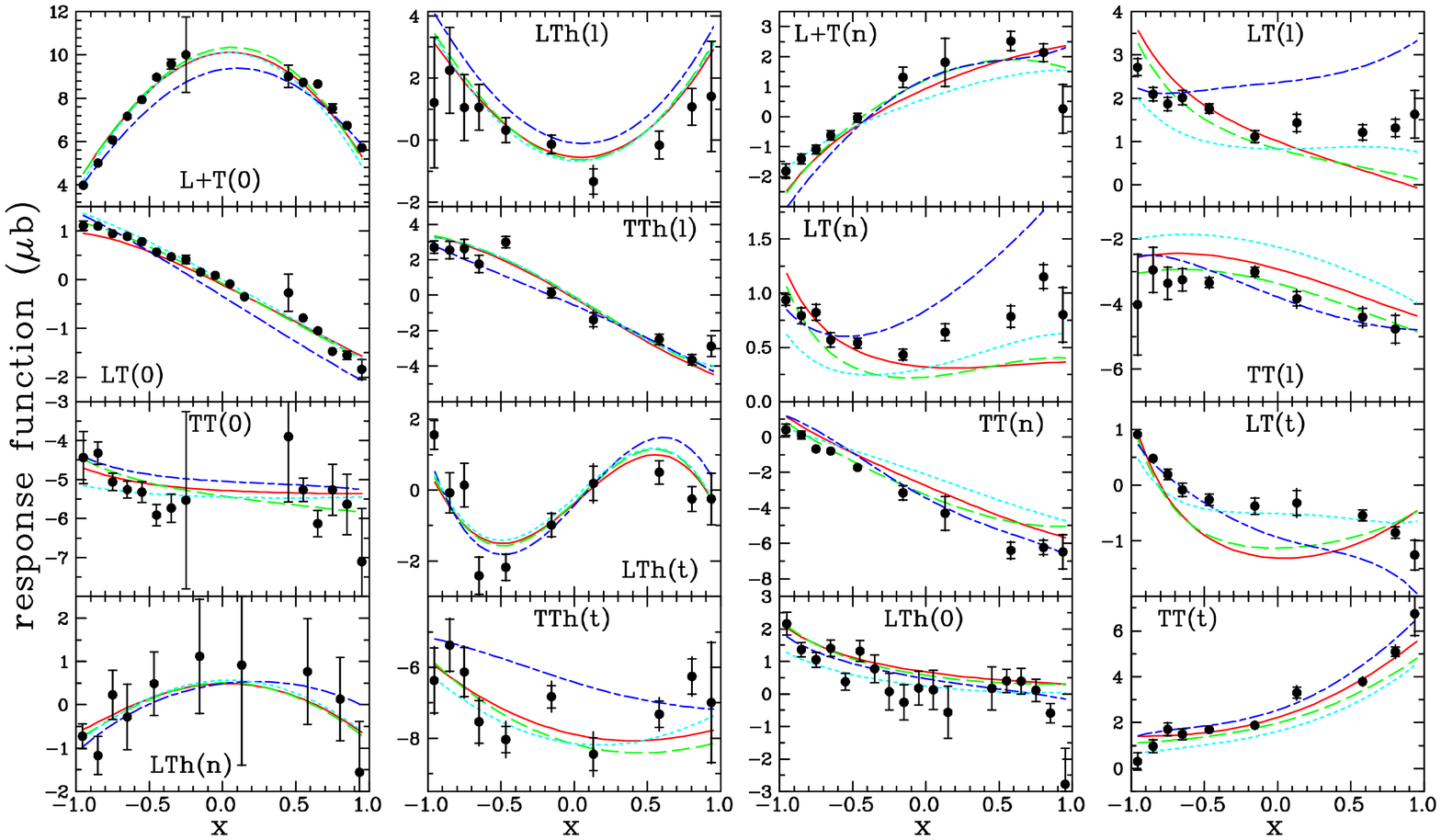}
\caption{(Color online) Data for response functions at $W=1.23$ GeV
are compared with selected models.
See Fig. \protect{\ref{fig:models_rsfns_1170}} for legend.}
\label{fig:models_rsfns_1230}
\end{figure*}

\begin{figure}
\centering
\includegraphics[angle=90,width=5in]{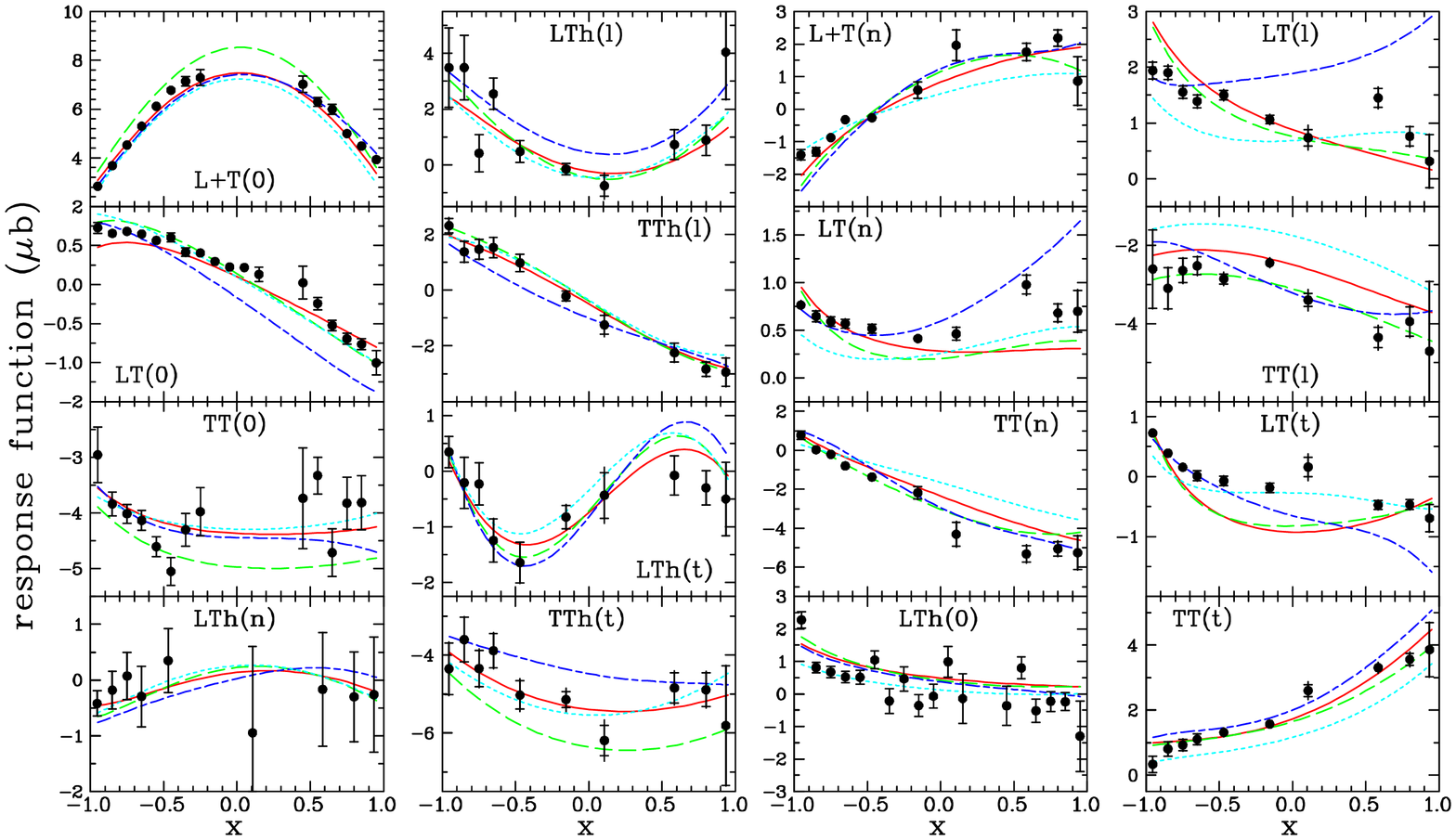}
\caption{(Color online) Data for response functions at $W=1.25$ GeV
are compared with selected models.
See Fig. \protect{\ref{fig:models_rsfns_1170}} for legend.}
\label{fig:models_rsfns_1250}
\end{figure}

\begin{figure*}
\centering
\includegraphics[angle=90,width=5in]{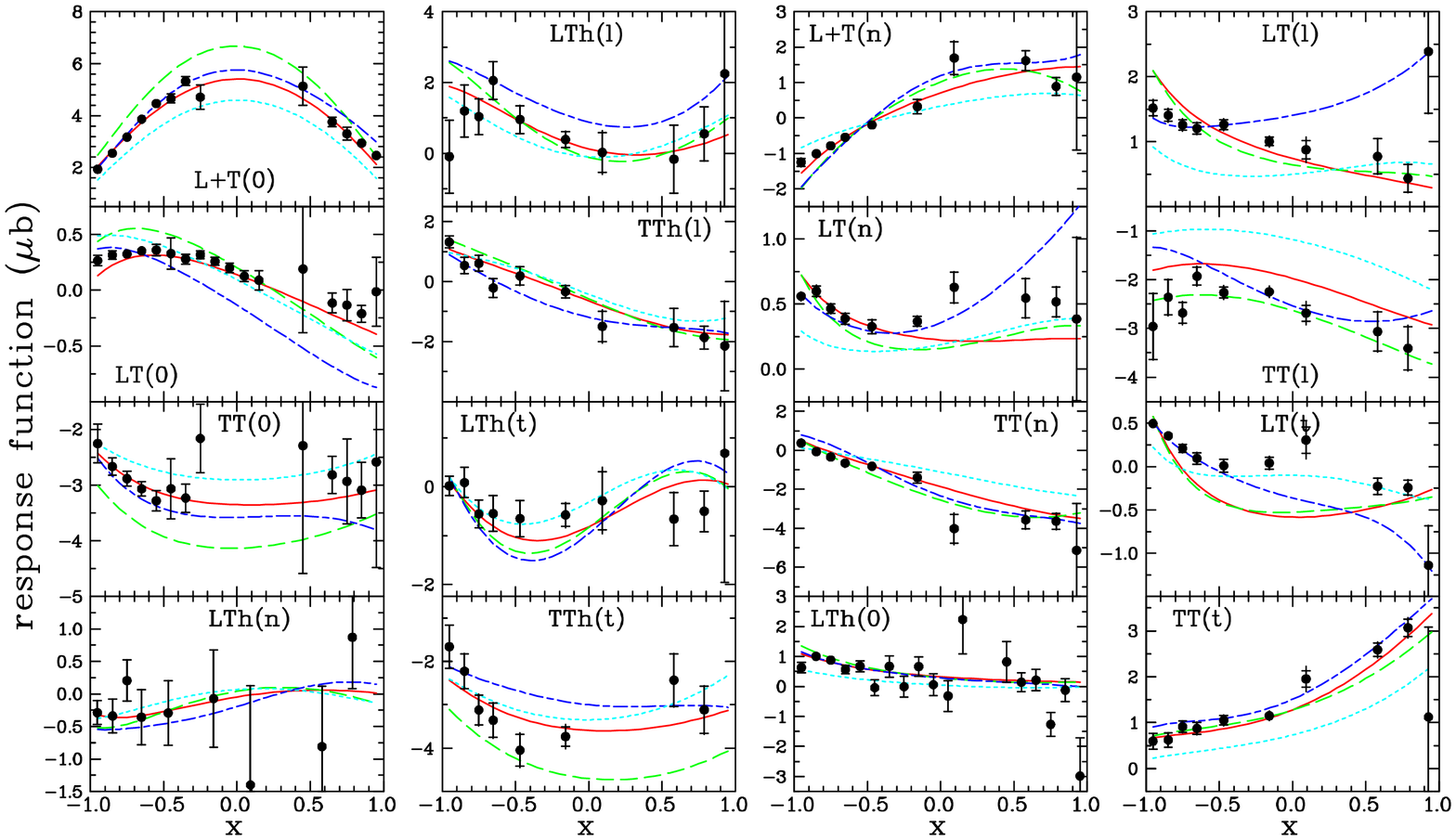}
\caption{(Color online) Data for response functions at $W=1.27$ GeV
are compared with selected models.
See Fig. \protect{\ref{fig:models_rsfns_1170}} for legend.}
\label{fig:models_rsfns_1270}
\end{figure*}

\begin{figure*}
\centering
\includegraphics[angle=90,width=5in]{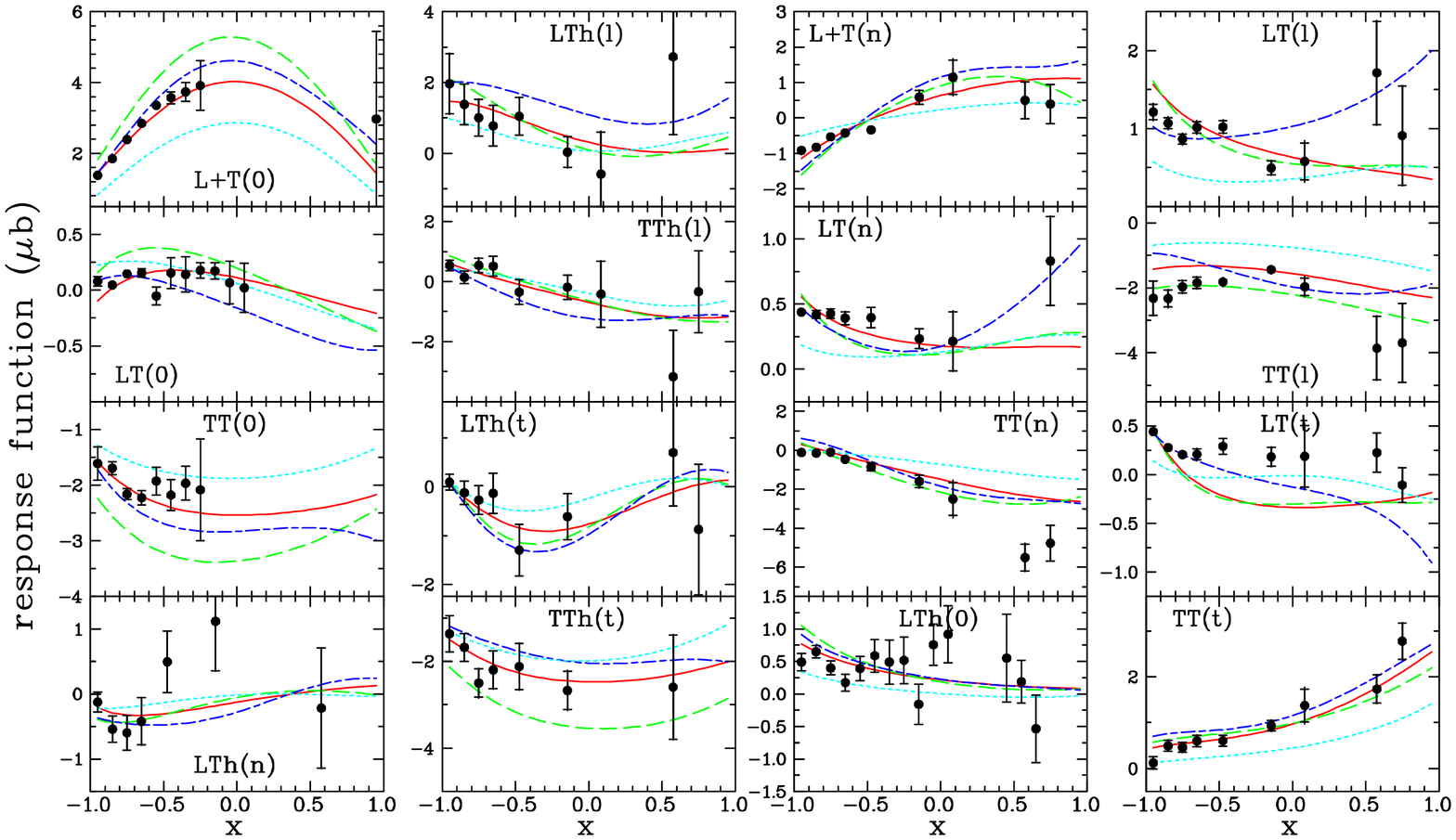}
\caption{(Color online) Data for response functions at $W=1.29$ GeV
are compared with selected models.
See Fig. \protect{\ref{fig:models_rsfns_1170}} for legend.}
\label{fig:models_rsfns_1290}
\end{figure*}

\begin{figure*}
\centering
\includegraphics[angle=90,width=5in]{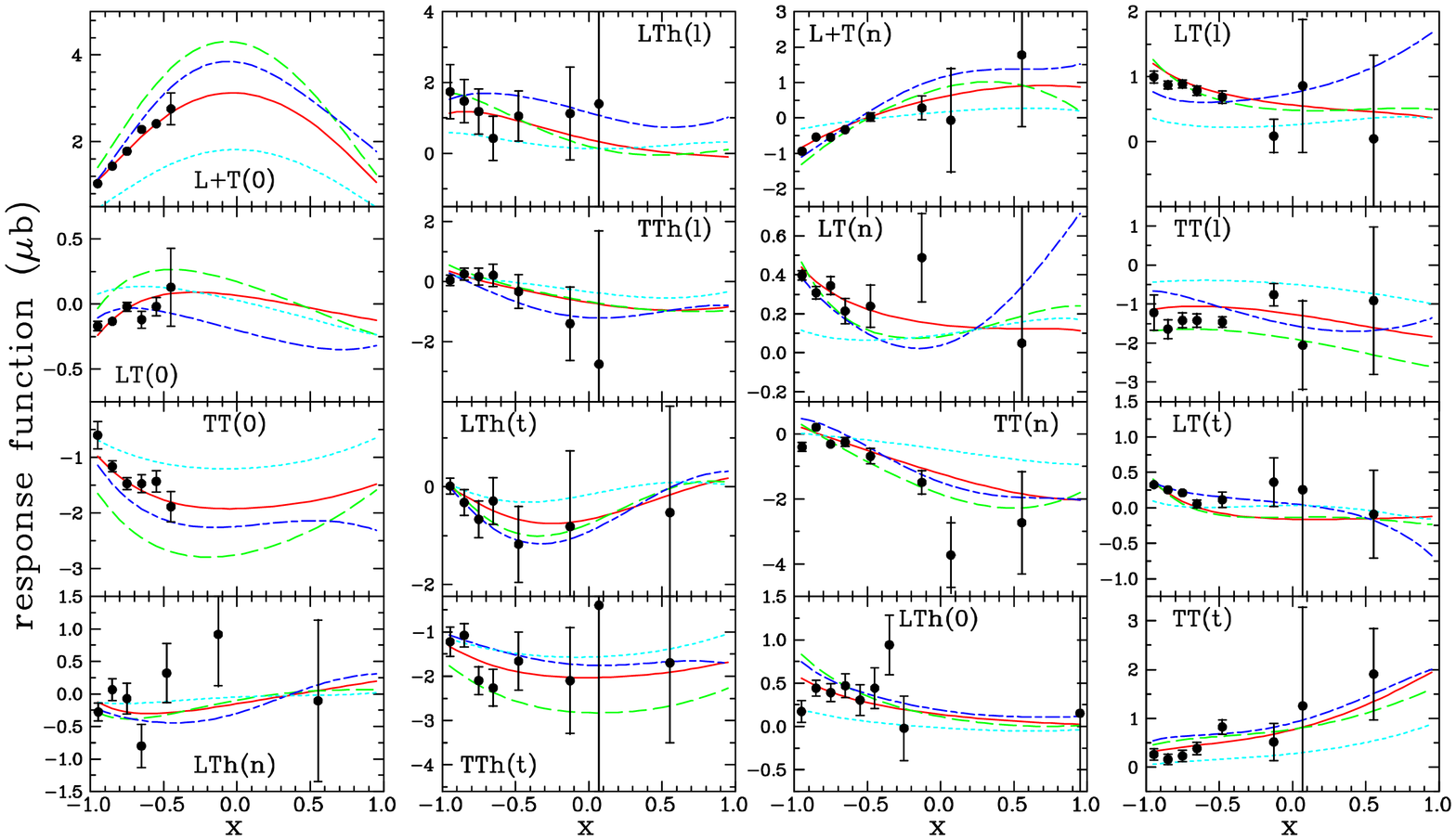}
\caption{(Color online) Data for response functions at $W=1.31$ GeV
are compared with selected models.
See Fig. \protect{\ref{fig:models_rsfns_1170}} for legend.}
\label{fig:models_rsfns_1310}
\end{figure*}

\begin{figure*}
\centering
\includegraphics[angle=90,width=5in]{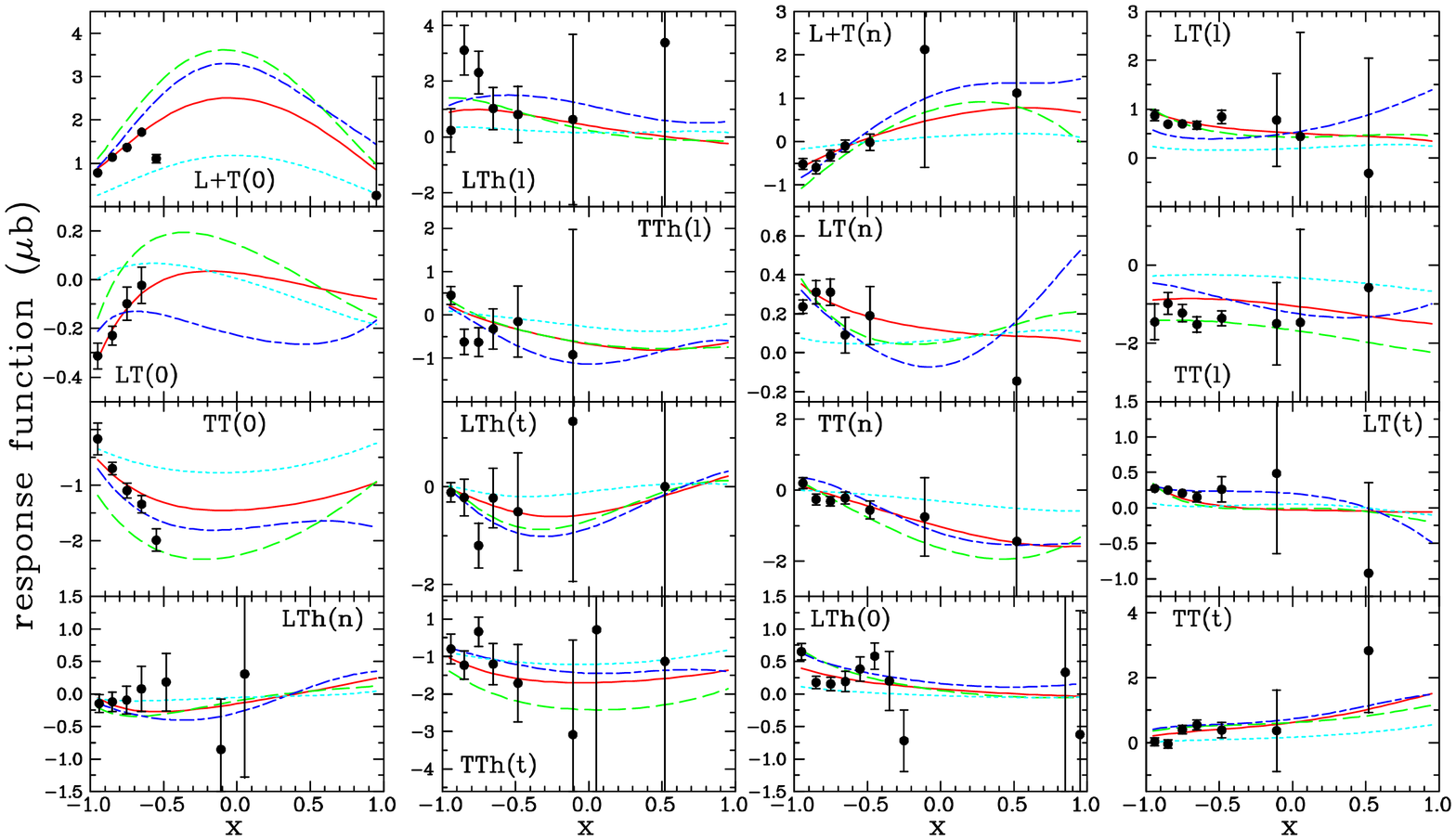}
\caption{(Color online) Data for response functions at $W=1.33$ GeV
are compared with selected models.
See Fig. \protect{\ref{fig:models_rsfns_1170}} for legend.}
\label{fig:models_rsfns_1330}
\end{figure*}

\begin{figure*}
\centering
\includegraphics[angle=90,width=5in]{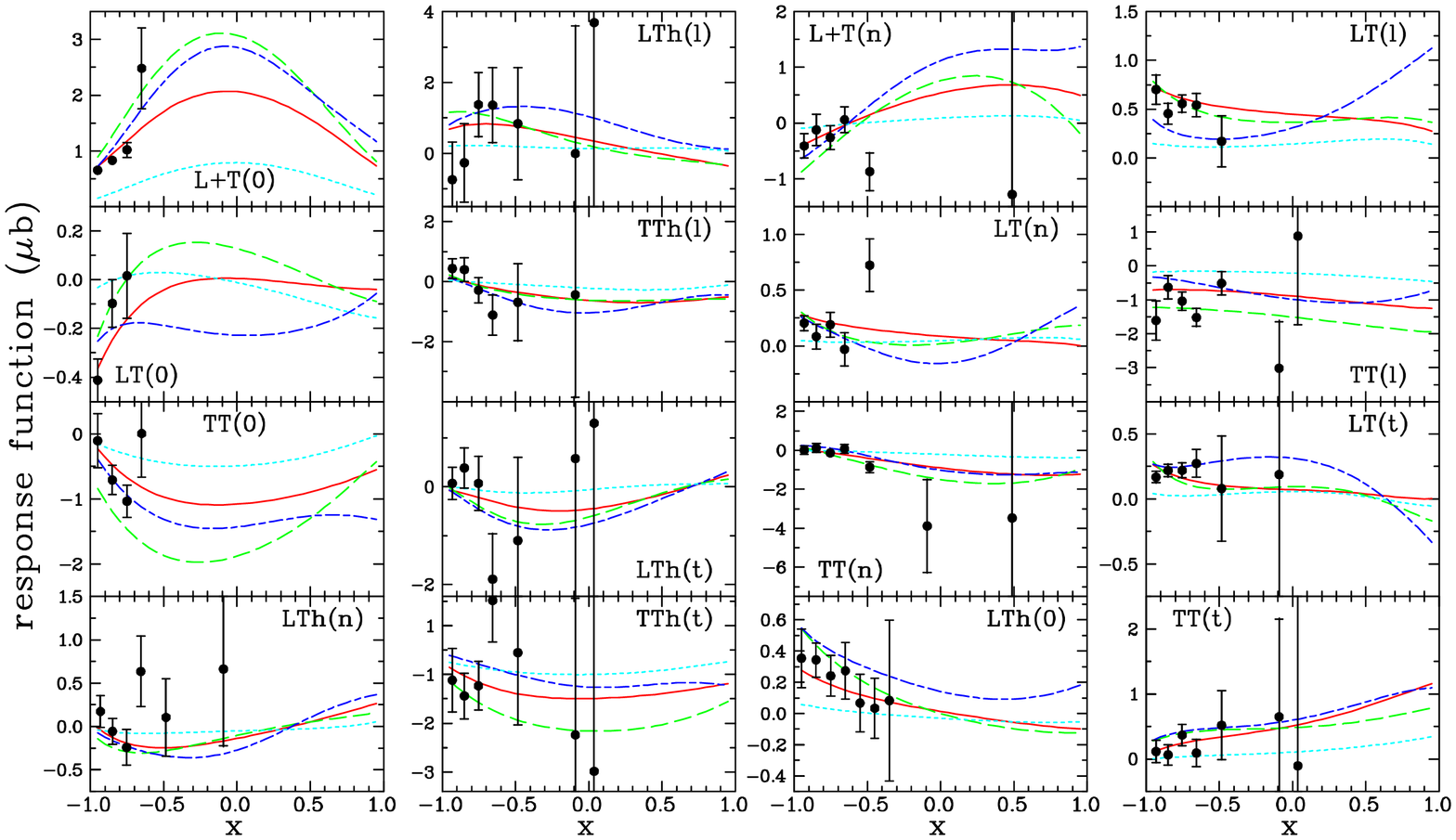}
\caption{(Color online) Data for response functions at $W=1.35$ GeV
are compared with selected models.
See Fig. \protect{\ref{fig:models_rsfns_1170}} for legend.}
\label{fig:models_rsfns_1350}
\end{figure*}

The data for response functions are compared in 
Figs. \ref{fig:models_rsfns_1170}-\ref{fig:models_rsfns_1350}
with calculations based upon these models.
The response functions in the first two columns, described as R-type, 
depend upon real parts of interference products while those in the last 
two columns, described as I-type, depend upon imaginary parts.
Although the first three response functions in column 1 and the last in
column 3 have been observed before, the other 12 response functions have been
observed for the first time in this experiment.
As a general rule we find that variations among the models are usually
greater for I-type than for R-type response functions, 
although $R_{TT}^{\prime t}$ also shows significant model dependence.
When $W \approx M_\Delta$, R-type responses are largely determined by the 
relatively well-known multipole amplitudes for the $\Delta$ resonance while 
I-type responses require interference with nonresonant background or tails of 
nondominant resonances that are constrained less well by previous data.
For both types the variations among models are typically smallest for $W$
near and below $M_\Delta$ and increase with $W$ above the $\Delta$ resonance.
By the time we reach $W \sim 1.3$ GeV, variations among models become large
even for R-type responses.
Above the $\Delta$ resonance the magnitudes for many of the SL response 
functions decrease faster than the data as $W$ increases, 
presumably due to neglect of higher resonances. 
Conversely, some of the DMT response become too strong as $W$ increases,
notably $R_{L+T}$, $R_{LT}$, and $R_{TT}$.
SAID appears to be the least accurate of these models, 
especially for LT response functions.
We speculate that this problem might be related to the use of
pseudoscalar coupling for Born amplitudes.
Among these models, MAID2003 seems to provide the best overall description
of the data, but none provides a uniformly good fit.

\subsection{Legendre Analysis}
\label{sec:legendre}

In the present representation, the response functions should be polynomials
in $x$ of relatively low order, especially if the assumption of $M_{1+}$ 
dominance is valid near the $\Delta$ resonance.
According to that assumption, one expects $R_{TT}$ to be constant,
$R_{LT}$, $R_{LT}^\prime$, $R_{L+T}^n$, $R_{TT}^n$, and $R_{TT}^\ell$ to
be linear, and only $R_{LT}^{\prime t}$, $R_{TT}^{\prime \ell}$ and
$R_{LT}^t$ to be cubic; the others are expected to be quadratic.
Indeed, at $W=1.23$ GeV we find that the data for $R_{TT}$ are almost 
constant and those for $R_{LT}$ are almost linear, 
with deviations from these simple behaviors that are qualitatively 
consistent with the departures of the models from $M_{1+}$ dominance.
Similarly, at $W=1.23$ GeV $R_{LT}^\prime$ and $R_{TT}^n$ appear to be 
consistent with linear behavior despite the somewhat larger experimental 
uncertainties.
However, model calculations for these responses show larger deviations with
respect to $M_{1+}$ dominance because imaginary parts of interference products
are more sensitive to nonresonant mechanisms and tails of nondominant 
resonances.
Finally, although $R_{LT}^{\prime t}$ displays cubic behavior, 
$R_{TT}^{\prime \ell}$ appears to be almost linear because the $|M_{1+}|^2$
contribution to its linear coefficient dominates the polynomial.
However, these simple rules deteriorate rapidly as $W$ increases 
and $R_{LT}$, $R_{L+T}^n$ and $R_{TT}^\ell$ data develop strong curvatures.
Furthermore, model calculations also show significant departures from
these simple behaviors; for example, the curvature of $R_{TT}$ calculations
is often appreciable.

\begin{figure*}
\centering
\includegraphics[angle=90,width=5in]{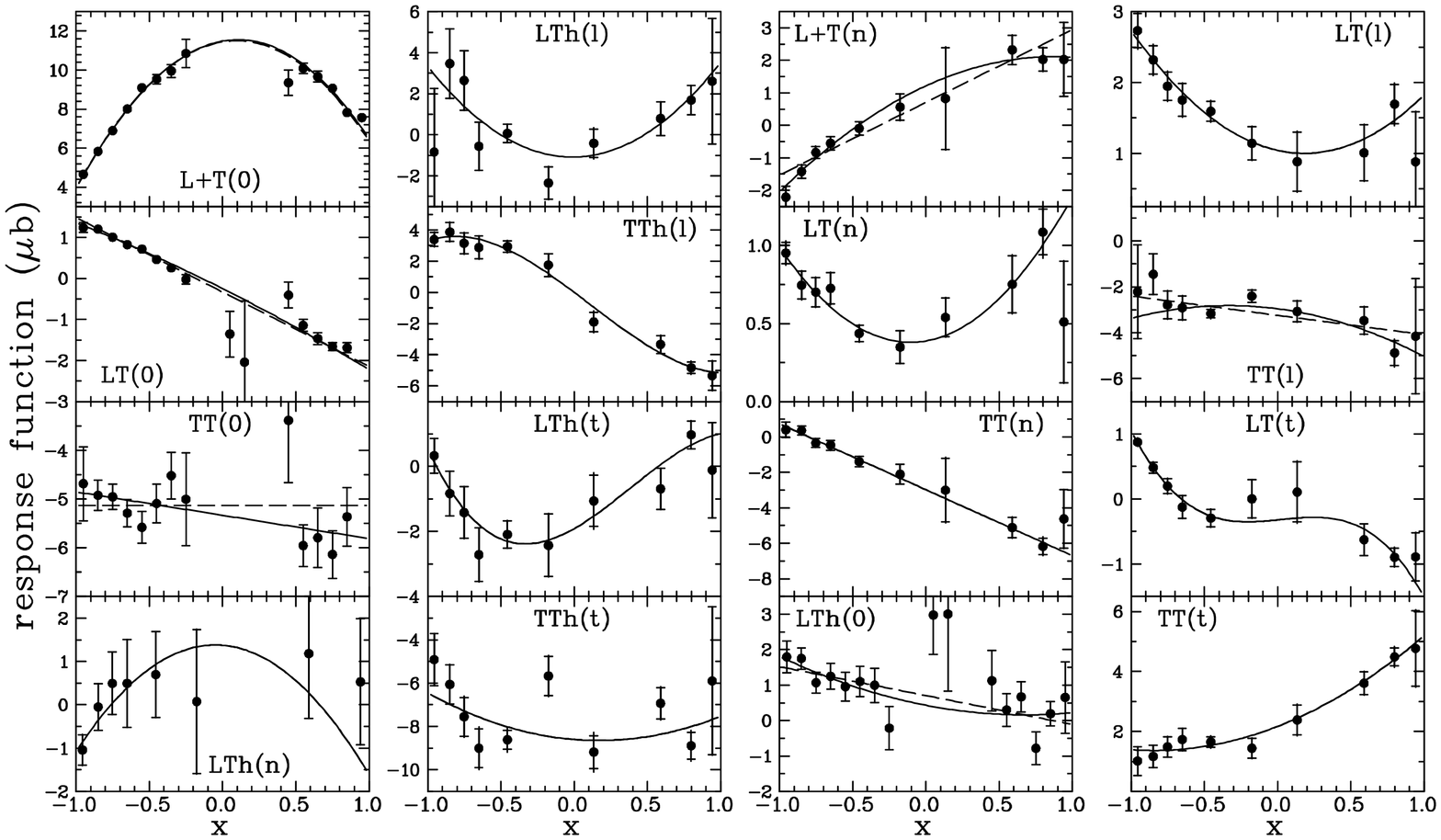}
\caption{Data for response functions at $W=1.21$ GeV
are compared with Legendre fits in the $sp$ truncation (dashed) and with
a few extra terms as needed (solid).
Inner error bars with endcaps are statistical; 
outer error bars without endcaps include systematic uncertainties.}
\label{fig:legfit_1210}
\end{figure*}

\begin{figure*}
\centering
\includegraphics[angle=90,width=5in]{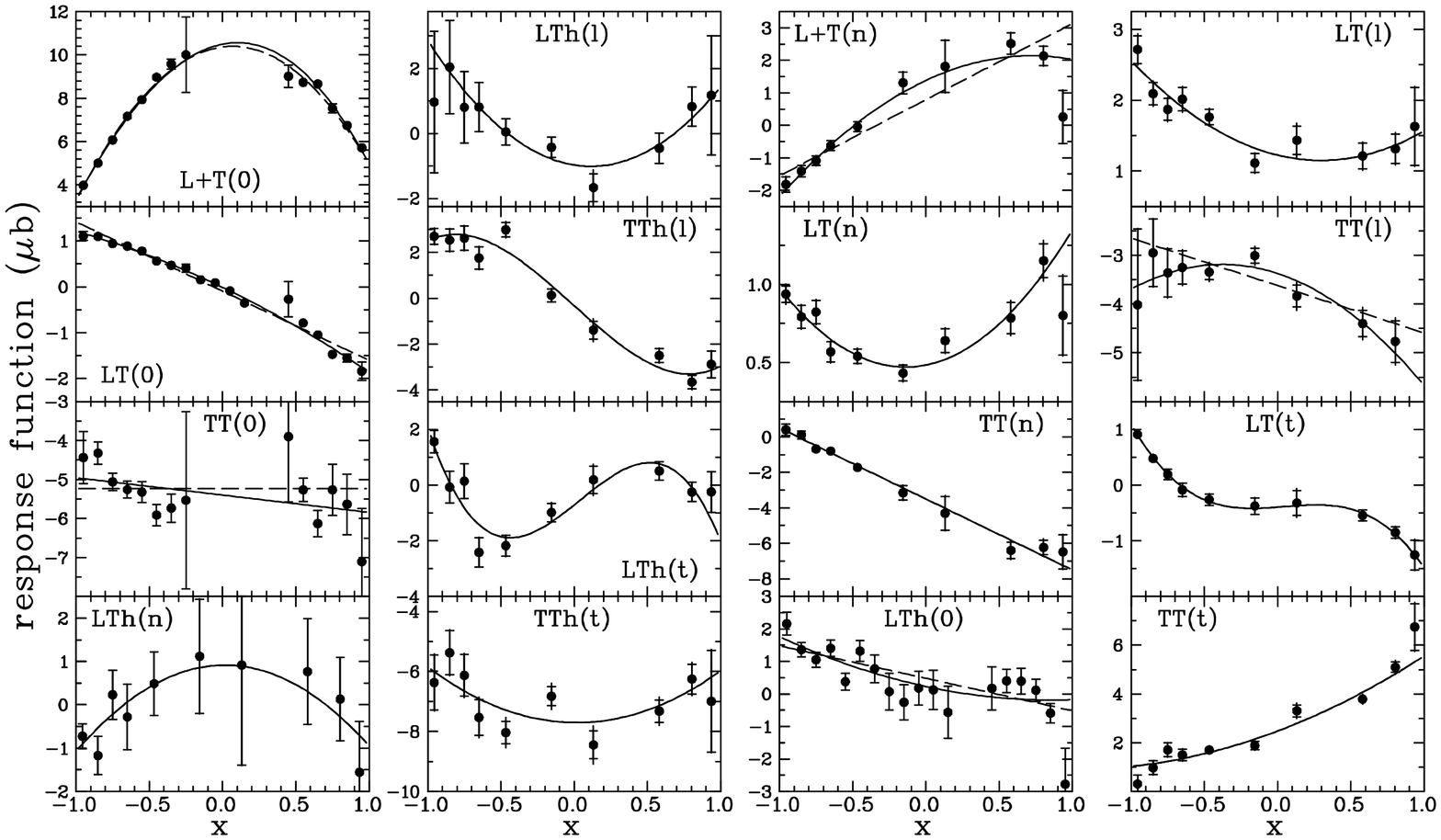}
\caption{Data for response functions at $W=1.23$ GeV
are compared with Legendre fits.
See Fig. \protect{\ref{fig:legfit_1210}} for legend.}
\label{fig:legfit_1230}
\end{figure*}

\begin{figure*}
\centering
\includegraphics[angle=90,width=5in]{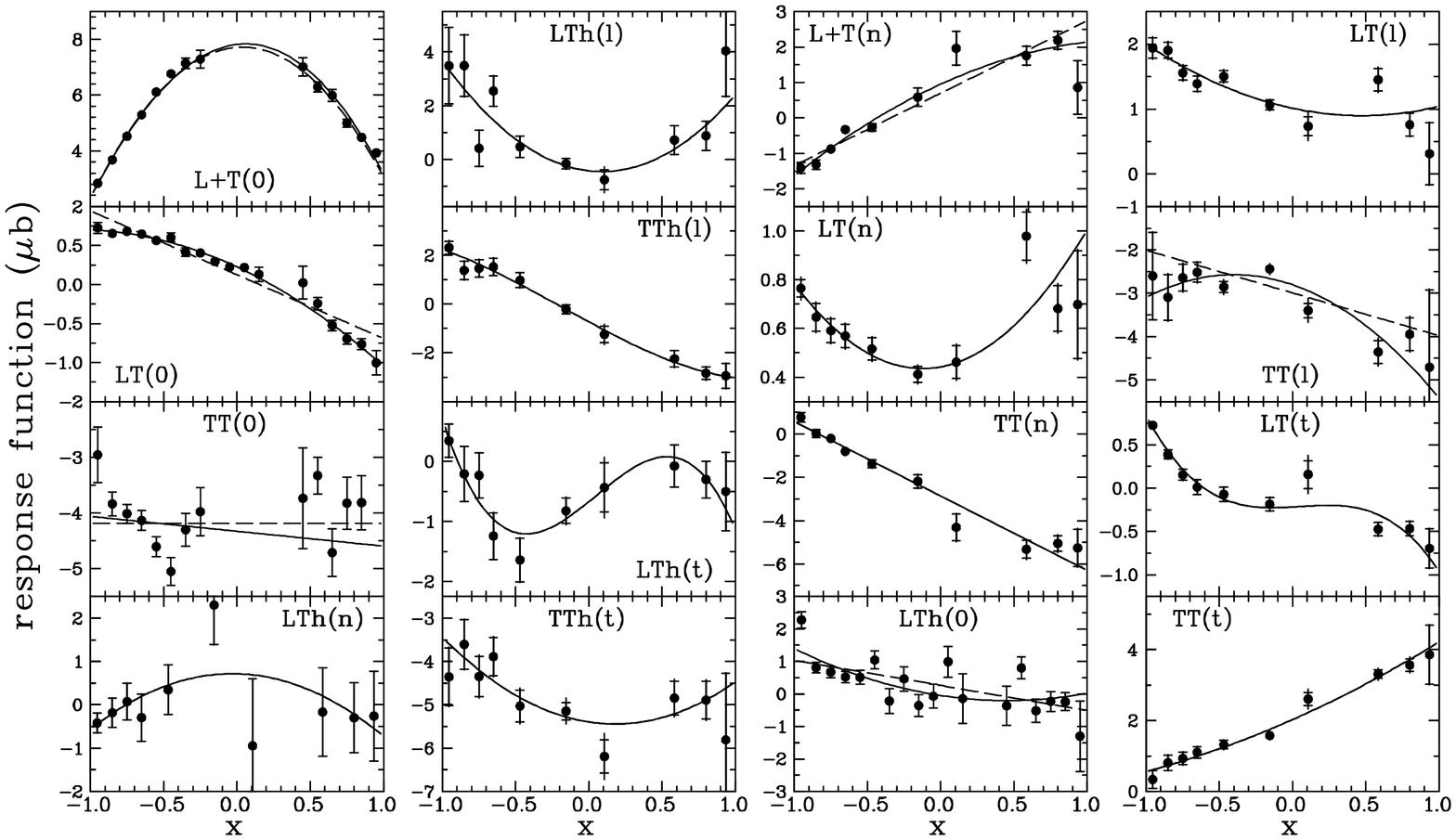}
\caption{Data for response functions at $W=1.25$ GeV
are compared with Legendre fits.
See Fig. \protect{\ref{fig:legfit_1210}} for legend.}
\label{fig:legfit_1250}
\end{figure*}

\begin{figure*}
\centering
\includegraphics[angle=90,width=5in]{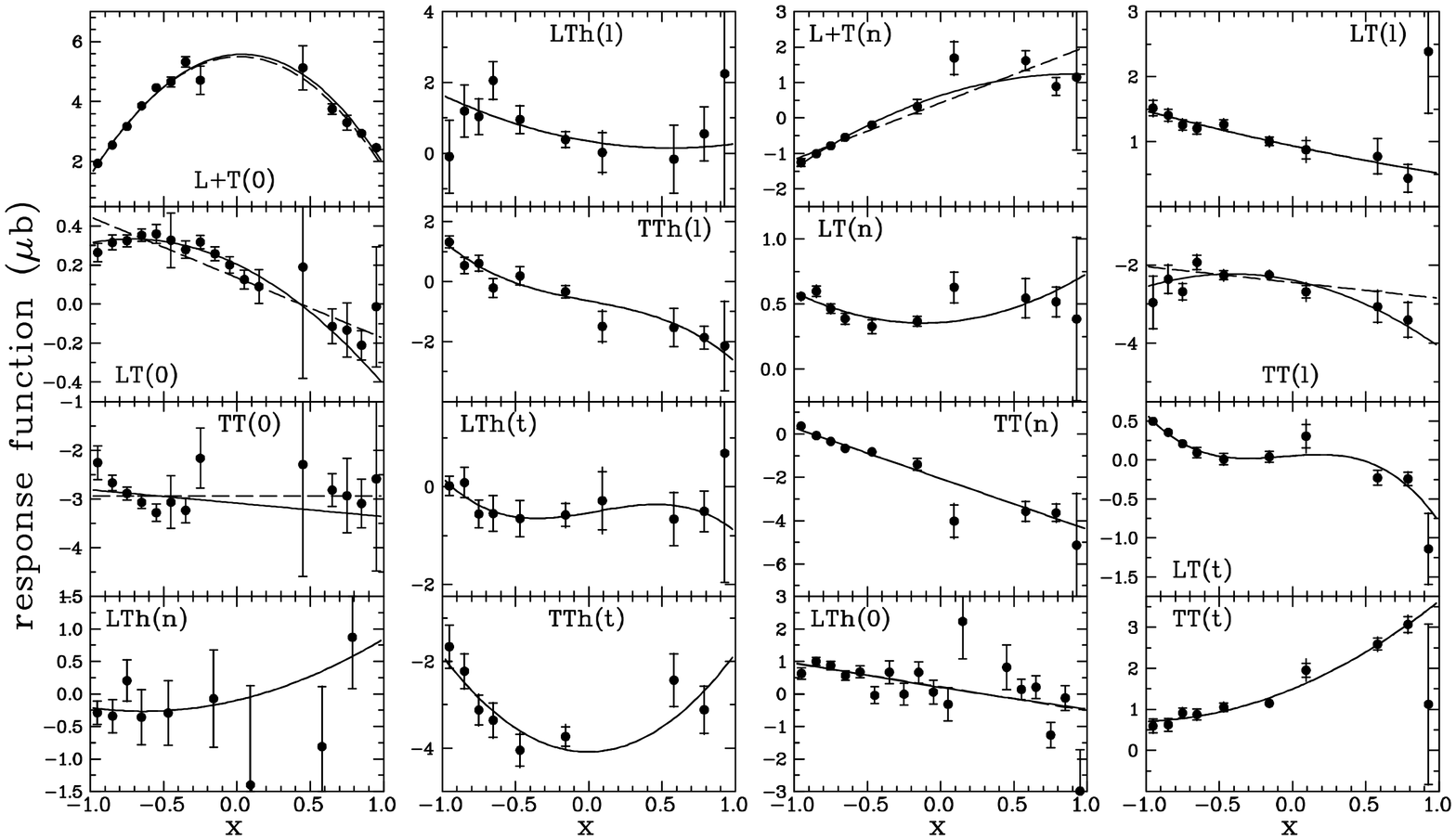}
\caption{Data for response functions at $W=1.27$ GeV
are compared with Legendre fits.
See Fig. \protect{\ref{fig:legfit_1210}} for legend.}
\label{fig:legfit_1270}
\end{figure*}

\begin{figure*}
\centering
\includegraphics[angle=90,width=5in]{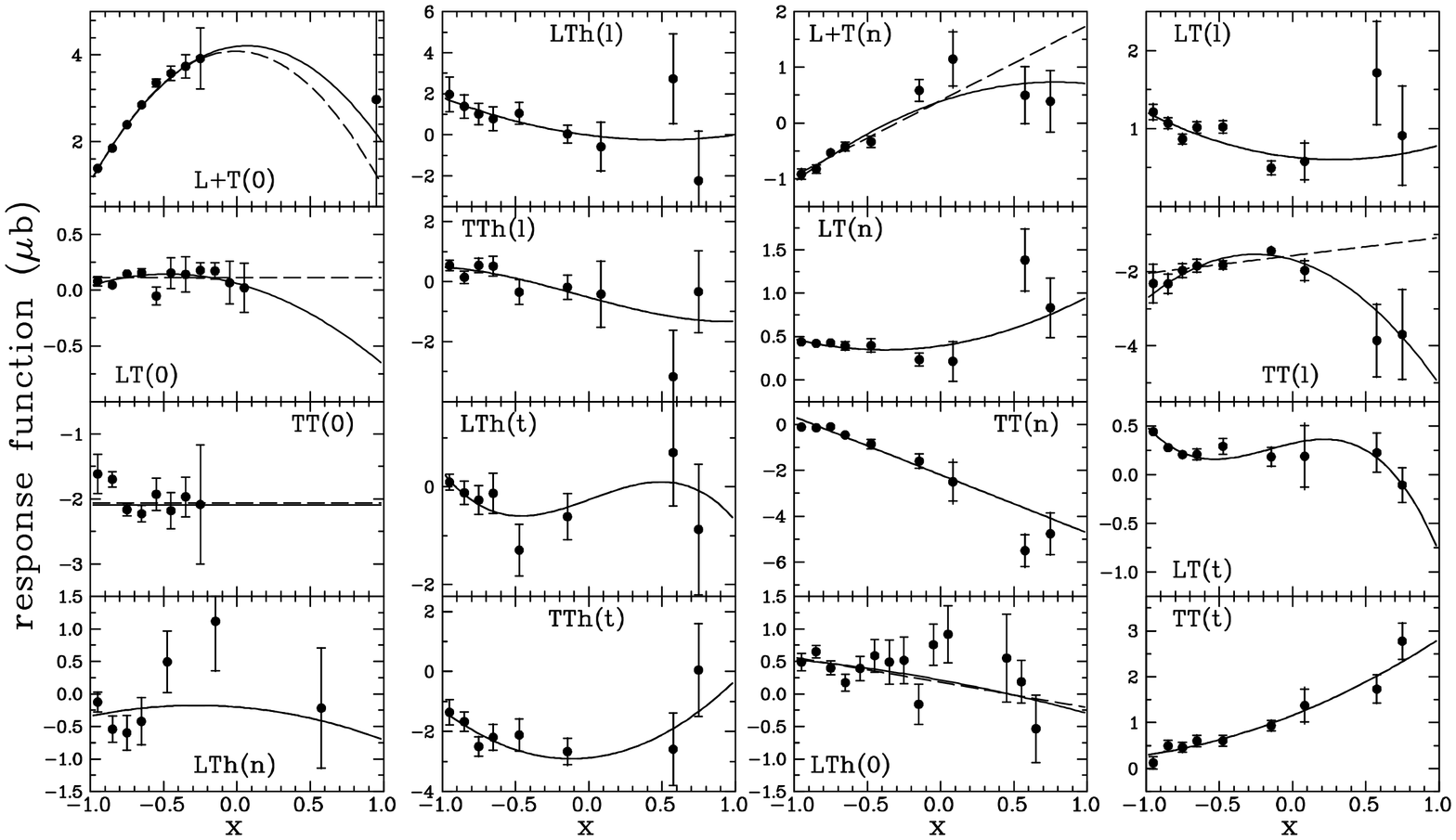}
\caption{Data for response functions at $W=1.29$ GeV
are compared with Legendre fits.
See Fig. \protect{\ref{fig:legfit_1210}} for legend.}
\label{fig:legfit_1290}
\end{figure*}

Representative Legendre fits are compared in 
Figs. \ref{fig:legfit_1210}-\ref{fig:legfit_1290} 
with data for response functions.
A more complete set of figures and comparisons between fitted
and predicted Legendre coefficients can be found in 
Ref. \cite{e91011_legfit}.
Note that these fits employed the $(x,\phi)$ distribution for the 
differential cross section and beam analyzing power together with the 
data for recoil-polarization response functions.
Thus, although $R_{L+T}$ and $R_{TT}$ could not be separated for 
$x \approx 0$ directly, the Legendre fits to those response functions 
are determined well in this region nonetheless.
The dashed curves are limited to the $sp$ truncation while solid curves
include additional terms in response functions for which the $sp$ fits 
appear to be systematically deficient over a range of $W$.
Most of the response functions can be fit well with the truncation based
upon $M_{1+}$ dominance, but $R_{LT}$, $R_{TT}$, $R_{L+T}^n$, and 
$R_{TT}^\ell$ generally require an extra term that reveals additional 
contributions.
However, it is not immediately obvious whether those additional 
contributions arise from $\ell_\pi \le 1$ terms that do not involve $M_{1+}$ 
or whether they require participation of higher partial waves.
It is also important to recognize that even when Legendre expansions limited
by $M_{1+}$ dominance do fit the data well, considerable violation of
this assumption may still be present.
Legendre fits are made to the data for each response function independently 
and ignore the correlations between response functions required by the 
expansions listed in 
Tables \ref{table:RtoLegendre1}-\ref{table:RtoLegendre3}.
More importantly, correct prediction of the number of significant
Legendre coefficients using this truncation scheme does not ensure that the 
relationships between the values of those coefficients and the underlying 
multipole amplitudes would be correct.
A detailed study of the truncation errors in the Legendre analysis 
of unpolarized response functions and their consequences for simplistic
extraction of multipole amplitudes is provided in Ref. \cite{Kelly05e}.
Therefore, a more rigorous analysis that fits the multipole amplitudes 
direcly, without the mediation of Legendre coefficients, is presented 
in Sec. \ref{sec:mpamps}.

Most previous extractions of $R_{EM}^{(3/2)}$ and $R_{SM}^{(3/2)}$ 
employed truncated Legendre analysis of the unpolarized differential
cross section.
To the extent that $M_{1+}$ dominance and $sp$ truncation apply for
$W \approx M_\Delta$, one can define
\begin{equation}
\label{eq:quad-trunc}
R_{EM}^{(p\pi^0)} \approx  \tilde{R}_{EM}^{(p\pi^0)} 
\hspace{1cm}
R_{SM}^{(p\pi^0)} \approx  \tilde{R}_{SM}^{(p\pi^0)}
\end{equation}
where
\begin{subequations}
\label{eq:quad-leg}
\begin{eqnarray}
\tilde{R}_{EM}  &=& 
\frac{3 A^{L+T}_2 - 2A^{TT}_0}{12 A^{L+T}_0} \\
\tilde{R}_{SM} &=& 
\frac{A^{LT}_1}{3 A^{L+T}_0}
\end{eqnarray}
\end{subequations}
are $W$-dependent combinations of Legendre coefficients for a particular
charge state.
To obtain the desired quantities for the $\Delta(1232)$ resonance, these 
quantities are evaluated at $W=M_\Delta$ and one needs to correct for the 
isospin-$1/2$ contamination of the $p\pi^0$ channel.
The isospin correction is expected to be small and has not been made.
The reliability of Eq. (\ref{eq:quad-trunc}) will be evaluated
in Sec. \ref{sec:Legendre-reliability}.

\begin{figure}
\centering
\includegraphics[width=3.0in]{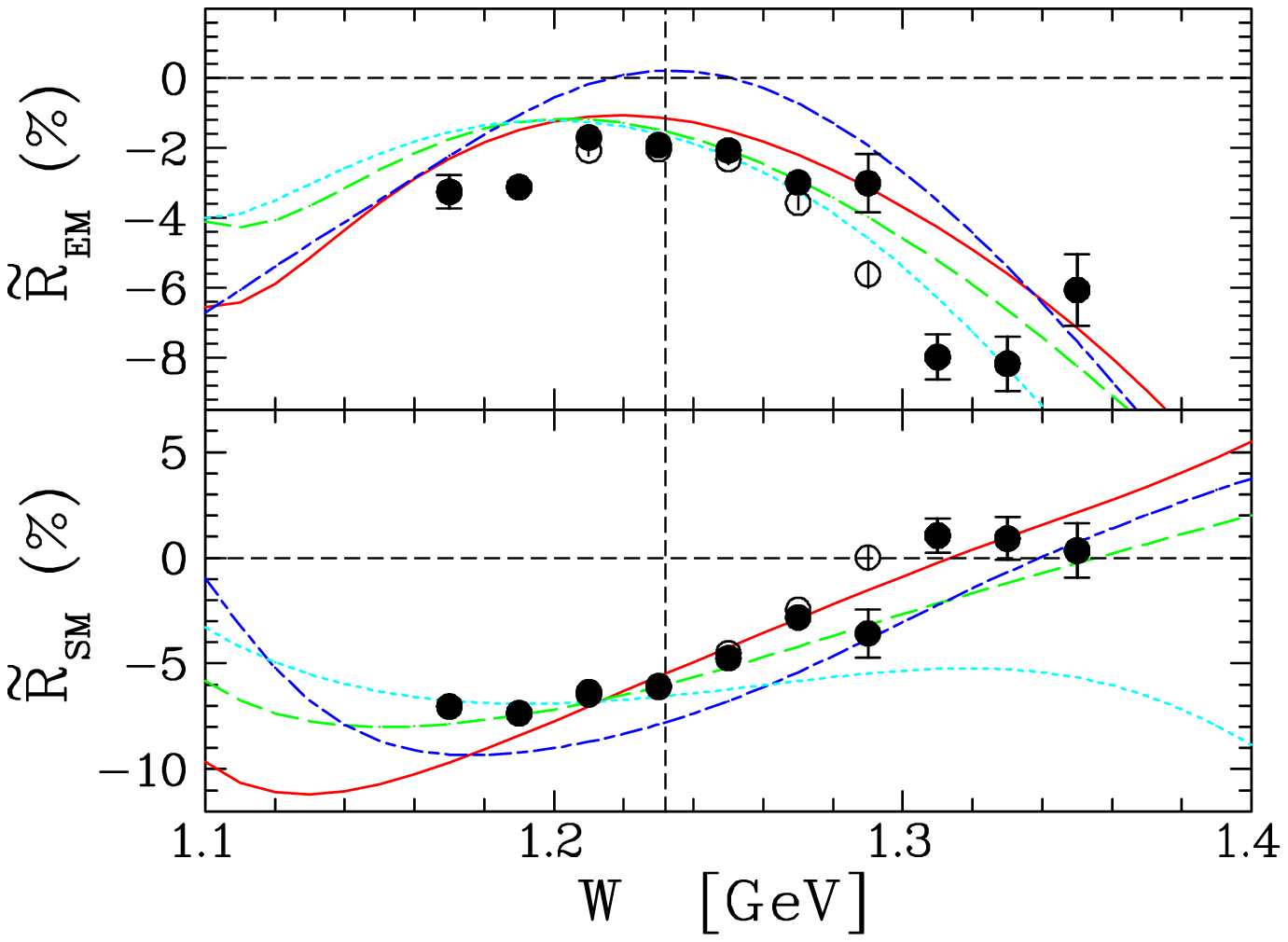}
\caption{(Color online) 
The truncated Legendre analysis for EMR and SMR at 
$Q^2=1.0$ (GeV/$c$)$^2$ is compared with MAID2003 (solid red), 
DMT (dashed green), SAID (dash-dot blue), and SL (dotted cyan).
Both theory and experiment employ Eq. (\protect{\ref{eq:quad-leg}})
and approximate EMR and SMR at $W=M_\Delta$, indicated by the vertical line.
Open circles are fit according to $sp$ truncation while filled
circles permit additional freedom in the Legendre analysis.
For filled circles, inner error bars with endcaps are statistical while 
outer error bars without endcaps include systematic uncertainties.}
\label{fig:legfit_quad}
\end{figure}

Figure \ref{fig:legfit_quad} compares values for 
$\tilde{R}_{EM}^{(p\pi^0)}$ and $\tilde{R}_{SM}^{(p\pi^0)}$ 
obtained from Legendre fits to the differential cross section data for 
$Q^2 = 1.0$ (GeV/$c$)$^2$ with model calculations obtained from 
Eq. (\ref{eq:quad-leg}) where the Legendre coefficients were
obtained by numerical integration.
Although these quantities approximate $R_{EM}^{(p\pi^0)}$ and 
$R_{SM}^{(p\pi^0)}$ only for $W \approx M_\Delta$, their $W$ dependence
offers some insight into the model dependence of the traditional
truncated Legendre expansion.
The open circles were obtained using the $M_{1+}$ truncation 
while the filled circles vary an additional term for each response
function in order to improve the fits for larger $W$.
Recall that Fig. \ref{fig:compare_clas} demonstrates that the
data are sensitive to at least one additional term per response
function and that model calculations predict significant Legendre 
coefficients beyond $M_{1+}$ dominance. 
Although the uncertainties increase for $W>1.3$ GeV because the
angular acceptance becomes too limited, we find that both analyses
are qualitatively consistent for an appreciable range of $W$ 
around $M_\Delta$.
The data for these quantities are relatively smooth with 
$W$ dependencies that are similar to model calculations of the
same quantities, whether or not these quantities are adequate
approximations to the desired quadrupole ratios.

The spread among models is smallest near the physical mass
but remains appreciable for $\tilde{R}_{EM}^{(p\pi^0)}$, for which SAID
differs significantly from both the data and the other models shown.
Although the SL slope is somewhat too small compared with data near 
$M_\Delta$, the other models provide a qualitatively consistent 
description of the $W$ dependence of $\tilde{R}_{SM}$. 
In the central $W$ region the experimental results are practically 
independent of truncation scheme and are in reasonable agreement with 
the MAID or DMT models, but the positive $\tilde{R}_{EM}$ values for SAID 
when $W \approx M_\Delta$ disagree sharply with the data.
For larger $W$ the SL calculation for $\tilde{R}_{SM}$ is much flatter 
than the data, probably because higher resonances are omitted.
Although it is more difficult to obtain unambiguous $\tilde{R}_{EM}$ 
results for $W \ge 1.31$ GeV, data based upon the $sp$ truncation remain 
in reasonable agreement with model calculations based upon the same
truncation scheme.

The results for $W = 1.23$ GeV are listed in Table \ref{table:legfit_quad}
and are practically independent of truncation scheme --- the slight
variation in $\tilde{R}_{EM}$ is within the estimated statistical uncertainty.
We also list in Table \ref{table:legfit_quad} the values obtained by
Joo {\it et al.} \cite{Joo02} using the $sp$ truncation,
averaging with respect to neighboring $Q^2$ bins.
Their results are consistent with ours for $\tilde{R}_{EM}^{(p\pi^0)}$ 
but are significantly larger for $\tilde{R}_{SM}^{(p\pi^0)}$.
Note that Joo {\it et al.} estimated that truncation errors in
determination of quadrupole ratios were no more than 0.5\% for SMR 
or 0.7\% for EMR in absolute terms.
While we agree that truncation of Legendre fits to the number of
terms in the $sp$ model has little effect upon fitted values for
$\tilde{R}_{EM}$ or $\tilde{R}_{SM}$, 
the discussion in Sec. \ref{sec:discussion} demonstrates that the
underlying assumptions of the traditional Legendre analysis do not
provide adequate approximations to the quadrupole ratios at the
present level of experimental precision.
Therefore, the next section presents a more rigorous analysis based 
upon multipole fits that does not assume $sp$ truncation or $M_{1+}$ 
dominance.

\begin{table}
\caption{Quadrupole ratios for $W =1.23$ GeV at $Q^2 = 1.0$ (GeV/$c$)$^2$.
The reduced chisquare for the entire data set is labeled $\chi^2_\nu$ 
while the chisquare per point for differential cross section data is 
labeled $\chi^2_N(\sigma)$. }
\label{table:legfit_quad}
\begin{ruledtabular}
\begin{tabular}{lccll}
method & SMR, \% & EMR, \% & $\chi^2_\nu$ & $\chi^2_N (\sigma)$ \\ \hline
$sp$            & $-6.07 \pm 0.11$ & $-2.04 \pm 0.13$ & 1.7 & 1.6 \\
$sp+$           & $-6.11 \pm 0.11$ & $-1.92 \pm 0.14$ & 1.5 & 1.4 \\
Joo {\it et al.} \footnote{Weighted average of $Q^2 = 0.9$ (GeV/$c$)$^2$ 
results for $E_i = 1.645$ and 2.445 GeV from \cite{Joo02}.} 
                 & $-7.4  \pm 0.4$  & $-1.8  \pm 0.4$  &     &     \\ 
\end{tabular}
\end{ruledtabular}
\end{table}

This analysis for the quadrupole ratios is based upon unpolarized cross 
sections and does not exploit any of the new recoil-polarization data. 
Examination of Tables \ref{table:RtoLegendre1}-\ref{table:RtoLegendre3}
shows that many other combinations of Legendre amplitudes could also
provide $\Re E_{1+}M_{1+}^\ast$ and $\Re S_{1+}M_{1+}^\ast$ --- 
if the $sp$ truncation is valid these quantities should be highly 
overdetermined.
Thus, one could obtain $\Re E_{1+}M_{1+}^\ast$ and $\Re S_{1+}M_{1+}^\ast$ 
using a least-squares analysis of the entire set of Legendre coefficients 
for $W=M_\Delta$ and measure the reliability of the truncation scheme
by $\chi^2$.
However, it is already clear from the Legendre fits that terms beyond
$sp$ truncation are needed for some of the response functions even for 
$W \approx M_\Delta$.
Furthermore, it is desirable to obtain the $W$ dependencies of both 
the real and the imaginary parts of the multipole amplitudes.
Therefore, we forgo further study of the consistency of $sp$ truncation
and proceed directly to a multipole analysis that exploits the new
response functions.

\subsection{Multipole Analysis}
\label{sec:mpamps}

Let 
\begin{equation}
\label{eq:mpfit}
A_i(W,Q^2) = A_i^{(0)}(W,Q^2) + \Delta A_i(W,Q^2)
\end{equation}
represent either the real or the imaginary part of one
of the multipole amplitudes ($M_{\ell j}$, $E_{\ell j}$, or $S_{\ell j}$)
where $A_i^{(0)}$ is a baseline amplitude obtained from a suitable model  
while $\Delta A_i$ is a variable to be fit to the data.
To minimize theoretical bias, our standard fits employ a baseline model
consisting of Born terms for pseudovector $\pi NN$ coupling plus 
$\rho$ and $\omega$ exchange; see Appendix \ref{app:Born} for details.
To test the sensitivity of fitted multipole amplitudes to neglect of
tails of nondominant resonances or to variations of nonresonant 
contributions, we have also performed fits using MAID2003, DMT, SL,
or SAID as baseline models.
Note that some of the $\Delta A_i$ parameters are relatively large
when using the Born baseline that contains no information about
the $\Delta(1232)$ resonance while the fitting parameters for more
sophisticated baseline models represent small corrections to the 
specified model. 
Nevertheless, we have demonstrated that fitted multipole amplitudes
are practically independent of the baseline model; 
see Ref. \cite{e91011_mpamps} for details and figures.
Both Legendre and multipole analyses were performed using the
{\tt EPIPROD} program \cite{epiprod}.

\begin{figure*}
\centering
\includegraphics[angle=90,width=5in]{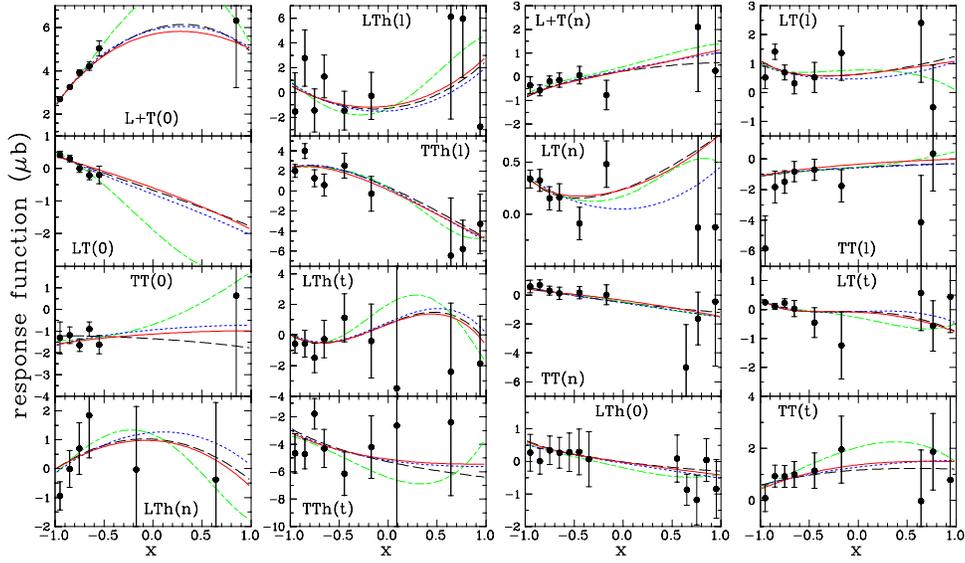}
\caption{(Color online)
Multipole fits for $W=1.17$ GeV at $Q^2=1.0$ (GeV/$c$)$^2$ using a 
Born baseline model.
Dashed curves fit corrections to all $s$- and $p$-wave amplitudes,
blue dotted curves also fit real parts of $2-$ multipoles, and
green dash-dotted curves fit all $s$-, $p$-, and $d$-wave amplitudes.
The solid red curves, considered the final fit, are similar to the
blue dotted curves except that $\Im M_{1-}$ is absent.
See text for further details.}
\label{fig:mpfit_1170}
\end{figure*}

\begin{figure*}
\centering
\includegraphics[angle=90,width=5in]{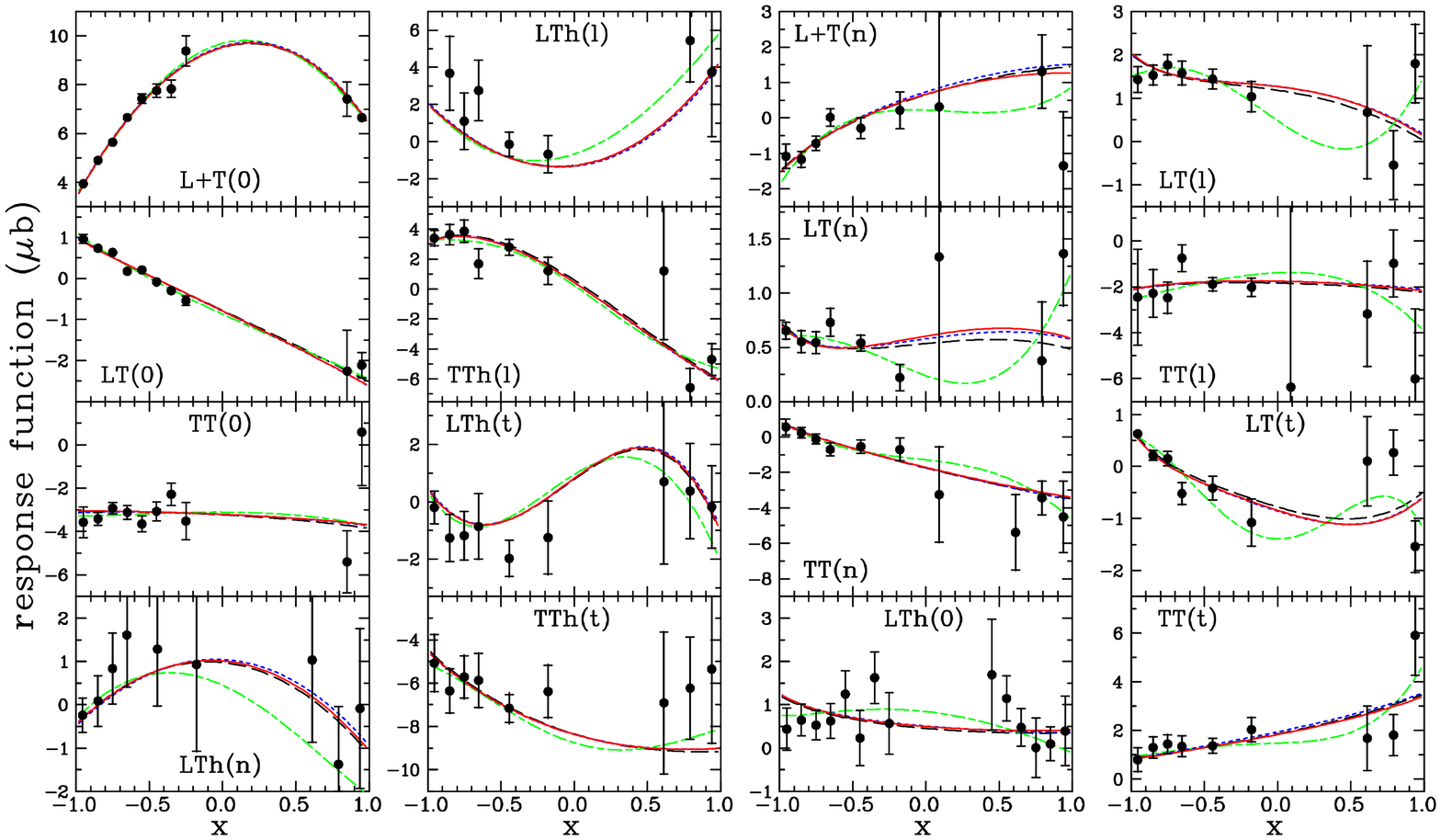}
\caption{(Color online)
Multipole fits for $W=1.19$ GeV at $Q^2=1.0$ (GeV/$c$)$^2$ using a 
Born baseline model.  
See Fig. \protect{\ref{fig:mpfit_1170}} for legend.}
\label{fig:mpfit_1190}
\end{figure*}

\begin{figure*}
\centering
\includegraphics[angle=90,width=5in]{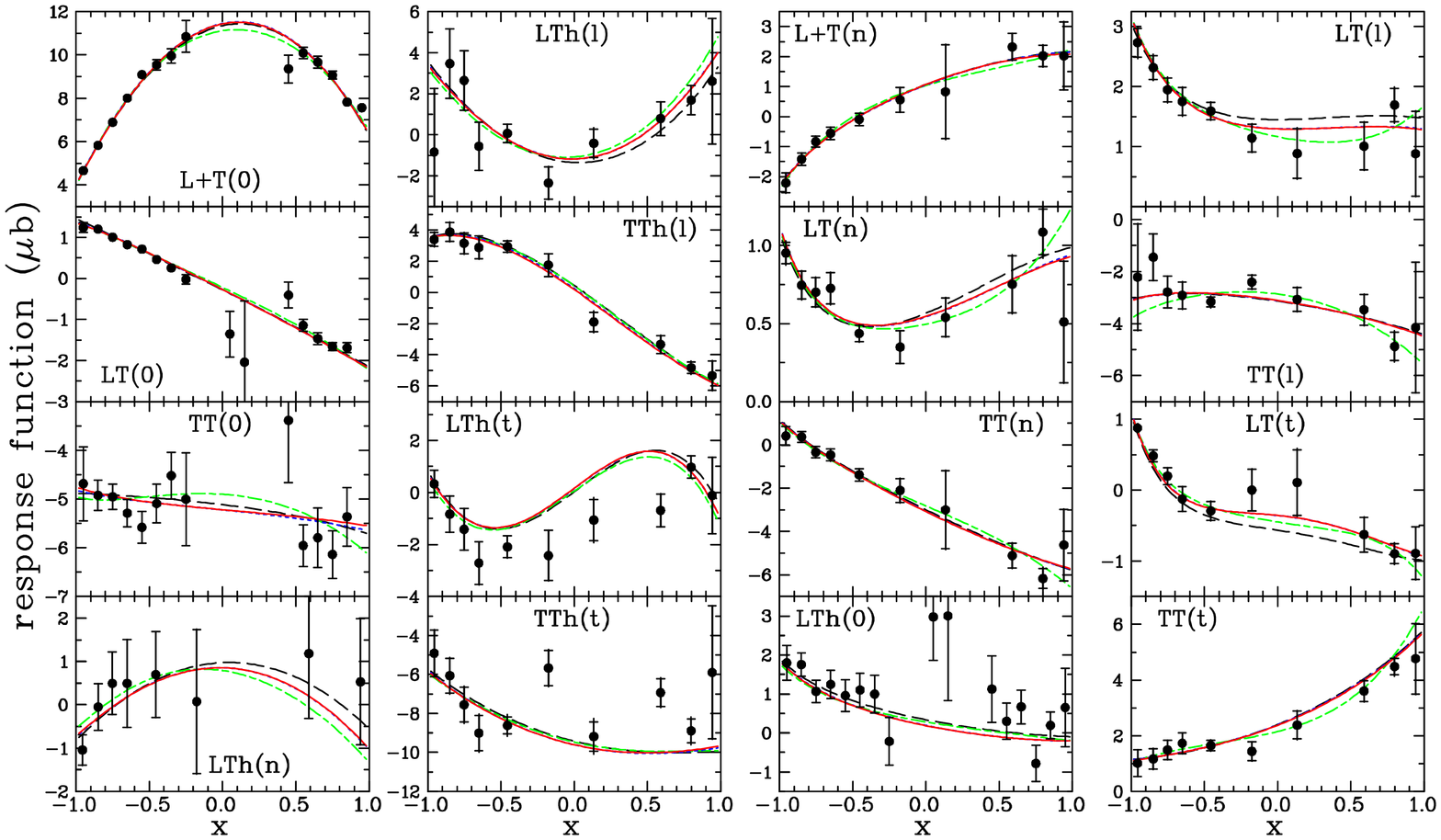}
\caption{(Color online)
Multipole fits for $W=1.21$ GeV at $Q^2=1.0$ (GeV/$c$)$^2$ using a 
Born baseline model.
See Fig. \protect{\ref{fig:mpfit_1170}} for legend.}
\label{fig:mpfit_1210}
\end{figure*}

\begin{figure*}
\centering
\includegraphics[angle=90,width=5in]{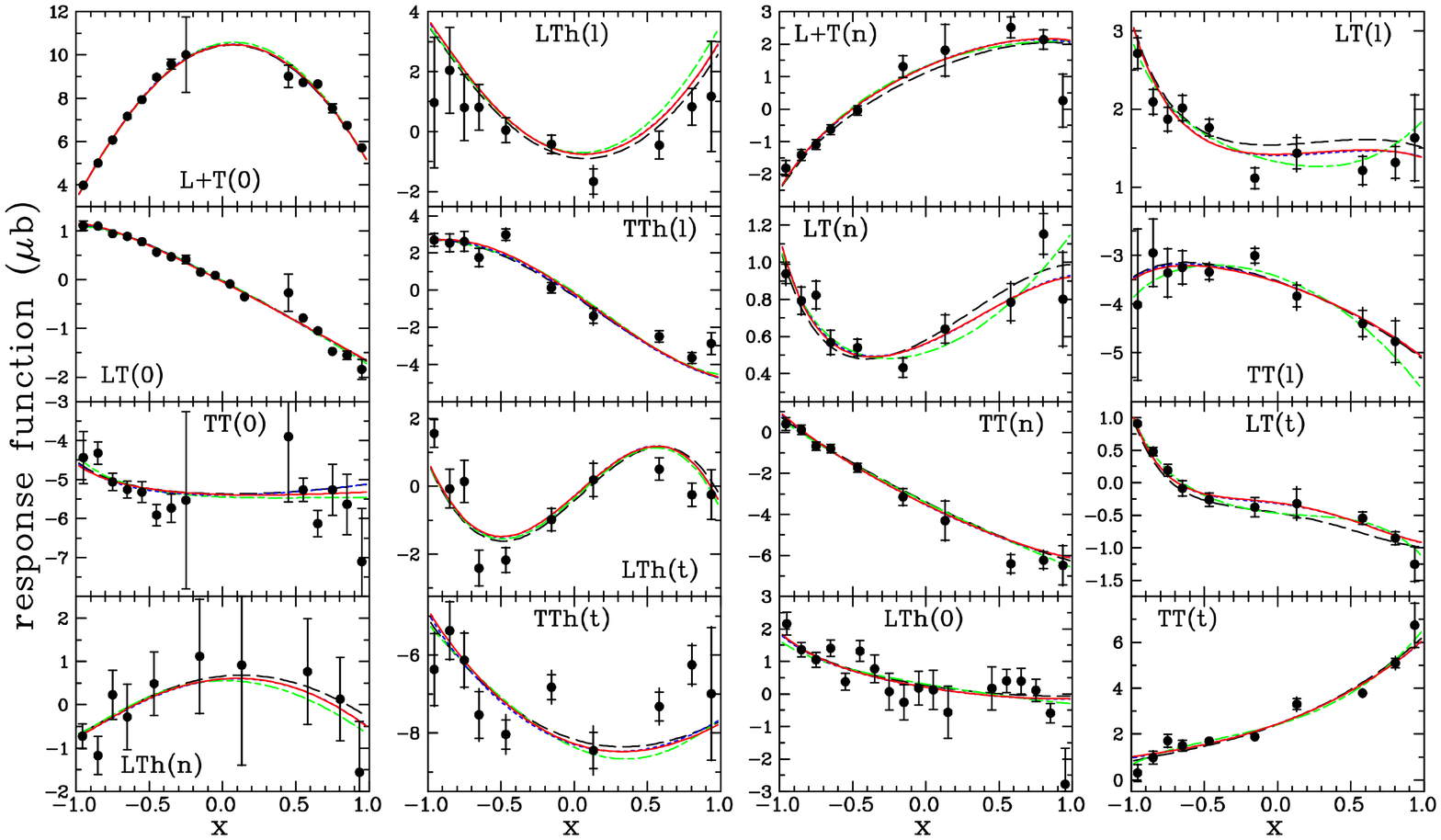}
\caption{(Color online)
Multipole fits for $W=1.23$ GeV at $Q^2=1.0$ (GeV/$c$)$^2$ using a 
Born baseline model.
See Fig. \protect{\ref{fig:mpfit_1170}} for legend.}
\label{fig:mpfit_1230}
\end{figure*}

\begin{figure*}
\centering
\includegraphics[angle=90,width=5in]{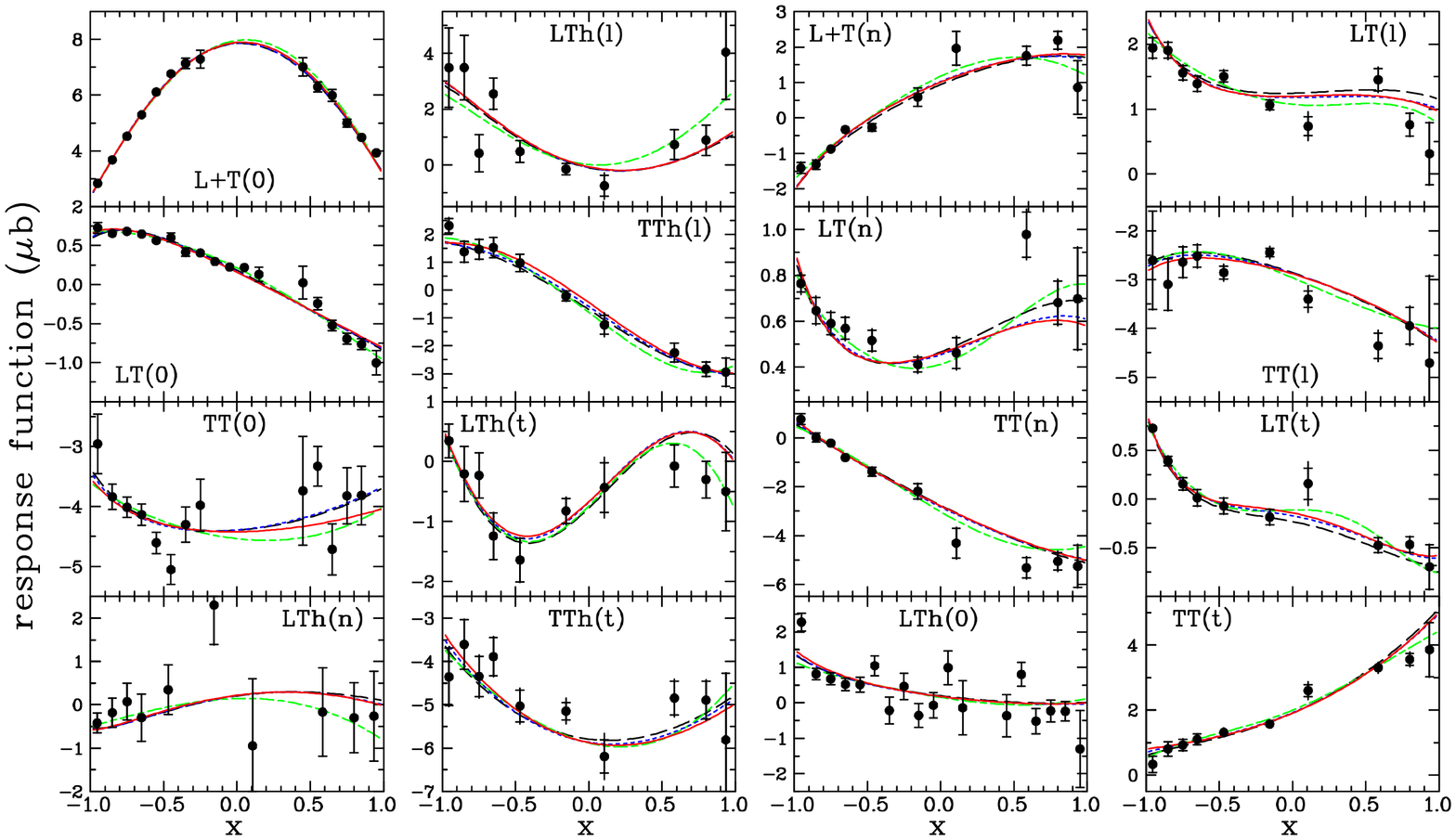}
\caption{(Color online)
Multipole fits for $W=1.25$ GeV at $Q^2=1.0$ (GeV/$c$)$^2$ using a 
Born baseline model.
See Fig. \protect{\ref{fig:mpfit_1170}} for legend.}
\label{fig:mpfit_1250}
\end{figure*}

\begin{figure*}
\centering
\includegraphics[angle=90,width=5in]{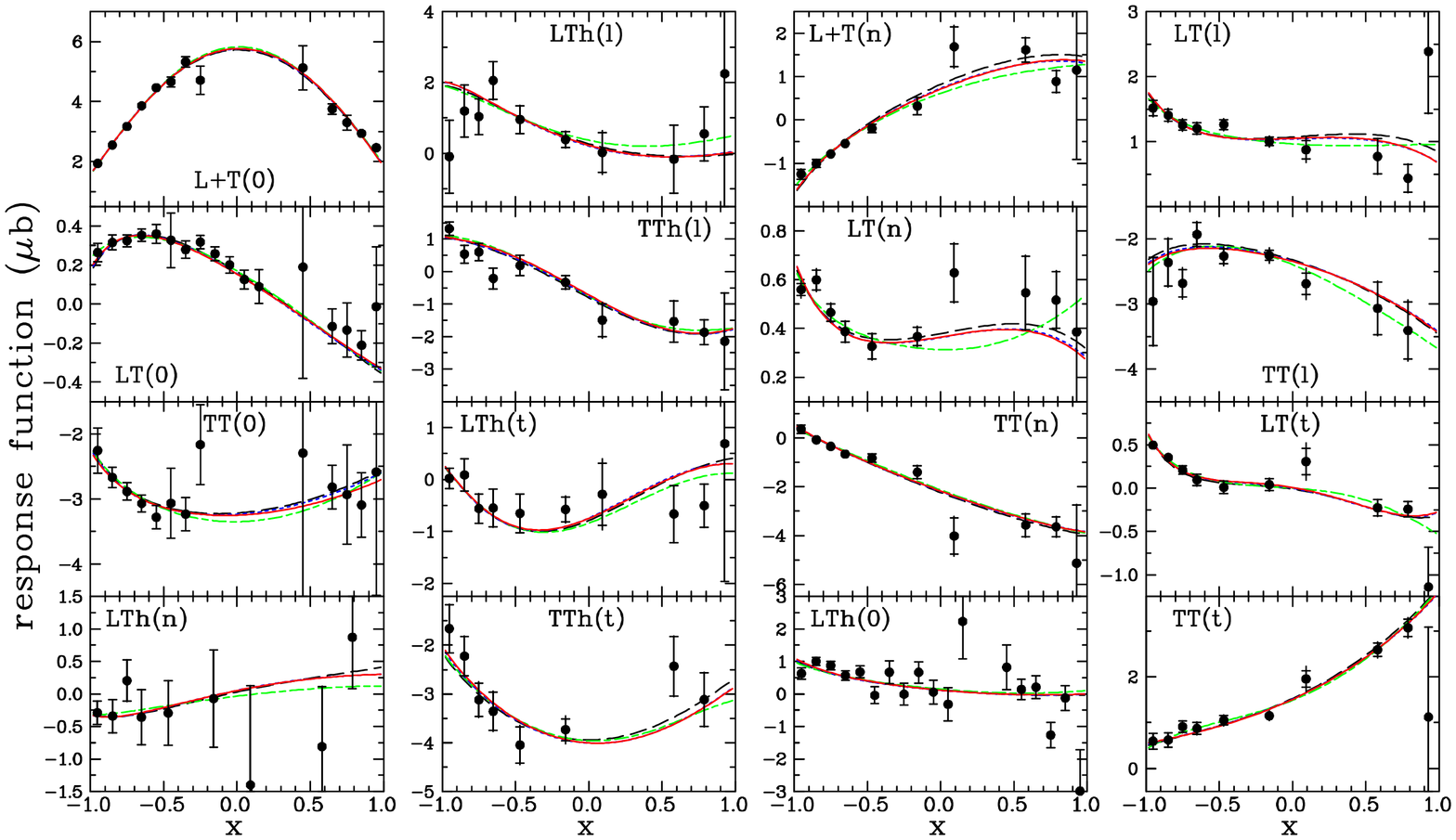}
\caption{(Color online)
Multipole fits for $W=1.27$ GeV at $Q^2=1.0$ (GeV/$c$)$^2$ using a 
Born baseline model.
See Fig. \protect{\ref{fig:mpfit_1170}} for legend.}
\label{fig:mpfit_1270}
\end{figure*}

\begin{figure*}
\centering
\includegraphics[angle=90,width=5in]{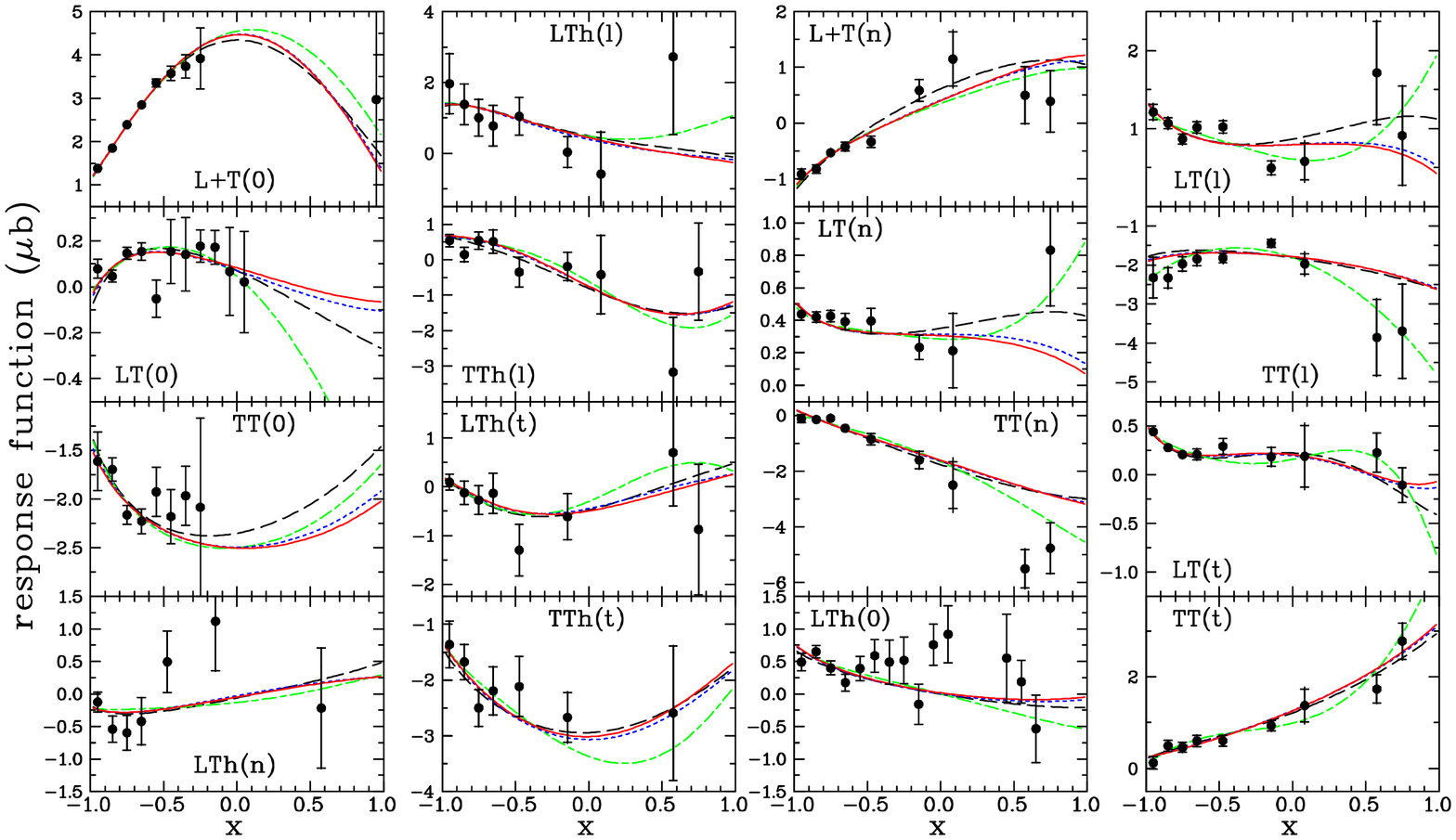}
\caption{(Color online)
Multipole fits for $W=1.29$ GeV at $Q^2=1.0$ (GeV/$c$)$^2$ using a 
Born baseline model.
See Fig. \protect{\ref{fig:mpfit_1170}} for legend.}
\label{fig:mpfit_1290}
\end{figure*}

\begin{figure*}
\centering
\includegraphics[angle=90,width=5in]{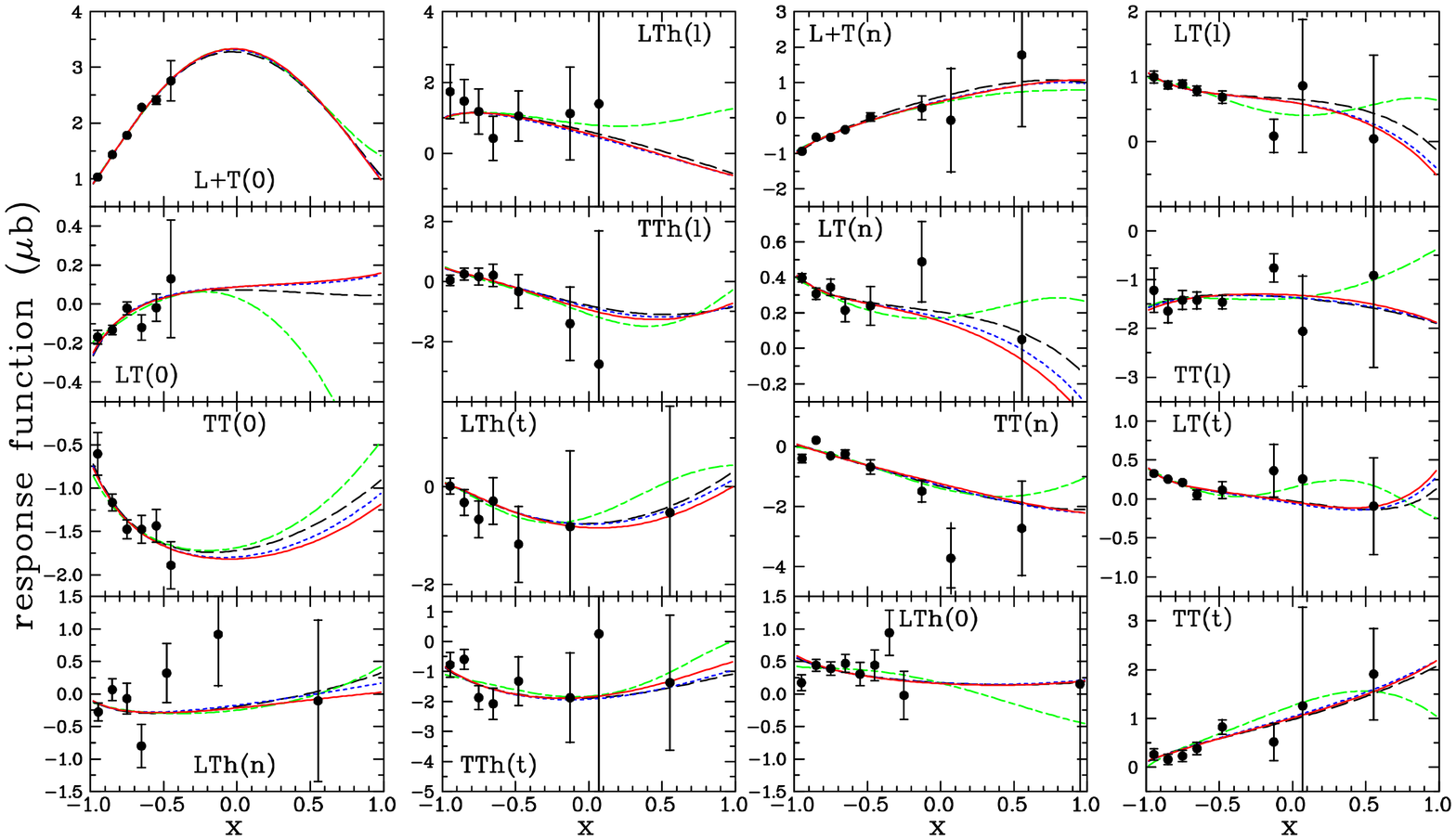}
\caption{(Color online)
Multipole fits for $W=1.31$ GeV at $Q^2=1.0$ (GeV/$c$)$^2$ using a 
Born baseline model.
See Fig. \protect{\ref{fig:mpfit_1170}} for legend.}
\label{fig:mpfit_1310}
\end{figure*}

\begin{figure*}
\centering
\includegraphics[angle=90,width=5in]{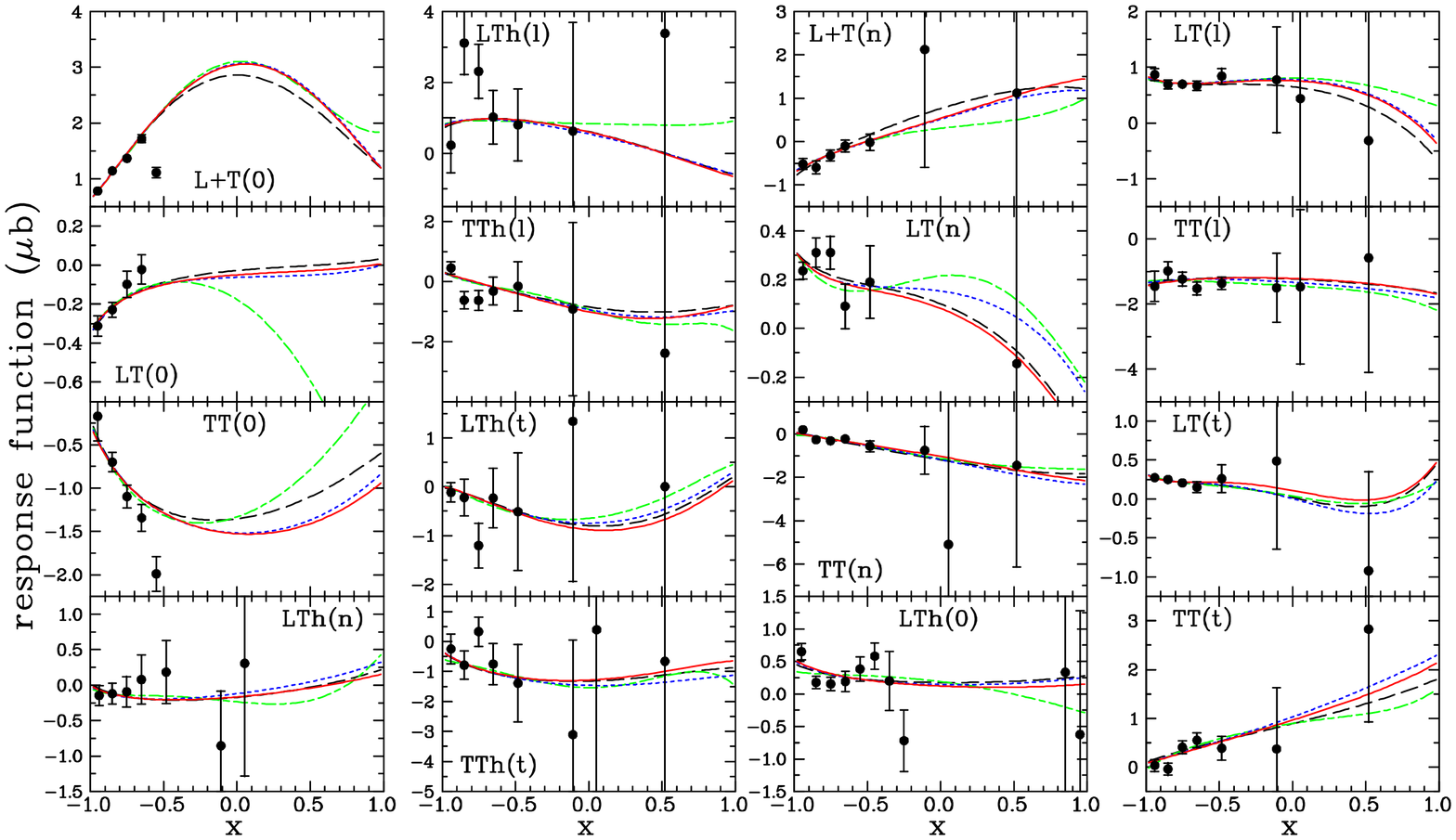}
\caption{(Color online)
Multipole fits for $W=1.33$ GeV at $Q^2=1.0$ (GeV/$c$)$^2$ using a 
Born baseline model.
See Fig. \protect{\ref{fig:mpfit_1170}} for legend.}
\label{fig:mpfit_1330}
\end{figure*}

\begin{figure*}
\centering
\includegraphics[angle=90,width=5in]{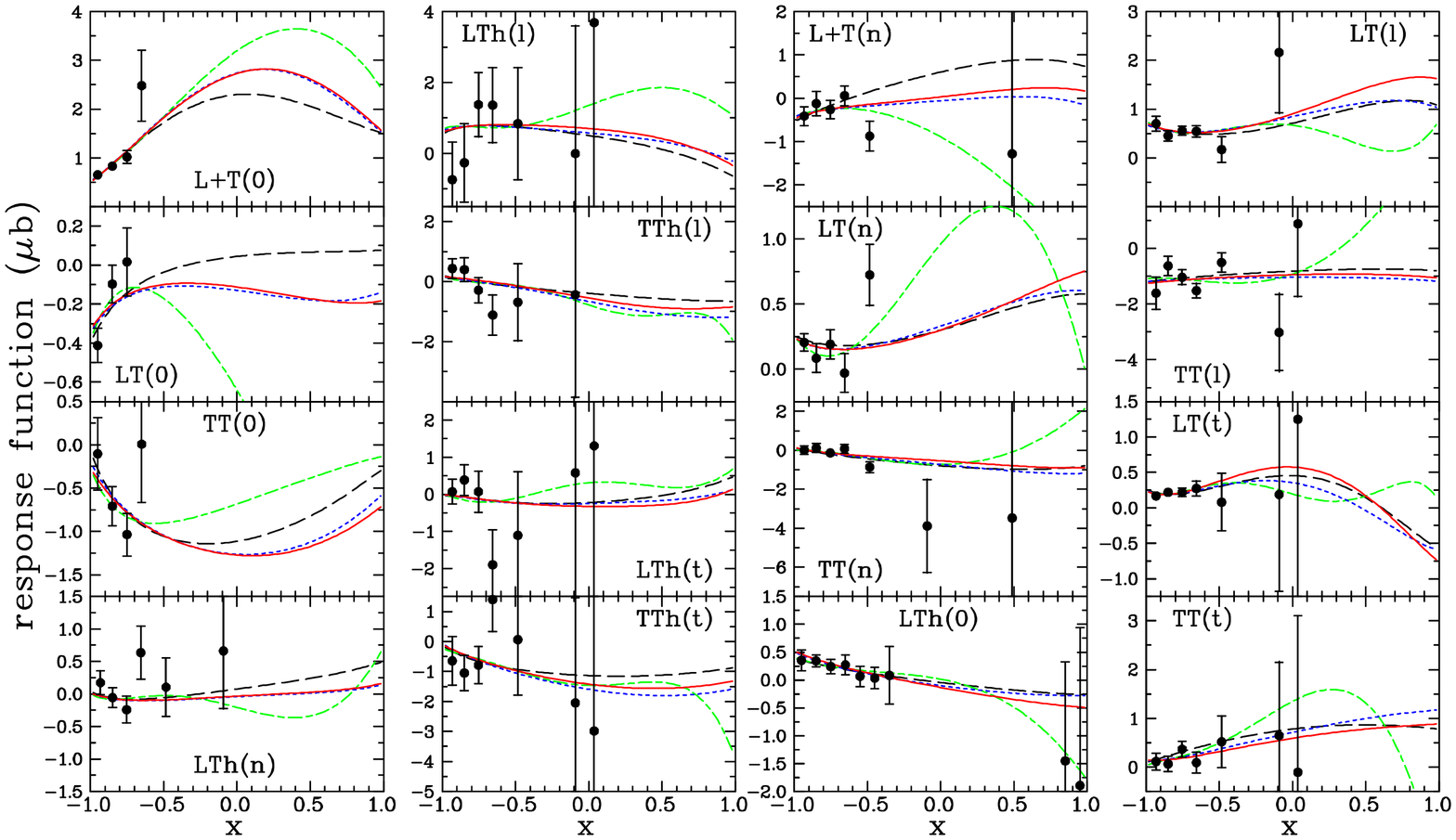}
\caption{(Color online)
Multipole fits for $W=1.35$ GeV at $Q^2=1.0$ (GeV/$c$)$^2$ using a 
Born baseline model.
See Fig. \protect{\ref{fig:mpfit_1170}} for legend.}
\label{fig:mpfit_1350}
\end{figure*}

Fits were performed for each $W$ bin to data consisting of 
the $(x,\phi)$ distributions of differential cross section and beam 
analyzing power plus $x$ distributions for 12 recoil-polarization 
response functions simultaneously.
Fits using Born amplitudes for pseudovector coupling as the baseline model
are shown in Figs. \ref{fig:mpfit_1170}-\ref{fig:mpfit_1350}.
Several fits were performed to determine the maximum number of parameters
that can be extracted without flattening the $\chi^2$ hypersurfaces
too severely or encountering uncontrollable correlations among parameters.
Dashed curves, designated $sp$, show fits that adjust real and imaginary 
parts of all $s$- and $p$-wave multipole amplitudes with higher partial 
waves constrained by the baseline model, 
here just real Born amplitudes without resonances.
The fits designated $spd3$ also vary the real parts of $2-$ multipoles 
and are shown as blue dotted curves.
Fits designated $spd$ vary real and imaginary parts of all amplitudes
with $\ell_\pi \leq 2$ and are shown as green dash-dotted curves. 
Finally, the fits designated ``final'' are similar to the $spd3$ fits
except that $\Im M_{1-}$ is held at baseline, here zero, for reasons
discussed below.
There are 14 free parameters for each $W$ in an $sp$ fit, 26 for 
an $spd$ fit, 17 for an $spd3$ fit, or 16 for the final fit. 
For comparison, Legendre fits vary 50 free parameters in the central
$W$ region.

These figures show that fitting just the $s$- and $p$-wave amplitudes, 
with a Born background, is already sufficient to obtain a good description 
of the data.
Although fitting $d$-wave amplitudes, or a subset thereof, sometimes provides
modest reductions of $\chi^2$, there is little systematic evidence that 
modification of $d$-wave or higher multipoles is really necessary.
However, it is also clear that variation of all $\ell_\pi \leq 2$ amplitudes
offers too much freedom --- the oscillations in green dash-dotted
curves for $W \ge 1.29$ GeV or $W \leq 1.19$ GeV are not needed to fit 
the data, are implausible in amplitude, and change too much from one 
$W$ to the next.

\begin{figure*}
\centering
\includegraphics[angle=90,width=5in]{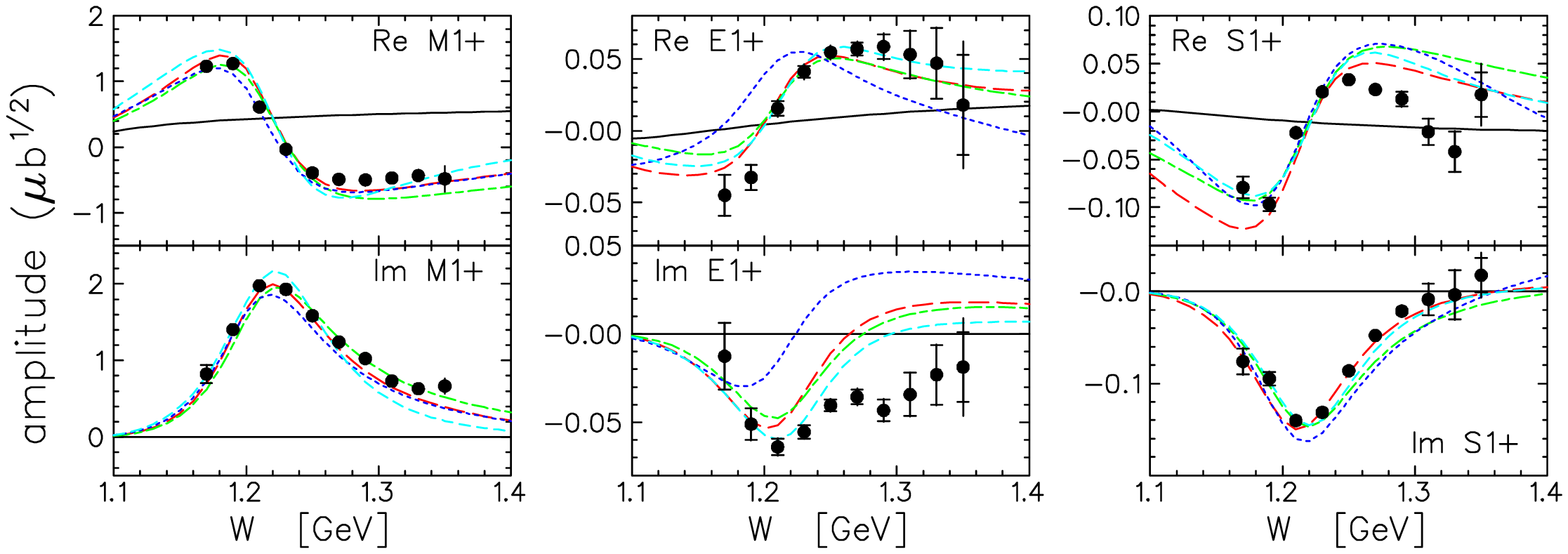}
\caption{(Color online)
Fitted $1+$ multipole amplitudes using Born baseline and adjusting
all $\ell_\pi$ amplitudes except $\Im M_{1-}$ plus real parts of
$2-$ amplitudes.
Inner error bars with endcaps are statistical;
outer error bars without endcaps include systematic uncertainties.
The baseline amplitudes are shown as black curves.
Several other recent models are shown also: MAID2003 (red dashed), 
DMT (green dot-dashed), SAID (blue dotted), and SL (cyan short-dashed).}
\label{fig:mpfit1p}
\end{figure*}

\begin{figure}
\centering
\includegraphics[angle=90,width=4in]{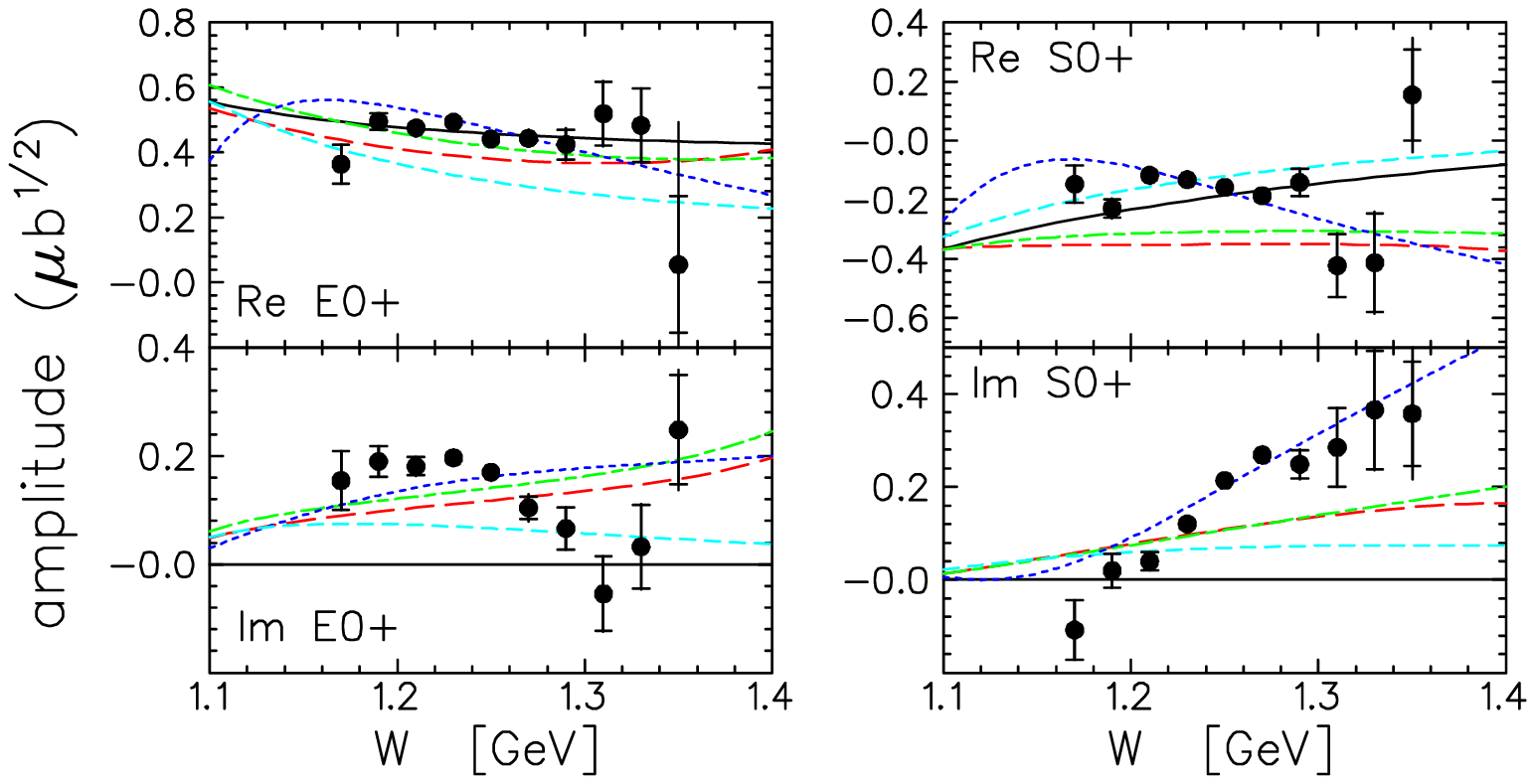}
\caption{(Color online)
Fitted $0+$ multipole amplitudes using Born baseline.
See Fig. \protect{\ref{fig:mpfit1p}} for legend.}
\label{fig:mpfit0p}
\end{figure}

\begin{figure}
\centering
\includegraphics[angle=90,width=4in]{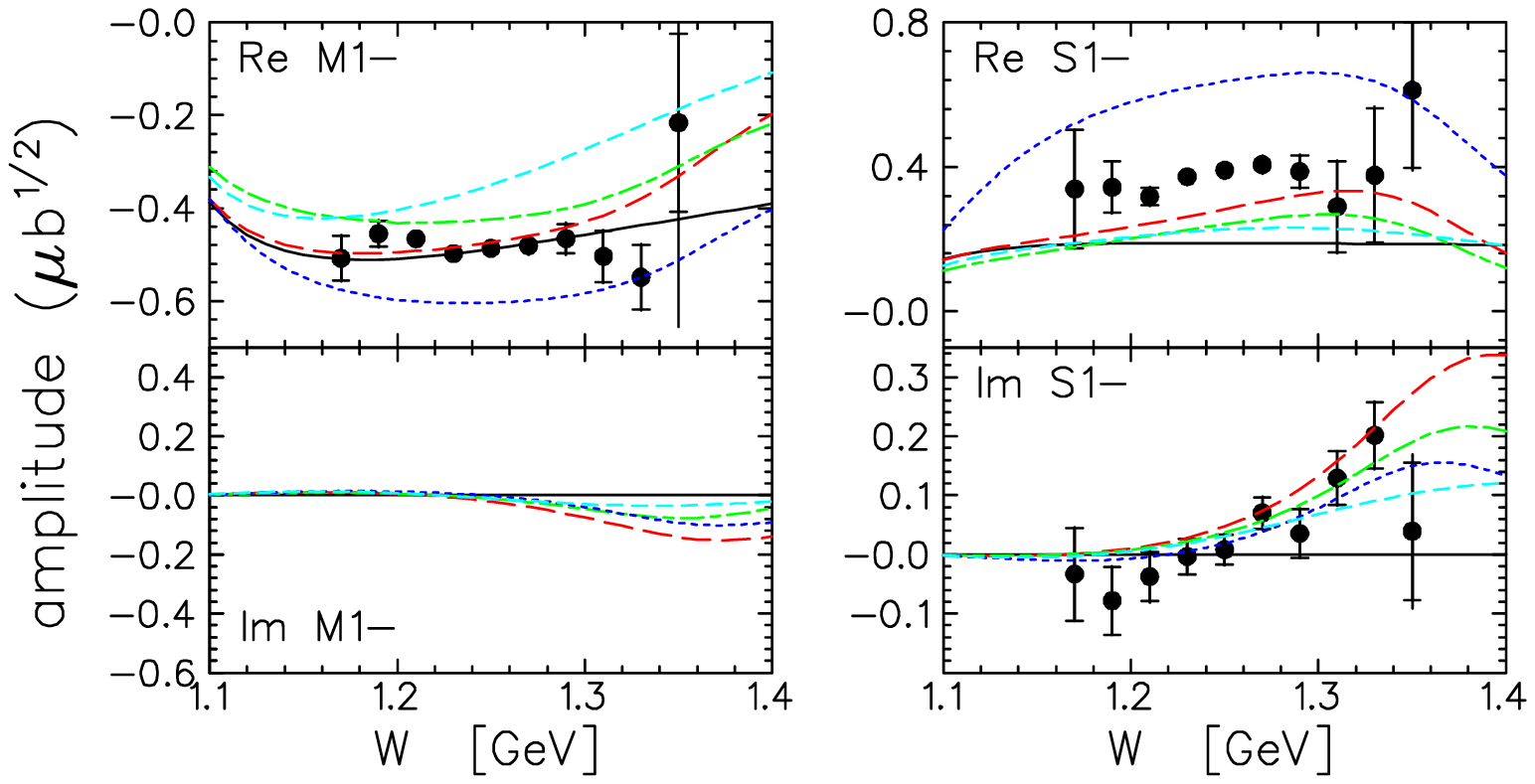}
\caption{(Color online)
Fitted $1-$ multipole amplitudes using Born baseline.
See Fig. \protect{\ref{fig:mpfit1p}} for legend.}
\label{fig:mpfit1m}
\end{figure}

\begin{figure*}
\centering
\includegraphics[angle=90,width=5in]{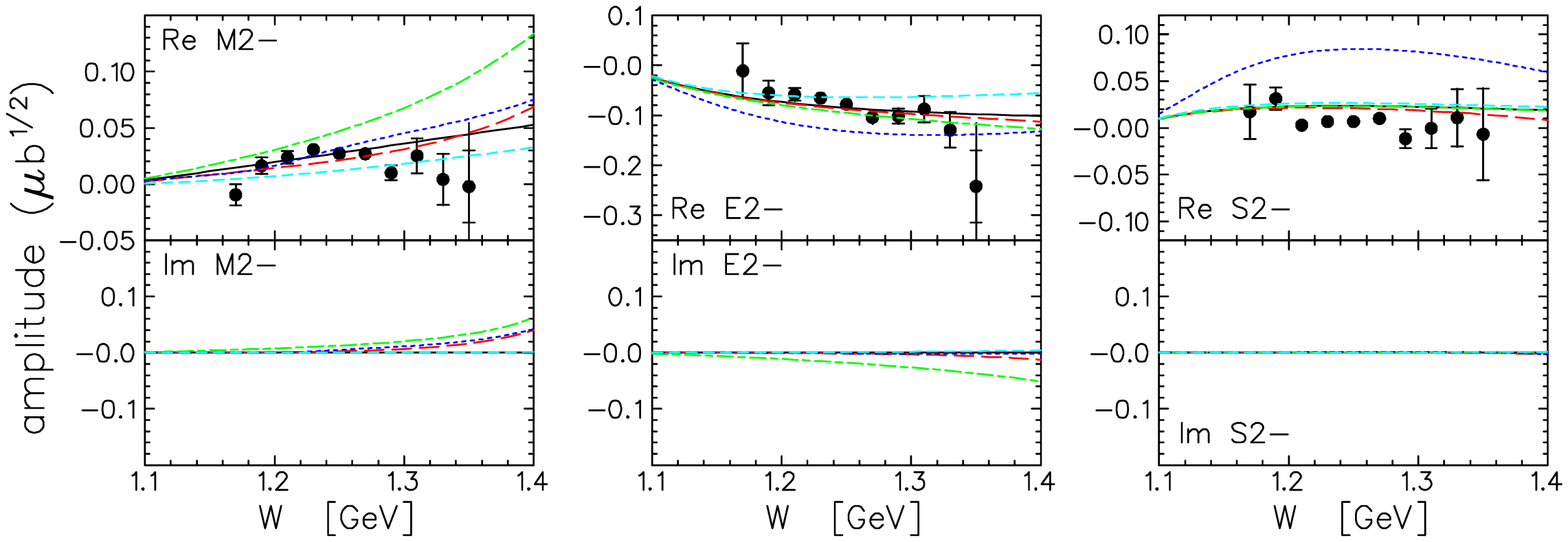}
\caption{(Color online)
Fitted $2-$ multipole amplitudes using Born baseline.
See Fig. \protect{\ref{fig:mpfit1p}} for legend.}
\label{fig:mpfit2m}
\end{figure*}

\begin{figure*}
\centering
\includegraphics[angle=90,width=5in]{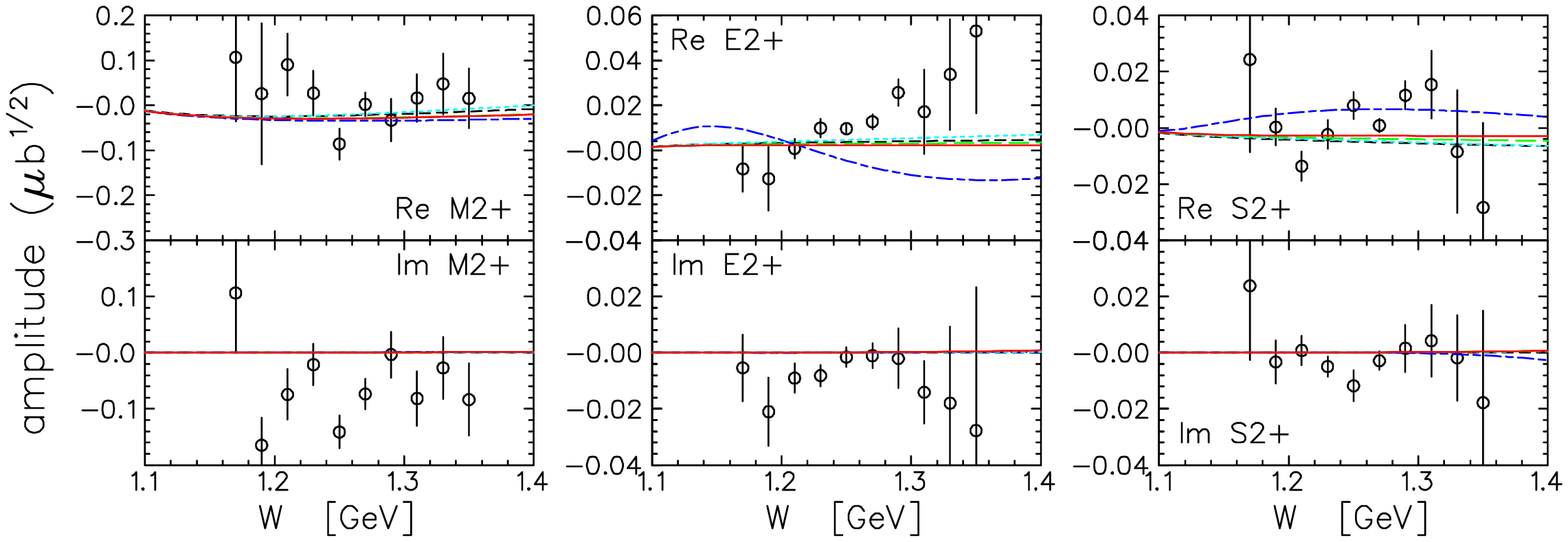}
\caption{(Color online)
Fitted $2+$ multipole amplitudes using Born baseline.
The open circles show results for the $spd$ analysis.
Only statistical errors are shown.
See Fig. \protect{\ref{fig:mpfit1p}} for curves.}
\label{fig:mpfit2p}
\end{figure*}

The fitted multipole amplitudes are compared in 
Figs. \ref{fig:mpfit1p}-\ref{fig:mpfit2p} with several recent 
models \cite{MAID,DMT,SAID,Sato01}.
In addition, the baseline Born amplitudes are shown by solid curves.
Note that all multipole amplitudes are real in this model and 
there are no resonances; therefore, the starting conditions 
are quite poor and large adjustments to the initial parameters
are required to fit the data.
Nevertheless, the fits describe the data well, the fitted 
parameters generally display smooth $W$ dependencies, and
the characteristic resonance profiles emerge in the $1+$ 
multipole amplitudes without coaching.
Note that the $sp$ and final fits began with baseline amplitudes 
but to improve stability the fits with more freedom began with 
the results of the final analysis.
The uncertainties increase with the number of free parameters
because the data do not adequately constrain multipoles for 
$\ell_\pi \ge 2$.
Thus, we reject the full $spd$ analysis because the uncertainties
in its parameters are large and the resultant oscillations in 
calculated response functions are not warranted by the available
data. 
The amplitudes and resulting fits for the other three analyses
tend to be very similar except that there is a rather strong
correlation between $\Im M_{1-}$ and $\Im S_{1-}$ for small $W$
that results in fitted values of opposite sign that are 
substantially larger than model predictions for $W \leq 1.21$ GeV.
This correlation also appears to affect imaginary parts of $1+$
amplitudes for $W=1.17$ GeV.
Evidently, the data for low $W$ do not distinguish between 
$\Im M_{1-}$ and $\Im S_{1-}$ well enough to fit both simultaneously.
Therefore, our final analysis eliminates $\Im M_{1-}$ because
models tend to predict stronger $\Im S_{1-}$ amplitudes and the 
$sp$ fit also produces rather small $\Im M_{1-}$ values.
The uncertainties in fitted multipoles is reduced and the $W$ 
dependencies are improved, especially for imaginary $1+$ amplitudes, 
upon elimination of this redundant parameter.
Furthermore, Figs. \ref{fig:mpfit_1170}-\ref{fig:mpfit_1350} demonstrate 
that elimination of $\Im M_{1-}$ does not visibly reduce the quality of
the fits to the data.
The ``final'' analysis varies both real and imaginary parts of
$S_{1-}$ and all $0+$ and $1+$ multipoles plus the real parts of
$M_{1-}$ and all $2-$ multipoles for a total of 16 free parameters
for each $(W,Q^2)$ bin.

There is rather little spread among models for $M_{1+}$ amplitudes
across the $\Delta(1232)$ resonance and our experimental amplitudes
agree well with model calculations even when the fit is based upon
a Born baseline model without resonances.
The variation among models is also relatively small for $S_{1+}$
amplitudes and good agreement is obtained with data for $\Im S_{1+}$, 
but for the real part the present data exhibit a steeper slope on 
the large $W$ side.
MAID2003, DMT, and SL calculations are similar for $E_{1+}$, but the
current SAID calculations are substantially different and disagree 
with the data.
Our results for $\Re E_{1+}$ agree relatively well with MAID2003,
DMT, or SL but the $\Im E_{1+}$ data do not show the node 
near $W \approx 1.27$ GeV predicted by those models; there is no
sign change for $W \leq 1.35$ GeV.

Among the models considered, MAID2003 tends to provide the best 
description of the recoil-polarization data, but it does not
reproduce the $R_{LT}^{\ell}$, $R_{LT}^t$, or $R_{LT}^n$ angular
distributions (see Figs. 
\ref{fig:models_rsfns_1210}-\ref{fig:models_rsfns_1290}).
These difficulties appear to arise primarily from the $S_{0+}$ 
amplitudes.
Whereas MAID2003 suggests a nearly constant $\Re S_{0+}$ 
amplitude in this $W$ range, we find less negative results that are
consistent with the negative slope in $W$ suggested by SAID.
Although the SL calculation crosses the $\Re S_{0+}$ data near the 
middle of the $W$ range, it has the opposite slope.
We also find a rather steep slope for $\Im S_{0+}$. 
It is interesting to observe that the SAID model agrees best with
the $0+$ amplitudes even though it is worst, among these models, 
for $E_{1+}$, $\Re 1-$, and $S_{2-}$.
All of the models agree pretty well with the $\Re E_{0+}$ data,
but none reproduce the $W$ dependence seen for $\Im E_{0+}$.

It is interesting to observe that there is a wide spread
among the models for $\Re M_{1-}$ but that the data are 
closest to the Born model that omits the Roper resonance,
which suggests that the transverse amplitude $_p A_{1/2}$ is small.
On the other hand, the fitted $\Re S_{1-}$ does differ from
the Born model and suggests that there is a nonnegligible 
longitudinal contribution from the Roper consistent with a 
radial excitation. 
These models agree fairly well with the $\Im S_{1-}$ data but,
with the exception of the large SAID prediction, are smaller
than the $\Re S_{1-}$ data.
Thus, it appears that excitation of the Roper resonance is
primarily longitudinal at $Q^2 = 1$ (GeV/$c$)$^2$.
We will return to this issue in Sec. \ref{sec:UIM} using
a unitary isobar model.

The real $2-$ amplitudes are small in this $W$ range, but
the slope for $\Re E_{2-}$ appears to be determined well by
these data and is in good agreement with Born, MAID2003, or 
DMT predictions.
However, the predictions of the SAID model are much larger 
than the data for $\Re S_{2-}$.
SAID also predicts significant oscillations in $2+$ amplitudes
that are absent in other models.
Although we cannot fit the $2+$ amplitudes accurately, 
the large $\ell_\pi=2$ amplitudes for SAID produce oscillations in 
many of the response functions that are not warranted by the data.


\subsection{Quadrupole ratios}
\label{sec:quad}

\begin{figure}
\centering
\includegraphics[width=3in]{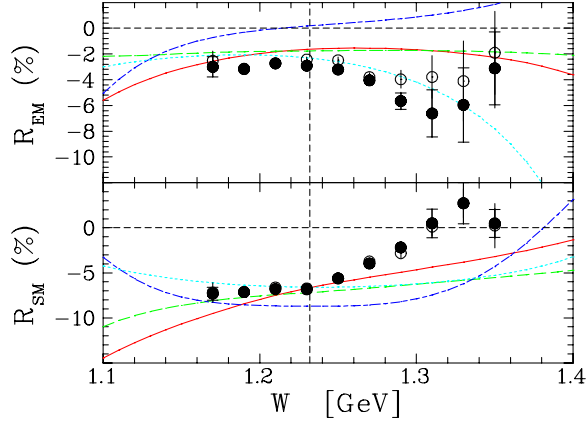}
\caption{(Color online)
Multipole analyses for EMR and SMR at $Q^2=1.0$ (GeV/$c$)$^2$ are 
compared with MAID2003 (solid red), DMT (dashed green), 
SAID (dash-dot blue), and SL (dotted cyan).
The vertical line shows the physical mass, $W=M_\Delta$.
Open circles adjust $\ell_\pi \leq 1$ multipoles while filled circles
represent the final fits.
For filled circles, inner error bars with endcaps are statistical while
outer error bars without endcaps include systematic uncertainties.}
\label{fig:mpfit_quad}
\end{figure}

The quadrupole deformation parameters can now be obtained directly from
the fitted multipole amplitudes using
\begin{subequations}
\label{eq:quad-mp}
\begin{eqnarray}
\Re \frac{E_{1+}}{M_{1+}} &=& \frac{\Re E_{1+} M_{1+}^\ast }{|M_{1+}|^2} =
\frac{\Re E_{1+} \Re M_{1+} + \Im E_{1+} \Im M_{1+}}
{\Re M_{1+} \Re M_{1+} + \Im M_{1+} \Im M_{1+}} \\
\Re \frac{S_{1+}}{M_{1+}} &=& \frac{\Re S_{1+} M_{1+}^\ast }{|M_{1+}|^2} = 
\frac{\Re S_{1+} \Re M_{1+} + \Im S_{1+} \Im M_{1+}}
{\Re M_{1+} \Re M_{1+} + \Im M_{1+} \Im M_{1+}}
\end{eqnarray}
\end{subequations}
where the multipole analysis provides the real and imaginary parts of
each amplitude separately.
The $W$ dependencies of quadrupole ratios for the $p\pi^0$ channel 
are shown in Fig. \ref{fig:mpfit_quad} and the results at $W=1.23$ GeV 
are listed in Table \ref{table:born_quad};
the correction for the small isospin $1/2$ contamination is discussed
in Sec. \ref{sec:discussion}.
The truncation dependence of the experimental results is relatively small for 
$R_{SM}^{(p\pi^0)}$ and most of the models are in good agreement with the data
for $W \approx M_\Delta$, but SMR is significantly stronger for SAID.
Although the truncation dependence is larger for $R_{EM}^{(p\pi^0)}$ data, 
the value at $M_\Delta$ still appears to be determined relatively well and
is consistent with all of the models except SAID, which gives
a much smaller value and at larger $Q^2$ sign opposite other models.
The model calculations spread more rapidly for the electric than for the 
scalar ratio as the distance from $M_\Delta$ increases.

Note, that if one defines $M_\Delta$ as the $W$ where $\Re M^{(3/2)}_{1+}=0$, 
then the quadrupole formulas in Eq. (\ref{eq:quad-mp}) reduce to
\begin{subequations}
\label{eq:quad-delta}
\begin{eqnarray}
R^{(3/2)}_{EM} &=& \Im E^{(3/2)}_{1+} / \Im M^{(3/2)}_{1+} \\
R^{(3/2)}_{SM} &=& \Im S^{(3/2)}_{1+} / \Im M^{(3/2)}_{1+}
\end{eqnarray}
\end{subequations}
for the isospin-3/2 channel.
However, these formulas are unsuitable for data analysis because comparable
$n\pi^+$ data are not available for isospin decomposition and because the
appropriate value of $M_\Delta$ is not known precisely or independently of
models.
It would also be necessary, in principle, to interpolate the multipole data 
with respect to $W$.
We employ Eq. (\ref{eq:quad-mp}) because it is independent of $W$, 
applies equally well to $p\pi^0$ or isospin-3/2, 
and does not require any model-dependent assumptions about $M_\Delta$.
Furthermore, because the energy dependence in Fig. \ref{fig:mpfit_quad} is
quite mild, Table \ref{table:born_quad} simply lists values for the bin 
closest to $M_\Delta$, namely $W=1.23$ GeV.
Small corrections for the energy dependence of these quantites are
evaluated in Sec. \ref{sec:quad-Wdep}

Table \ref{table:born_quad} evaluates the sensitivity of quadrupole
ratios to the selection of adjustable multipoles.
The uncertainties increase when higher partial waves that are not 
well-constrained by the data are permitted to vary.  
Above we argued that the best compromise is obtained by varying
$0+$, $1+$, $1-$, and real parts of $2-$ multipole amplitudes 
with $2+$ and higher partial waves constrained by the baseline model.
Elimination of $\Im M_{1-}$ further reduces the uncertainties in SMR,
without affecting the quality of the fit, by suppressing its unresolvable
correlation with $\Im S_{1-}$.
As previously argued, we believe that elimination of $\Im M_{1-}$ is
justified by the prediction, by all models considered, that it is 
negligible in this energy range.
The results in the last two lines of Table \ref{table:born_quad}
are practically identical for SMR, though with reduced uncertainty in
the final line, while the change in EMR is within the estimated 
uncertainites.

\begin{table}
\caption{Quadrupole ratios for $W=1.23$ GeV at $Q^2 = 1.0$ (GeV/$c$)$^2$
using the pseudovector Born baseline model.
Only statistical uncertainties from fitting are given.}
\label{table:born_quad}
\begin{ruledtabular}
\begin{tabular}{lllll}
variables &  SMR, \% & EMR, \% & $\chi^2_\nu$ \\ \hline
$0+,1+,1-$                & $-6.73 \pm 0.24$ & $-2.43 \pm 0.19$ & 1.69 \\
$0+,1+,1-,2-,2+$          & $-6.95 \pm 0.49$ & $-3.19 \pm 0.79$ & 1.64 \\
$0+,1+,1-,\Re 2-$         & $-6.85 \pm 0.27$ & $-2.73 \pm 0.20$ & 1.65 \\
above except $\Im M_{1-}$ & $-6.84 \pm 0.15$ & $-2.91 \pm 0.19$ & 1.65 \\
\end{tabular}
\end{ruledtabular}
\end{table}

\subsection{Sensitivity to baseline model}
\label{sec:baselines}

\begin{figure*}
\centering
\includegraphics[angle=90,width=5in]{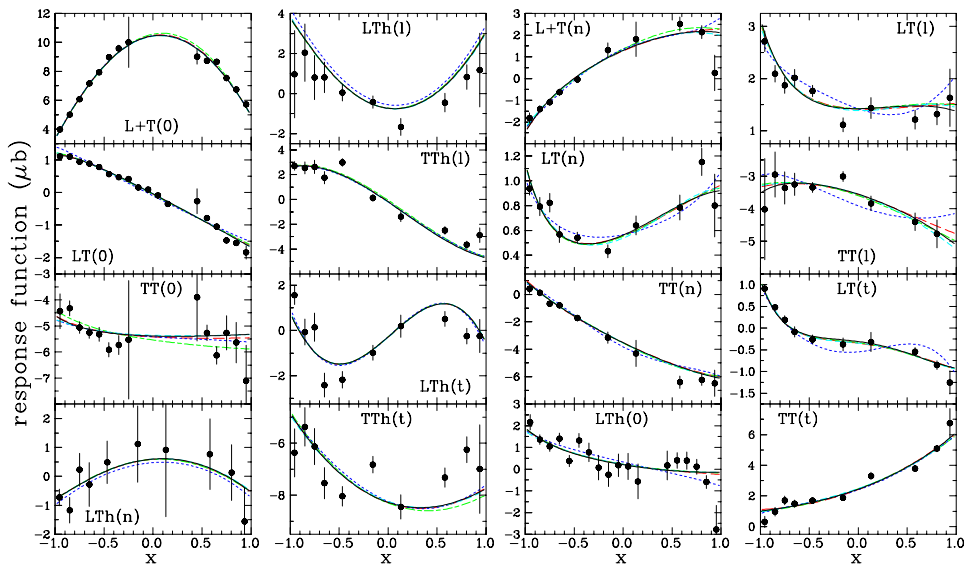}
\caption{(Color online)
Sensitivity of multipole fits for $W=1.23$ GeV at $Q^2=1.0$ 
(GeV/$c$)$^2$ to the choice of baseline model:
Born (black solid), MAID2003 (red dashed), 
DMT (green dash-dotted), SAID (blue dotted), or
SL (cyan short-dotted).}
\label{fig:base_1230}
\end{figure*}

As mentioned above, the multipole fits are rather insensitive to the
choice of baseline model.
To illustrate this, Fig. \ref{fig:base_1230} compares fits to the
response functions for $W=1.23$ GeV based upon several baseline
models; figures for other $W$ bins are available in Ref. \cite{e91011_mpamps}.
The fits based upon Born, MAID2003, DMT, or SL models are practically
indistinguishable.
The fits based upon SAID display a more oscillatory structure that
is not supported by the data in the middle of the $W$ range where the
precision is best.
The oscillations are presumably due to relatively large $\Re E_{2+}$ 
and $\Re S_{2+}$ amplitudes that are not ameliorated by the current
truncation scheme.
Therefore, the data clearly require smaller $2+$ amplitudes than
predicted by the SAID model.

Similarly, the fitted multipole amplitudes are also rather insensitive 
to the choice of baseline model.
Even the fits based upon SAID, starting from rather different initial
conditions and with significantly larger fixed $2+$ amplitudes, 
converge upon essentially the same final results.
For example, Table \ref{table:base_quad} lists the quadrupole ratios for 
$W=1.23$ GeV based upon several choices of baseline models and using 
the ``final'' parameter space.
All of the results are consistent except those using the SAID
baseline, for which SMR is substantially higher and EMR lower
than for other baseline models.
However, the quality of the fit is also noticeably inferior even though
the differences in $\chi^2_\nu$ are not impressive.
Therefore, we conclude that this version of the SAID model does not
provide a suitable baseline for multipole analysis and judge the
sensitivity to uncertainties in the baseline model to be similar to
the tabulated fitting uncertainties.

\begin{table}
\caption{Dependence of quadrupole ratios for $W=1.23$ GeV at 
$Q^2 = 1.0$ (GeV/$c$)$^2$ upon baseline model.
All $s$- and $p$-wave amplitudes, except $\Im M_{1-}$, plus
real $2-$ amplitudes were fit with other amplitudes given
by the specified baseline model.}
\label{table:base_quad}
\begin{ruledtabular}
\begin{tabular}{lllll}
baseline &  SMR, \% & EMR, \% & $\chi^2_\nu$ \\ \hline
Born     & $-6.84 \pm 0.15$ & $-2.91 \pm 0.19$ & 1.65 \\
MAID2003 & $-6.90 \pm 0.15$ & $-2.79 \pm 0.19$ & 1.67 \\
DMT      & $-6.82 \pm 0.15$ & $-2.70 \pm 0.19$ & 1.67 \\
SL       & $-6.79 \pm 0.15$ & $-2.81 \pm 0.19$ & 1.64 \\
SAID     & $-7.38 \pm 0.15$ & $-2.53 \pm 0.20$ & 1.85 \\
\end{tabular}
\end{ruledtabular}
\end{table}

\section{Discussion}
\label{sec:discussion}

\subsection{Reliability of traditional Legendre analysis}
\label{sec:Legendre-reliability} 

Both the Legendre and multipole analyses fit the data well, but they
yield significantly different estimates of the $N \rightarrow \Delta$
quadrupole ratios.
The Legendre results are listed in the first line of 
Table \ref{table:quad_trunc} and those for the multipole analysis
in the second and fifth columns of the second line.
Subsequent lines show model calculations for quadrupole ratios 
based upon several truncation schemes.
The second and fifth columns are the proper ratios of multipole
amplitudes while the remaining columns use the traditional estimators 
given by Eq. (\ref{eq:quad-leg}) with Legendre coefficients that 
were computed by numerical integration of response functions obtained
from the indicated truncation of the multipole amplitudes with respect
to $\ell_\pi$.
Thus, the third and sixth columns represent the $sp$ truncation while the
fourth and seventh columns are practically complete with 
respect to $\ell_\pi$.
We placed the experimental Legendre results in the $\ell_\pi \leq 5$
columns because truncation is not possible experimentally.  
The model Legendre coefficients were computed without using $M_{1+}$ 
dominance but the corresponding traditional quadrupole estimators, 
$\tilde{R}_{SM}$ and $\tilde{R}_{EM}$, employ combinations that were
derived under that assumption.
The values of $\tilde{R}_{SM}$ and $\tilde{R}_{EM}$ for $\ell_\pi \leq 5$ 
obtained from the fitted multipole amplitudes are similar to those 
obtained from the fitted Legendre coefficients but are distinctly
smaller than the fitted values for $R_{SM}$ and $R_{EM}$ even though
the fits to the cross section data are practically identical.
The differences between $\ell_\pi \leq 1$ and $\ell_\pi \leq 5$ model 
calculations demonstrate that $sp$ truncation is often a poor 
approximation to $\tilde{R}_{SM}$ and $\tilde{R}_{EM}$, 
especially for the latter.
Although the $sp$ truncation of $\tilde{R}_{SM}$ is reasonable for 
the SAID and SL models and for the present multipole fit, it is 
inaccurate for the MAID2003 and DMT models.
However, $sp$ truncation of $\tilde{R}_{EM}$ is quite poor for 
all models considered.
Furthermore, the correspondence between the traditional Legendre
estimators and the actual quadrupole ratios also depends upon the
requirement that $M_{1+}$ appears in every term of the multipole
expansion of Legendre coefficients.
The differences between the $\ell_\pi \leq 5$ results and the actual
quadrupole ratios demonstrate that the assumption of $M_{1+}$
dominance is not sufficiently accurate either.

A more detailed study of truncation errors in the traditional Legendre 
analysis of $N \rightarrow \Delta$ quadrupole ratios has been provided
in Ref. \cite{Kelly05e}.
Truncation errors are especially severe for $\tilde{R}_{EM}$ where the
contribution of $\Re M_{1-}E_{1+}^*$ alone is approximately $-40\%$ of 
the leading term using MAID2003 $p\pi^0$ multipoles for our kinematics.
Many other neglected terms are significant and the convergence is 
slow and model dependent.
Furthermore, the contributions of $A_0^L$ and $A_2^L$ to 
Eq. (\ref{eq:quad-leg}) are not negligible, as assumed using 
$M_{1+}$ dominance.
The contribution of $A_2^L$ can, in fact, have a large effect upon 
delicate cancellations within the numerator of $\tilde{R}_{EM}$.
Thus, Rosenbluth separation should be performed before using the
Legendre method, especially for $\tilde{R}_{EM}$, but none of the recent 
Legendre analyses \cite{Kalleicher97,Frolov99,Joo02,Sparveris05} have done 
so, including the present experiment.
The convergence of $\tilde{R}_{SM}$ is better, but its relative accuracy as 
an estimate of $R_{SM}$ is still no better than about $20\%$ \cite{Kelly05e}. 
Therefore, although the details are model dependent, it is clear
that neither assumption of the traditional Legendre analysis 
is sufficiently accurate at the present levels of experimental
precision and completeness.

\begin{table}
\caption{Calculated quadrupole ratios for 
$W =1.23$ GeV at $Q^2 = 1.0$ (GeV/$c$)$^2$.
Columns with ranges of $\ell_\pi$ are based upon traditional estimators 
given by Eq. (\ref{eq:quad-leg}). }
\label{table:quad_trunc}
\begin{ruledtabular}
\begin{tabular}{lrrrrrr}
model & 
$R_{SM}^{(p\pi^0)}$, \% & $\tilde{R}_{SM}^{(p\pi^0)}$, \% & $\tilde{R}_{SM}^{(p\pi^0)}$, \% & 
$R_{EM}^{(p\pi^0)}$, \% & $\tilde{R}_{EM}^{(p\pi^0)}$, \% & $\tilde{R}_{EM}^{(p\pi^0)}$, \% \\ 
&  & $\ell_\pi \leq 1$ & $\ell_\pi \leq 5$ &  & $\ell_\pi \leq 1$ & $\ell_\pi \leq 5$  \\ \hline
Legendre fit    &          &         & $-6.11$ &         &         & $-1.92$ \\ 
multipole fit   &  $-6.84$ & $-6.46$ & $-6.00$ & $-2.91$ & $-1.54$ & $-2.18$ \\
MAID2003        &  $-6.73$ & $-6.37$ & $-5.63$ & $-1.65$ & $-0.57$ & $-1.12$ \\
DMT             &  $-7.21$ & $-6.77$ & $-6.10$ & $-1.77$ & $-0.70$ & $-1.47$ \\
SAID            &  $-8.71$ & $-7.78$ & $-7.88$ & $+0.17$ & $+1.96$ & $+0.22$ \\
SL              &  $-6.59$ & $-6.69$ & $-6.58$ & $-2.29$ & $-1.29$ & $-1.58$ \\ 
\end{tabular}
\end{ruledtabular}
\end{table}

\subsection{Isospin-1/2 contamination of EMR and SMR}

Separation of the isospin $1/2$ and $3/2$ contributions to the multipole
amplitudes would require comparable data for the $n\pi^+$ channel, including
angular distributions for either recoil or target polarization,
which are presently unavailable.
Fortunately, the isospin $1/2$ contamination is expected to have relatively 
little effect upon the determination of isospin $3/2$ quadrupole ratios.
According to MAID2003, one expects $(R_{SM},R_{EM}) = (-6.71\%,-1.62\%)$
at $(W,Q^2)=(1.23,1.0)$ 
for isospin $3/2$ compared with $(-6.73\%,-1.65\%)$ for the $p\pi^0$ 
channel \cite{MAID}.
Similarly, the quadrupole ratios for the pure $N \rightarrow \Delta$ 
contribution become $(-6.73\%,-1.53\%)$ in the absence of background.  
Finally, if one attributes the correction terms for $1+$ multipoles 
in Eq. (\ref{eq:mpfit}) entirely to the $\Delta$ resonance, 
assuming that the Born baseline model is accurate, we would estimate
\begin{subequations}
\begin{eqnarray}
R_{SM}^{(3/2)} &\approx& \Re \frac{\Delta S_{1+}}{\Delta M_{1+}} = -6.81\% \\
R_{EM}^{(3/2)} &\approx& \Re \frac{\Delta E_{1+}}{\Delta M_{1+}} = -3.12\% 
\end{eqnarray}
\end{subequations}
in good agreement with the full results for the $p\pi^0$ channel
that include background.
However, the fact that changes in $R_{EM}$ due to neglect of background
are opposite for MAID2003 and the experimental multipole fit suggests
that part of the fitted $\Delta E_{1+}$ should probably be attributed
to the background in the baseline model.
Nevertheless, it appears that corrections for isospin $1/2$ contamination
are probably smaller than the present error bars. 
Therefore, this model-dependent correction has not been made. 

\subsection{$W$ dependence of EMR and SMR}
\label{sec:quad-Wdep}

\begin{figure}
\centering
\includegraphics[width=2.5in]{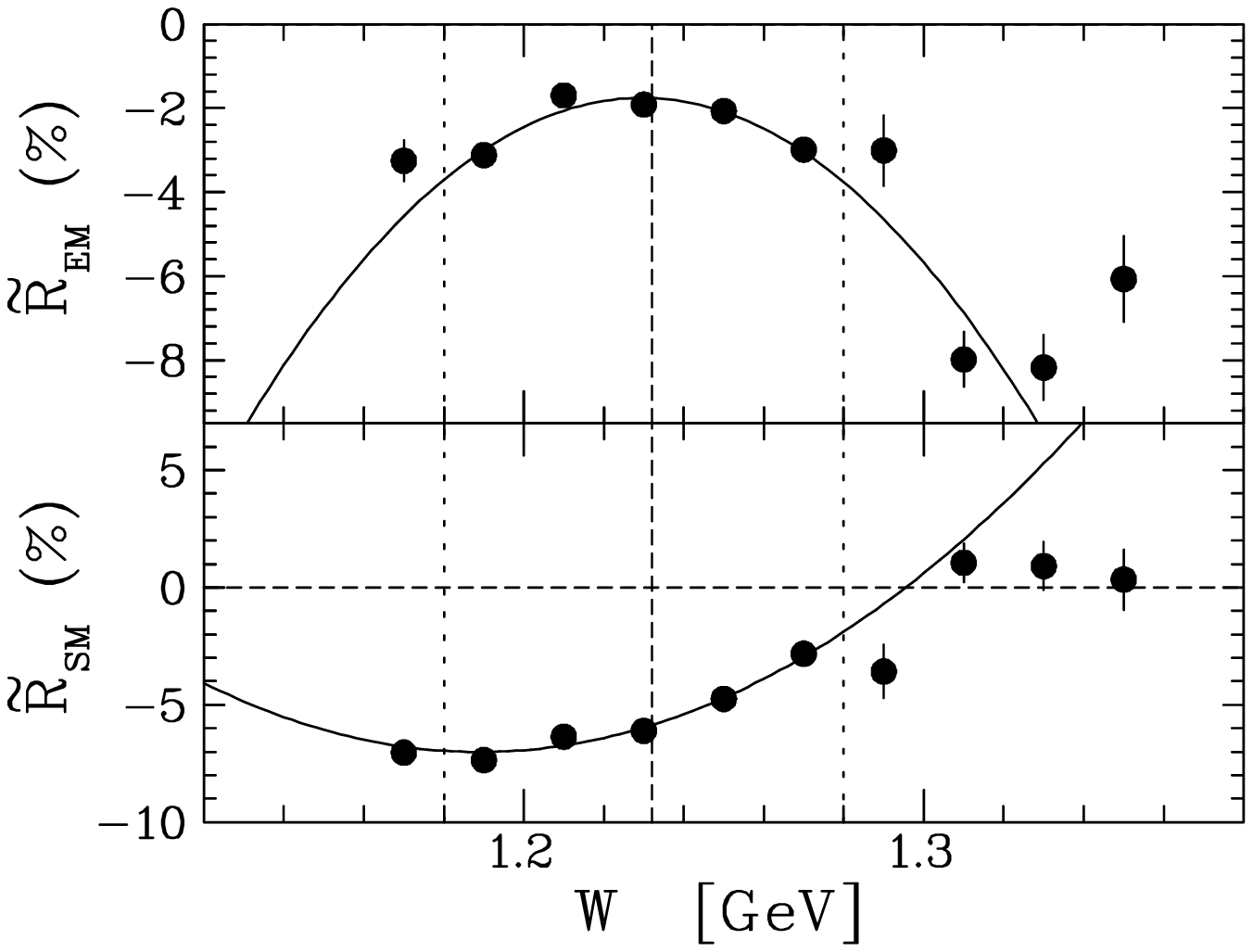}
\includegraphics[width=2.5in]{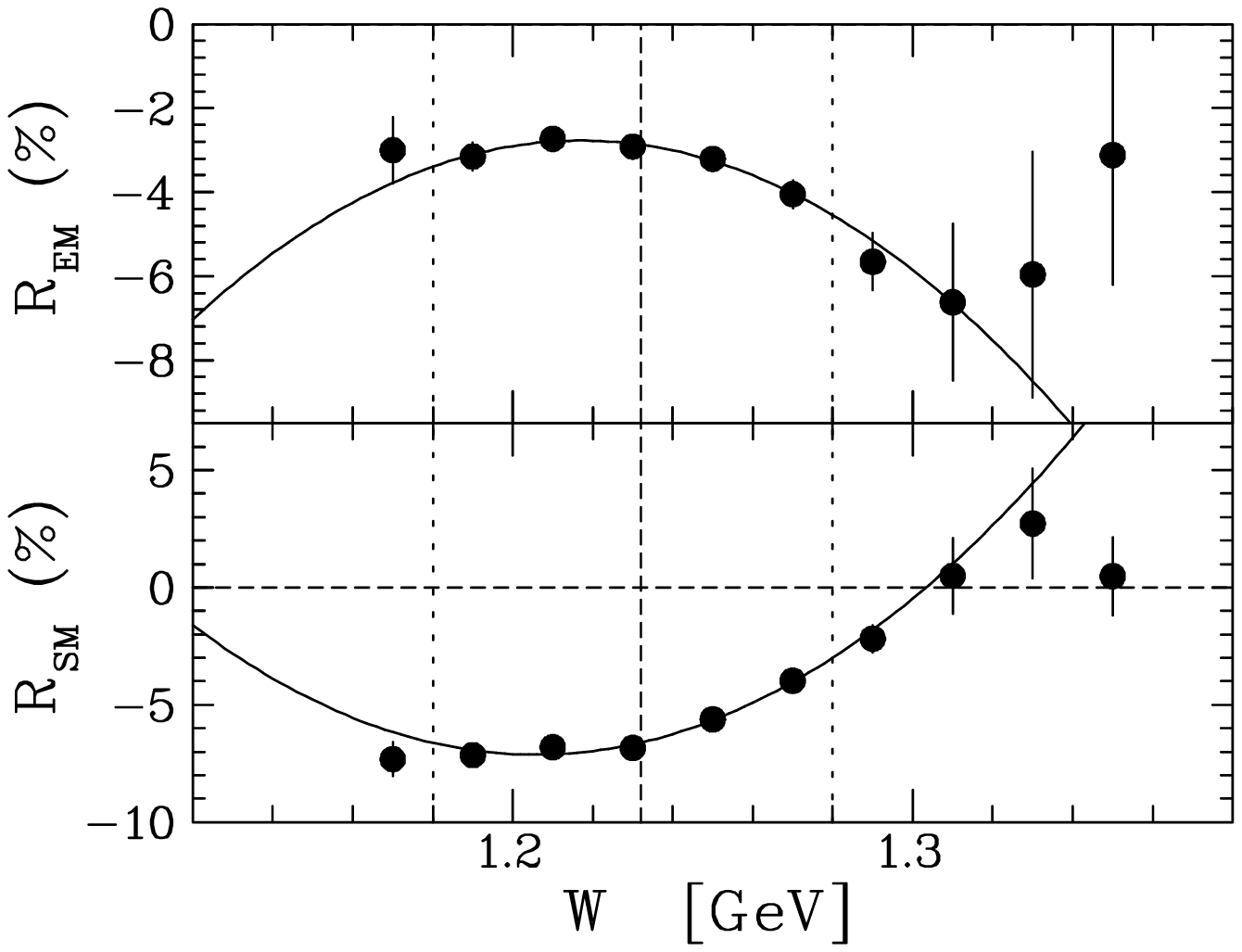}
\caption{Parabolic fits to the $W$ dependence of quadrupole ratios
from Legendre (left) and multipole (right) analyses.
Vertical dashed and dotted lines show $M_\Delta$ and the fitted
ranges of $W$.}
\label{fig:quad-Wdep}
\end{figure}

The $W$ dependencies of quadrupole ratios obtained using both Legendre and
multipole analyses are compared in Fig. \ref{fig:quad-Wdep} with
parabolic fits of the form
\begin{equation}
y = \sum_{k=0}^{2} a_k (W - M_\Delta)^k
\end{equation}
using a nominal value of $M_\Delta = 1.232$ GeV.
The fits were confined to the central region, indicated by dotted vertical
lines, where this simple parametrization should suffice for interpolation.
Both fits describe the data for the central region well.
It is important to remember that the quadrupole ratios for the two types of 
analysis are different quantities and need not have the same shapes ---
the Legendre estimators are affected by $\epsilon$ and by all partial 
waves while those from multipole analysis are not.
The parabolic fits appear to extrapolate better for the multipole analysis 
than for the Legendre analysis but should still only be used in the central 
region.

The expansion coefficients fitted using the weighted linear least-squares 
method are given in Table \ref{table:quad-Wdep}, where
the data for quadrupole ratios are expressed in percent and where the
multipole amplitudes are based upon the Born baseline.
The $a_0$ parameters represent the best estimates of the quadrupole
ratios at $M_\Delta$ but fits with only 2 degrees of freedom do not 
necessarily provide realistic uncertainties.
Instead, we quote the largest of $\delta R(1.23)$, $\delta a_0$, 
and $\delta a_0 \sqrt{\chi_\nu^2}$ where $\delta R(1.23)$ is the 
uncertainty in the single-energy fit to data for $W=1.23$ GeV, 
$\delta a_0$ is the uncertainty in the value of $a_0$ according to the
linear-least squares fit to the energy dependence, and 
$\chi_\nu^2$ is the reduced chisquare for that fit in the central region.
The second-order terms are negligible between $M_\Delta$ and the nearest
$W$ bin and changes due to the linear terms are less than one 
standard deviation.
For example, the fitted EMR and SMR values for the multipole analysis
are $-2.85$ and $-6.69\%$ at $W = 1.23$ GeV.
The remaining differences between these values and those listed on
the last line of Table \ref{table:born_quad} are less than one 
standard deviation.
Although the quadrupole ratios in Table \ref{table:quad-Wdep} 
are slightly smaller, they are consistent with those we reported in 
Ref. \cite{Kelly05c}.
The analysis in Ref. \cite{Kelly05c} considered only a single energy,
$W=1.23$ GeV, while the present analysis fits the energy dependence 
within the central region.
The small differences are partly due to the change in $W$ from 1.23 to  
1.232 GeV and partly due to statistical fluctuations of the data for 
1.23 GeV relative to the average trends represented by the curves in  
Fig. \ref{fig:quad-Wdep}.
Therefore, we consider the interpolated values in Table \ref{table:quad-Wdep} 
to be the best estimates of the quadrupole ratios for $W=M_\Delta$.

\begin{table}
\caption{Power series for quadrupole ratios in terms of $(W-M_\Delta)$ in 
units of GeV.  
The best estimate of the quadrupole ratios at $W=M_\Delta=1.232$ GeV is given 
for each analysis method by $a_0$.}
\label{table:quad-Wdep}
\begin{ruledtabular}
\begin{tabular}{lcrr|crr}
& \multicolumn{3}{c}{EMR} & \multicolumn{3}{c}{SMR} \\
method & $a_0$ & $a_1$ & $a_2$ & $a_0$ & $a_1$ & $a_2$ \\ \hline
Legendre  & -1.76 $\pm$ 0.19 & -3.77 & -793 & -5.87 $\pm$ 0.20 & 53.2 & 618 \\ 
multipole & -2.87 $\pm$ 0.19 & -13.3 & -450 & -6.61 $\pm$ 0.18 & 39.3 & 749 \\
\end{tabular}
\end{ruledtabular}
\end{table}

\subsection{Relationship between $G_{En}$ and $R_{SM}$}
\label{sec:buchmann}

\begin{figure}
\centering
\includegraphics[width=3.0in]{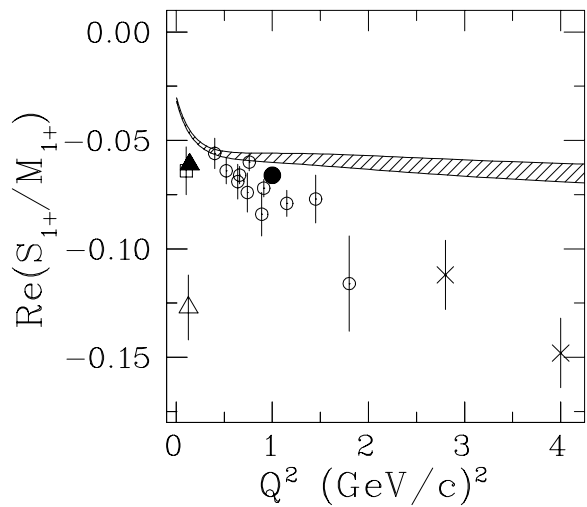}
\caption{Comparison between $R_{SM}$ data and Buchmann's formula (band)
using fitted neutron form factors.
Data: open square \protect{\cite{Pospischil01}}, 
filled triangle \protect{\cite{Sparveris05}},
open triangle \protect{\cite{Kalleicher97}}, 
open circles \protect{\cite{Joo02}},
crosses \protect{\cite{Frolov99}}; 
the filled circle is the present result.
Small horizontal displacements are used to reduce clutter.
Error bars include statistical and systematic but not model uncertainties.}
\label{fig:buchmann}
\end{figure}

Buchmann \cite{Buchmann04} has derived a relationship 
\begin{equation}
\label{eq:buchmann}
R_{SM} = \frac{q M_N}{2 Q^2} \frac{G_{En}}{G_{Mn}}
\end{equation}
between $R_{SM}$ for the $N \rightarrow \Delta$ transition and the
neutron electric and magnetic Sachs form factors, $G_{En}$ and $G_{Mn}$.
Here $q$ is virtual photon momentum in the cm frame.
Deviations from this relationship were attributed to three-quark and 
higher-order currents and were estimated to be at the level of $1/N_c^2$, 
or about $10\%$.
Figure \ref{fig:buchmann} compares this prediction with recent data
\cite{Pospischil01,Sparveris05,Joo02,Frolov99} 
where the band is based upon fitted neutron form factors from 
Ref. \cite{Kelly04b}.
Note that the growth of the band for $Q^2 > 1.5$ (GeV/$c$)$^2$ where
$G_{En}$ data are presently unavailable, is artificially limited by the
use of a model with only two parameters.
The Buchmann formula underestimates most of the $R_{SM}$ data.
The discrepancy of about $15\%$ at $Q^2 = 1$ (GeV/$c$)$^2$ is similar to 
the estimated theoretical uncertainty, but this model predicts a nearly
constant quadrupole ratio for larger $Q^2$ while the data show a steep slope.
Note that for the high-$Q^2$ data we chose the effective Lagrangian analysis 
instead of the Legendre analysis from Ref. \cite{Frolov99} because truncation 
errors in the Legendre method are expected to increase with 
$Q^2$ \cite{Kelly05e} and the effective Lagrangian results are consistent 
with the MAID and DMT analyses of the same data \cite{Kamalov01}.
Although the $R_{SM}$ slope is described well by dynamical models of
$\pi N$ rescattering, perturbative QCD predicts that $R_{SM}$ should 
become constant asymptotically.
Therefore, it would be of interest to extend measurements of $G_{En}$ 
to higher $Q^2$ and to use model-independent multipole analysis of
new polarization data for pion electroproduction to verify the 
apparent slope in $R_{SM}$.

\subsection{Sensitivity of Legendre coefficients to specific multipole 
amplitudes}

\begin{figure}
\centering
\includegraphics[width=3in]{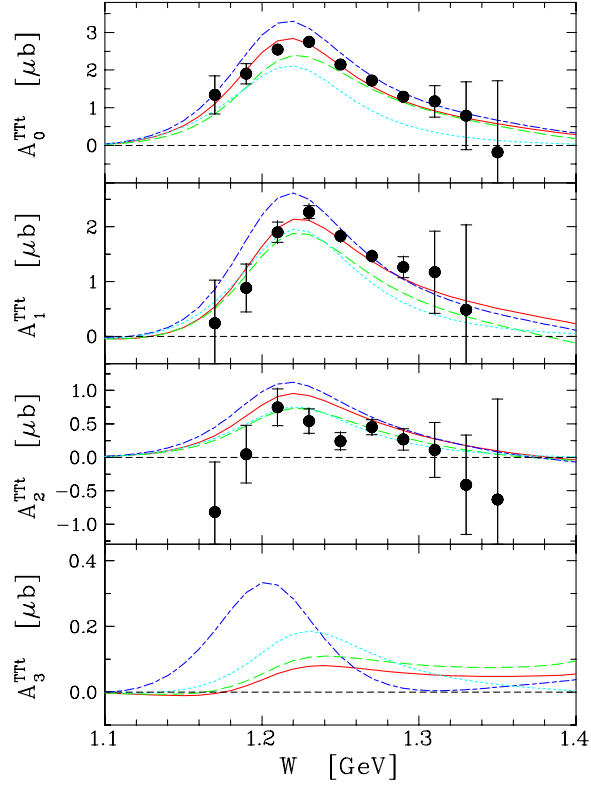}
\caption{(Color online)
Fitted Legendre coefficients for $R_{TT}^t$ are compared with 
MAID2003 (red solid), DMT (green dashed), SAID (blue dash-dotted), 
and SL (cyan dotted).
In the simplest approximation, the first and second panels are  
$-3 \Re M_{1-}\Im M_{1+}$ and $3 \Re E_{0+}\Im M_{1+}$, respectively.
Inner error bars with endcaps are statistical;
outer error bars without endcaps include systematic uncertainties.}
\label{fig:RTTt}
\end{figure}

Even though $sp$ truncation and $M_{1+}$ dominance are not sufficiently
accurate for quantitative analysis of the quadrupole ratios, that truncation
can still provide qualitative insight into the sensitivity of selected
Legendre coefficients to particular multipole amplitudes.
For example, Table \ref{table:RtoLegendre3} shows us that
$A^{TTt}_0 \approx 3 \Im M_{1+}^\ast M_{1-}$ while Fig. \ref{fig:mpfit1m}
shows that most models predict that $M_{1-}$ is nearly real and slowly
varying over this range of $W$.
Consequently, one expects the $W$ dependence of $A^{TTt}_0$ to strongly
resemble $\Im M_{1+}$ and its amplitude to be proportional to $\Re M_{1-}$
and opposite in sign.
Figure \ref{fig:RTTt} shows that these expectations are realized by
the Legendre fit.
The observation that the $A^{TTt}_0$ data are smaller than SAID, 
larger than DMT and SL, and in good agreement with MAID2003 predictions
is consistent with the same pattern seen in Fig. \ref{fig:mpfit1m} 
for $\Re M_{1-}$ and with response function figures in 
Sec. \ref{sec:comparisons}.
A similar correspondence is also observed for $A^{L+Tn}_1$, but the
truncation is not as reliable because Rosenbluth separation is
not available and model calculations have greater shape differences.
The $\Re M_{1-}$ amplitude also appears in $A^{TTn}_1$ but is again diluted.
Therefore, the best sensitivity to $\Re M_{1-}$ is provided by the
$R_{TT}^t$ response function.

\begin{figure}
\centering
\includegraphics[width=3in]{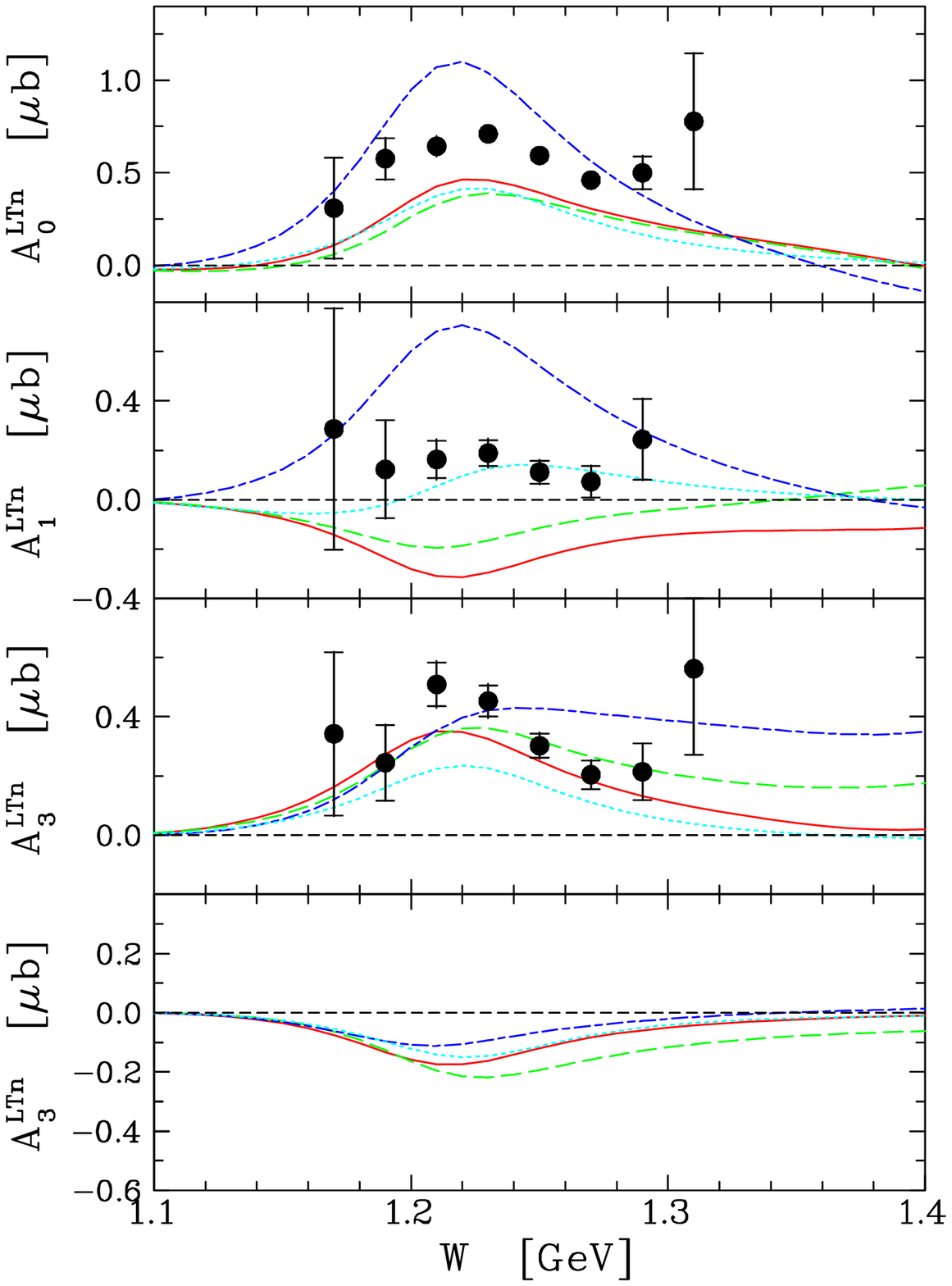}
\caption{(Color online)
Fitted Legendre coefficients for $R_{LT}^n$ are compared with 
MAID2003 (red solid), DMT (green dashed), SAID (blue dash-dotted), 
and SL (cyan dotted). 
In the simplest approximation, the top panel is $\Re S_{1-}\Im M_{1+}$
and the second is $\Im S_{0+}^\ast M_{1+}$.
Inner error bars with endcaps are statistical;
outer error bars without endcaps include systematic uncertainties.}
\label{fig:RLTn}
\end{figure}

Similarly, model calculations suggest that $S_{1-}$ also varies relatively
slowly and is nearly real in this $W$ range, although neither feature 
is quite as accurate as for $M_{1-}$.
Inspection of Table \ref{table:RtoLegendre3} then suggests that the
best sensitivity to $\Re S_{1-}$ is offered by the $R_{LT}^n$ response
function through its $A^{LTn}_0$ Legendre coefficient.
The shape differences show that $M_{1+}$ dominance is not as
accurate for this Legendre coefficient, but Fig. \ref{fig:RLTn} shows
that its $W$ dependence does resemble that of $\Im M_{1+}$ nonetheless.
Thus, we find that the SAID prediction for $A^{LTn}_0$ is considerably
too strong while those of MAID2003, DMT, and SL are too weak and that the
same pattern is observed in Fig. \ref{fig:mpfit1m} for $\Re S_{1-}$.

Within the truncated Legendre expansion, $\Re E_{0+}$ is isolated 
by $A^{TTn}_0$, $A^{TT\ell}_0$, or $A^{TTt}_1$, 
which should be equal modulo signs if $E_{0+}$ were real and $M_{1+}$ 
dominance accurate.
The $A^{TTt}_1$ coefficient is included in Fig. \ref{fig:RTTt} but
the other figures are omitted and can be found in Ref. \cite{e91011_legfit}.
We do observe the expected pattern of signs and all are similar in
shape to $\Im M_{1+}$ but their magnitudes do not conform to these
simplistic predictions.
Nevertheless, the relationships between data and model calculations for 
the Legendre coefficients are similar to those for $\Re E_{0+}$.

The most complicated situation is $S_{0+}$ because both model predictions
and fitted amplitudes show important imaginary contributions to this
nonresonant partial wave.
Hence, measurement of $S_{0+}$ amplitudes requires $LT$ response
functions of both R-type and I-type.
The $S_{0+}$ contributions to the truncated Legendre expansion are 
isolated by 5 R-type and 5 I-type coefficients but, 
although each group displays a relatively uniform shape with respect to $W$, 
there are significant differences in detail that show that the simple 
truncation is not especially accurate for these coefficients.
Nevertheless, the fitted multipole amplitudes provide good fits to all 
of the $LT$ response functions simultaneously. 
Therefore, polarization data provide the phase of $S_{0+}$.

\begin{figure}
\centering
\includegraphics[width=3.0in]{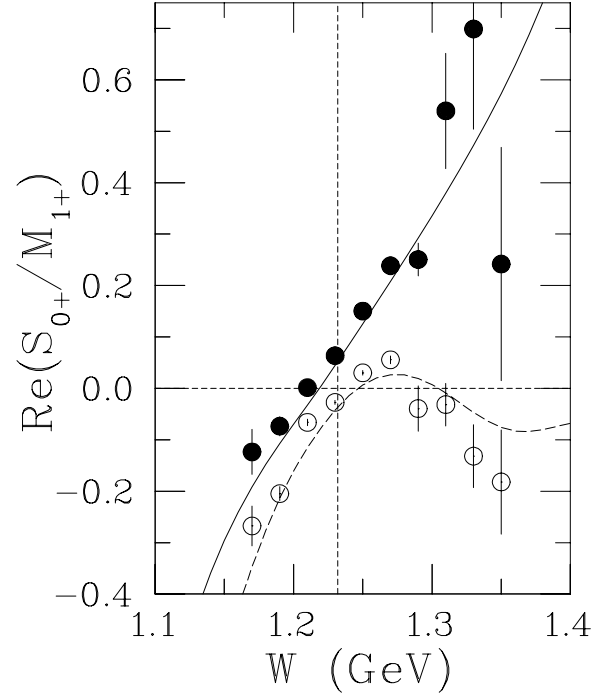}
\caption{ Comparison between multipole and Legendre analyses of 
$\Re(S_{0+}/M_{1+})$.
Solid and dashed curves show MAID2003 calculations using multipole
amplitudes or Legendre coefficients, respectively.
Solid (open) circles show experimental ratios based upon
multipole (Legendre) fits.
The vertical dashed line indicates $W=1.232$ GeV.}
\label{fig:S0}
\end{figure}

Schmieden \cite{Schmieden01} speculated that the disagreement between 
SMR values for $Q^2 \sim 0.13$ (GeV/$c$)$^2$ obtained by 
Kalleicher {\it et al}. \cite{Kalleicher97} using the forward pions versus 
those obtained using cross sections \cite{Mertz01} 
or polarizations \cite{Warren98,Pospischil01} for forward protons  
might be explained using $\Re(S_{0+}/M_{1+}) \approx -0.14$.
However, recent models predict smaller positive values with relatively
slow $Q^2$ dependence that are similar to $-R_{SM}$.
Although it might appear that one could estimate this quantity using
\begin{equation}
\label{eq:S0/M1}
\Re \frac{S_{0+}}{M_{1+}} \approx \frac{2A_0^{LT}}{A_0^T} 
\approx \frac{2A_0^{LT}}{A_0^{L+T}}\; ,
\end{equation}
based upon $M_{1+}$ dominance and $sp$ truncation,
Figure \ref{fig:compare_clas} shows that $A_0^{LT}$ has a node near
$M_\Delta$ and is very small for larger $W$.
By contrast, $A_1^{LT}$ peaks near $M_\Delta$.
Thus, it is likely that truncation errors in the multipole expansion
of Legendre coefficients will be more serious for $S_{0+}$ than for $S_{1+}$.
This problem is illustrated in Fig. \ref{fig:S0} which compares
fitted values for $\Re(S_{0+}/M_{1+})$ from the multipole analysis with 
those based upon Eq. (\ref{eq:S0/M1}) where, because Rosenbluth separation
is unavailable, we assume that $A_0^T \approx A_0^{L+T}$ because 
$M_{1+}$ dominance predicts $A_0^L=0$.
We also show MAID2003 calculations for both quantities.
We find that MAID2003 describes both the steep slope in $\Re(S_{0+}/M_{1+})$ 
and the more complicated shape of $2 A_0^{LT}/A_0^{L+T}$ fairly well, 
but that these quantities are rather different even in the
immediate vicinity of $W=M_\Delta$ --- the Legendre analysis does not 
even give the correct sign for this multipole ratio at $W=1.232$ GeV.  
The sign difference between these quantities using MAID2003 calculations
for $Q^2 = 1$ (GeV/$c$)$^2$ was previously noted in Ref. \cite{Kelly05e} 
and here the same problem is observed in data.
The analysis of a recent experiment for $Q^2 = 0.2$ (GeV/$c$)$^2$ 
that measured left-right cross section asymmetries for $\theta_\pi = 20^\circ$
and $160^\circ$ also observed a large difference between ratios based
upon $M_{1+}$ dominance and $sp$ truncation and those obtained by scaling 
MAID2003 $S_{0+}$ and $S_{1+}$ multipoles to fit the data \cite{Elsner05}.
With a much more complete data set, our multipole analysis does not rely 
upon models like MAID and gives
$\Re(S_{0+}/M_{1+}) = (6.4 \pm 0.7)\%$ at $(W,Q^2)=(1.23,1.0)$ directly.
Assuming that $\Re M_{1+} \approx 0$ for $W \approx M_\Delta$, the
ratio $\Re(S_{0+}/M_{1+}) \approx \Im S_{0+} / \Im M_{1+}$ shows that
$\Im S_{0+}$ is positive and somewhat larger than the MAID2003 prediction
for $W=1.23$ GeV, as shown in Fig. \ref{fig:mpfit0p}. 
Although there is nothing special about $M_\Delta$ for $S_{0+}$, 
we can use the observed slope to estimate 
$\Re(S_{0+}/M_{1+}) = (7.1 \pm 0.8)\%$ at $(W,Q^2)=(1.232,1.0)$ 
for comparison with similar analyses purportedly at $W=M_\Delta$; 
however, the energy dependence is steep enough that kinematical
uncertainties could become important.
Recognizing that the $Q^2$ dependence is mild in most models, neither the 
present result nor that of Ref. \cite{Elsner05} supports Schmieden's 
hypothesis of a large negative value for this ratio.

\subsection{Interpretation of $\chi^2_\nu$}

The Legendre and multipole analyses employ data for differential cross
section, beam analyzing power, and recoil-polarization response functions
with uncertainties that are primarily statistical.
The cross section data include uncertainties in acceptance that can
also be considered statistical because they are estimated from the 
flatness of a yield/simulation plateau.
The uncertainties for recoil-polarization response functions are based
upon diagonal elements of the covariance matrix for the maximum likelihood
method.
However, the fact that reduced chi-square values, $\chi^2_\nu$, are 
consistently larger than unity for both Legendre and multipole analyses 
suggests that uncertainties in extracted quantities may be underestimated.
These statistics are listed in Table \ref{table:chi-square} for each $W$.
There are several possible explanations for this observation.
First, the various recoil-polarization response functions in a given
$(x,W,Q^2)$ bin are correlated with each other, but those correlations
are not considered by the Legendre or multipole analyses because we 
have no efficient means to account for them.
Thus, the same fluctuation can affect several data points and artificially
increase its contribution to $\chi^2$ without necessarily affecting the
quality of the fit.
Second, systematic uncertainties that vary between kinematical bins 
were not included in the uncertainties that were used in the Legendre and 
multipole analyses because their effects upon various response functions 
are also highly correlated.
Third, inaccuracies in baseline calculations of fixed amplitudes would 
impose a lower limit on $\chi^2$ even if all experimental correlations
could be handled properly.
Finally, no corrections have been made for polarized radiative corrections.

Radiative corrections for the beam asymmetry in the 
$p(\vec{e},e^\prime p)\pi^0$ reaction have been evaluated for 
$Q^2 = 0.4$ (GeV/$c$)$^2$ by Afanasev {\it et al.}
\cite{Afanasev02b} and found to be quite small across the 
$\Delta(1232)$ resonance.
Radiative corrections for polarized target asymmetries are presently under 
investigation and generally appear to be small also \cite{Afanasev05a}, 
but procedures for recoil polarization are not yet available in a form 
suitable for the present analysis.
In principle, external radiation permits additional kinematical 
dependencies that cannot be accommodated by the response 
function expansions given in Eq. (\ref{eq:obs}).
Analysis of such effects probably requires an iterative procedure that
begins with the current results to obtain model response functions, 
then calculates radiatively-corrected polarizations for each experimental
event as input to an extended version of the likelihood analysis that
would use a more general representation of the $\phi$ dependence. 
In the future it may be possible to improve upon the current
multipole results by iteration within a model of radiative corrections, 
and hopefully reduce $\chi^2_\nu$, 
but that is obviously a very ambitious project.

The simplest method of correcting for underestimates of experimental
uncertainties is to multiply the uncertainties in extracted quantities
by $\sqrt{\chi^2_\nu}$.
We have not performed that operation here because it is somewhat 
arbitrary, assuming that neglected errors are random and uniform, 
but we provide Table \ref{table:chi-square} for the user's convenience.
However, if that procedure is applied, the systematic uncertainties 
should probably be reduced to avoid double-counting of random errors 
presently labeled systematic.

\begin{table}
\caption{$\chi^2_\nu$ for Legendre and multipole analyses.}
\label{table:chi-square}
\begin{ruledtabular}
\begin{tabular}{lrr}
$W$ (GeV) & $\chi^2_\nu$, Legendre & $\chi^2_\nu$, multipole  \\ \hline
1.17 & 1.32 & 1.24 \\
1.19 & 1.69 & 1.67 \\
1.21 & 1.32 & 1.39 \\
1.23 & 1.50 & 1.65 \\
1.25 & 1.87 & 1.94 \\
1.27 & 1.59 & 1.58 \\
1.29 & 1.53 & 1.52 \\
1.31 & 1.61 & 1.42 \\
1.33 & 1.53 & 1.32 \\
1.35 & 1.41 & 1.30 \\ 
\end{tabular}
\end{ruledtabular}
\end{table}

\section{Unitary Isobar Model}
\label{sec:UIM}

Further insight can be obtained by comparing the fitted multipole
amplitudes with calculations based upon the unitary isobar model (UIM).
We use a unitarization prescription suggested by Olssen \cite{Olssen74}
in which the Born amplitudes are interpreted as contributions to the 
$K$-matrix while isobar contributions are unitarized separately using 
an empirical phase.
This procedure is applied to the multipole amplitudes for isospin states.
Thus, each multipole amplitude is expressed in the form
\begin{equation}
\label{eq:unitarization}
A = (1+i t)B + R e^{i \psi}
\end{equation}
where $B$ is the Born contribution (including $\omega$ and $\rho$ exchange),
$t$ is the $\pi N$ partial-wave amplitude, $R$ is a Breit-Wigner resonance,
and $\psi$ is an energy-dependent phase.
Angular momentum and isospin labels have been suppressed for brevity.
The unitarity phase is adjusted according to the Fermi-Watson theorem 
\cite{Watson54}, which requires
\begin{equation}
R e^{i \psi} = \pm |R| e^{i \delta}
\end{equation}
where $\delta$ is the $\pi N$ elastic phase shift and is assumed
to be real throughout our energy region.
These phases differ for each multipole and depend upon the background
parametrization.
The sign ambiguity is resolved locally and is significant only for
the $P_{11}$ partial wave at low energies.

We used the SAID FA02 solutions for $\pi N$ partial-wave amplitudes
\cite{Arndt02}. 
Resonances were parametrized in Breit-Wigner form, 
Eq. (\ref{eq:MAID-resonance}), with energy-dependent widths based upon the 
MAID model \cite{Drechsel99}.
Only the $\Delta$ and Roper resonances are significant in our $W$ range.
Widths and branching ratios were taken from the most recent compilation 
from the Particle Data Group \cite{PDG04}, from which
widths at resonance were taken as 120 MeV for the $\Delta$ and
350 MeV, with a $\pi N$ branch of $65\%$, for the Roper.

\begin{figure*}
\centering
\includegraphics[angle=90,width=6in]{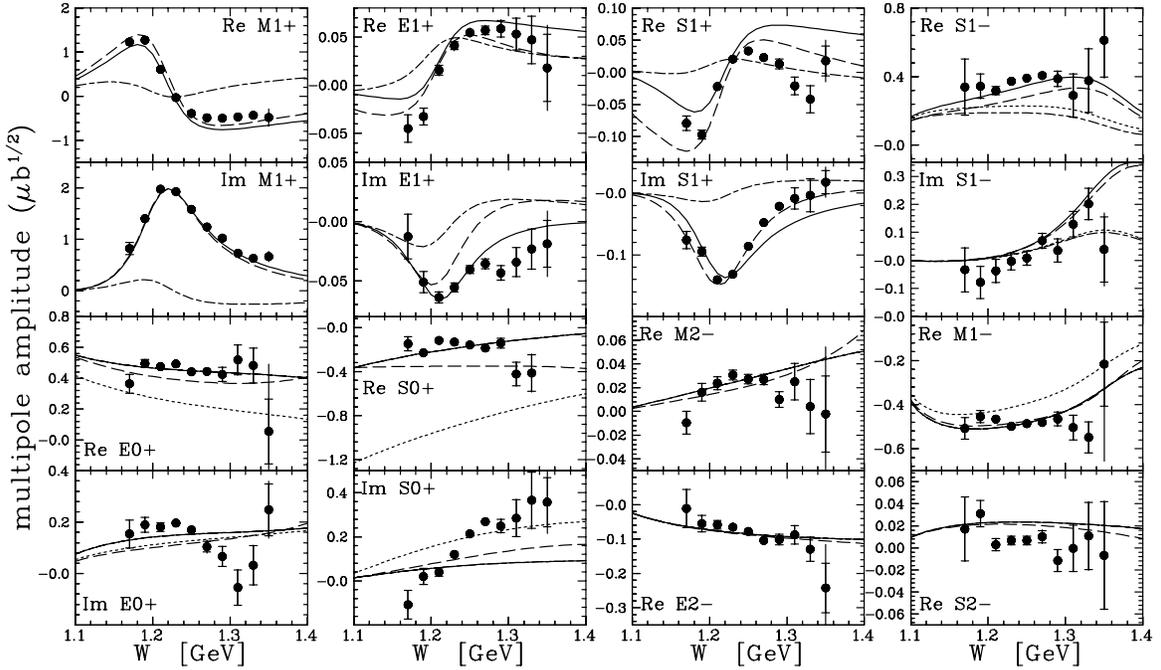}
\caption{Comparison of fitted multipoles with UIM calculations.
Inner error bars with endcaps are statistical;
outer error bars without endcaps include systematic uncertainties.
Solid curves show our UIM results while dashed curves show MAID2003.
Dash-dotted versus dotted curves compare backgrounds for 
pseudovector versus pseudoscalar $\pi NN$ coupling, but dash-dotted 
and solid curves for $0+$ and $\Re M_{1-}$ amplitudes coincide.}
\label{fig:uim}
\end{figure*}

UIM calculations are compared with fitted multipole amplitudes in
Fig. \ref{fig:uim}.
The dash-dotted  versus dotted curves compare background amplitudes 
for pseudovector (PV) versus pseudoscalar (PS) $\pi NN$ coupling and 
differ only for $0+$ and $1-$ multipoles.
The solid curves show the total UIM multipoles for pseudovector
coupling, which coincide with dash-dotted curves for $0+$ and 
$\Re M_{1-}$ amplitudes because no resonances were included for 
those amplitudes.
The dashed curves show MAID2003, which uses a similar unitarization
prescription but interpolates between pseudovector coupling for small
$W$ and pseudoscalar coupling for large $W$ \cite{Drechsel99}. 
The background amplitudes acquire their imaginary parts from the
$\pi N$ phase shift.
This effect is especially important for the $1+$ and $0+$ multipoles
and is responsible for the differences between the real parts
of background amplitudes shown in Fig. \ref{fig:uim} and the baseline
amplitudes shown in Figs. \ref{fig:mpfit1p}-\ref{fig:mpfit2m}.
The effect of the $P_{33}$ resonance upon $M_{1+}$ Born amplitudes
is particularly strong.
The best sensitivity to the $\pi NN$ coupling is in real parts of $0+$
multipoles, especially $\Re S_{0+}$.
The data for $\Re S_{0+}$ clearly favor pseudovector over pseudoscalar
coupling; the admixture used by MAID2003 flattens the $W$ dependence
but is not supported by these data.
The data for $\Re E_{0+}$ also prefer pseudovector coupling but
could accommodate the MAID2003 admixture if a sufficiently strong
$S_{11}(1535)$ contribution were included; we did not investigate that
possibility because our data are limited to relatively small $W$.

Neither our version of the UIM nor that of MAID2003 reproduces the
imaginary parts of $0+$ amplitudes well.
Both approximate the average value for $\Im E_{0+}$ but neither 
reproduces its $W$ dependence.
Note that $\Im E_{0+}$ is especially sensitive to the axial form
factor through the unitarization factor.
We used a dipole form factor with $M_A = 1.0$ (GeV/$c$)$^2$. 
The discrepancy is more severe for $\Im S_{0+}$ which grows much
more rapidly than either UIM, at least for PV coupling.
One might be tempted to include a strong $S_{11}$ contribution and to 
adjust the PS/PV admixture to maintain the fit to $\Re S_{0+}$, 
but we are loath to attempt such a fit without data across, 
or at least closer to, the $S_{11}(1535)$ resonance.
Therefore, our full calculations, shown as solid lines, assume pure 
pseudovector $\pi NN$ coupling and do not include an $S_{11}$ resonance.
The discrepancy in $\Im S_{0+}$ bears further investigation.

The present UIM reproduces $\Im M_{1+}$ very well, but appears to be
missing a small positive contribution to $\Re M_{1+}$.
Nevertheless, the accuracy is sufficient to quote a resonance contribution
of $\bar{M}_{1+}^{(3/2)} = (2.91 \pm 0.15)$ $\mu b^{1/2}$ where the
estimated uncertainty is qualitative.
The $E_{1+}$ and $S_{1+}$ calculations use the quadrupole ratios
from Sec. \ref{sec:quad} without further adjustment.
These calculations are qualitatively consistent with the data 
but do not reproduce the $W$ dependence accurately.
Interestingly, the present UIM is more successful for $\Im E_{1+}$
while MAID2003 is more successful for $\Im S_{1+}$.
It might be possible to improve these fits by adding phenomenological
background contributions to the $B$ terms of 
Eq. (\ref{eq:unitarization}).

Although our $W$ acceptance does not span the Roper resonance, its
width is broad enough to permit estimation of the electromagnetic helicity 
amplitudes based upon data for its low-$W$ side.  
Having selected pure PV coupling, we find little room in $\Re M_{1-}$
for transverse excitation of the Roper resonance, such that $_p A_{1/2}$ 
is consistent with zero at $Q^2 = 1$ (GeV/$c$)$^2$.
That conclusion obviously depends upon the PS/PV admixture, with more
PS coupling permitting larger $_p A_{1/2}$.
Conversely, reproduction of $\Re S_{1-}$ requires appreciable
longitudinal Roper excitation, larger for PV than for PS coupling.
With PV coupling, we estimate $_p S_{1/2} = (0.05 \pm 0.01)$ GeV$^{-1/2}$ 
where the uncertainty is again qualitative, based upon simultaneous
fits to both real and imaginary parts.
Similarly, Laveissi\`ere {\it et al.} \cite{Laveissiere04} fit cross section 
data for $Q^2 = 1.0$ (GeV/$c$)$^2$ by reducing the MAID2003 estimate of 
$_p A_{1/2}$ to a value consistent with zero but their estimate of 
$_p S_{1/2} = 0.019 \pm 0.010$ GeV$^{-1/2}$ is smaller than ours; 
however, it should be noted that their angular range was quite limited.
These findings are also qualitatively consistent with those of
Aznauryan {\it et al.} \cite{Aznauryan05} based upon cross section and 
$R_{LT}^\prime$ data for the $p\pi^0$ and $n\pi^+$ channels at $Q^2 = 0.4$ 
and $0.65$ and $p\eta$ cross section data for $Q^2 = 0.375$ and $0.75$ 
(GeV/$c$)$^2$ with $W$ spanning the second resonance region.
They found that $_p A_{1/2}$ is small and appears to change sign
near 0.5 (GeV/$c$)$^2$.
Our value for $_p S_{1/2}$ is somewhat larger than theirs but some models,
such as those by Capstick and Keister \cite{Capstick95}, 
feature a peak in $_p S_{1/2}$ for $Q^2 \sim 0.7$.
Furthermore, unlike the data set employed by Ref. \cite{Aznauryan05}, 
recoil polarization provides sufficient phase information to separate
multipole amplitudes explicitly.
It is obviously desirable to acquire new data for either recoil or target 
polarization that reach larger $W$.
Nevertheless, the current results and those of Ref. \cite{Aznauryan05}
appear to exclude the hybrid baryon model of the Roper resonance for 
which electromagnetic excitation would be purely transverse \cite{Li91,Li92}.

Clearly it would be of interest to perform an energy-dependent analysis 
of the entire data simultaneously that uses $\pi N$ phase shifts to enforce 
unitarity.
However, the unitarization procedure is not unique \cite{Davidson91} and 
requires background models for both $p\pi^0$ and $n\pi^+$.
Furthermore, Born diagrams at tree level do not necessarily represent 
background amplitudes with sufficient accuracy.
An even more ambitious analysis could use dispersion relations to improve
the background model. 
However, such an analysis should consider all available data and is beyond 
the scope of the present work.
Therefore, we have not attempted to optimize the UIM parameters, except
for $\Delta$ and Roper strengths.

The present multipole analysis can be described as energy-independent 
because each $W$ is fit independently.
It also uses the minimum possible theoretical information, 
simply the small Born amplitudes for partial waves with $\ell_\pi >2$.
No attempt has been made to enforce unitarity in the multipole analysis 
and some concern has been expressed that the steep slope we find for 
$\Im S_{0+}$ might be inconsistent with the unitary isobar model 
\cite{Aznauryan05a}.
On the other, the good agreement between SAID calculations and the data
for both real and imaginary parts of $S_{0+}$ shows that it is possible 
to describe the $p\pi^0$ data for this multipole in a manner that is 
consistent with unitarity (see Fig. \ref{fig:mpfit0p}). 
An experimental test of unitarity that is independent of models for
resonant and nonresonant amplitudes would require comparable polarization 
data for the $n\pi^+$ reaction, but such data are presently unavailable.

\section{Summary and Conclusions}
\label{sec:conclusions}

We measured angular distributions for differential cross section, 
beam analyzing power, and recoil polarization in
the $p(\vec{e},e^\prime \vec{p})\pi^0$ reaction at $Q^2 = 1$ 
(GeV/$c$)$^2$ with $1.17 \leq W \leq 1.35$ GeV across the $\Delta$ 
resonance and have obtained 14 separated response functions and 2 Rosenbluth
combinations, of which 12 have been measured for the first time.

We compared the data for response functions with calculations
for four recent models: MAID, DMT, SAID, and SL.
Variations among these models are relatively small at $W \approx M_\Delta$
for quantities that depend upon real parts of interference products, 
but increase with $W$.
Variations among models are much larger for quantities dependent upon 
imaginary parts that are more sensitive to background amplitudes.
MAID and DMT are similar and in relatively good agreement with data
for $W \approx M_\Delta$, but neither provides a uniformly good 
description of the data for larger $W$.
The SL model, which does not include higher resonances, underpredicts 
the cross section for larger $W$ while DMT is too strong.
The SAID model has considerable difficulty with helicity-independent $LT$ 
response functions that are probably caused mostly by its rather strong
$\Re S_{1-}$ amplitude.

We performed a multipole analysis that fits both real and imaginary parts 
of the multipole amplitudes for low partial waves while those for higher 
partial waves are constrained by either Born terms or by the best 
available model calculations.
Fitted multipole amplitudes based upon Born, MAID, DMT, or SL models
are practically indistinguishable, but the available version of SAID 
does not provide a suitable baseline because some of its $\ell_\pi \geq 2$ 
amplitudes are too strong.
The final analysis is based upon the Born model to minimize bias.
We chose not to vary $\Im M_{1-}$ in the final analysis because it is 
predicted to be negligible in our energy range but its fitted values are 
strongly correlated with those of $\Im S_{1-}$ for the present data set.
We were able to extract consistent results for all $\ell_\pi \leq 1$ 
amplitudes, except $\Im M_{1-}$, plus the real parts of $2-$ multipoles.
The most significant differences between fitted and model amplitudes are 
found in $0+$ and $1-$ multipoles.
The data also show that $\Im S_{0+}$ grows faster than predicted by
MAID, DMT, or SL, but is described reasonably well by SAID.
Good sensitivity to $\Re M_{1-}$ is provided by the $R_{TT}^t$ response 
function; there is a wide spread among models and MAID2003 fits the 
$\Re M_{1-}$ data best and is close to the Born baseline.
Similarly, the best sensitivity to $\Re S_{1-}$ is provided by $R_{LT}^n$ 
but none of the models is accurate --- SAID is much too strong while
MAID, DMT, and SL are too weak for that amplitude.
The data are substantially stronger than the Born amplitude, 
suggesting significant longitudinal Roper contributions arising from 
a radial excitation.

We find that truncation errors in the traditional Legendre analysis 
of $N \rightarrow \Delta$ quadrupole ratios can be significantly larger 
than statistical errors.
Using parabolic fits to the energy dependence, we obtain
$\tilde{R}_{SM}^{(p\pi^0)} = (-5.87 \pm 0.20)\%$ and 
$\tilde{R}_{EM}^{(p\pi^0)} = (-1.76 \pm 0.19)\%$ 
from the traditional analysis or
$R_{SM}^{(p\pi^0)} = (-6.61 \pm 0.18)\%$ and 
$R_{EM}^{(p\pi^0)} = (-2.87 \pm 0.19)\%$ from the multipole analysis
for $W=1.232$ GeV and $Q^2 = 1.0$ (GeV/$c$)$^2$.
These results are consistent with the single-energy analysis 
published previously \cite{Kelly05c}.
The model dependence of the multipole analysis is small and the Legendre
fits are stable with respect to the number of fitted terms, yet 
the differences between these analyses are several standard deviations.
We have demonstrated that the multipole analysis is more reliable because
it does not depend upon $M_{1+}$ dominance or $sp$ truncation.
Both model calculations and the multipole analysis of data demonstrate
that neither assumption is reliable and that multipole products omitted by 
that truncation scheme make important contributions to the Legendre 
coefficients that spoil the accuracy of the simple estimators of quadrupole 
ratios employed by the traditional Legendre analysis.
Truncation errors are especially severe for $\tilde{R}_{EM}$. 

We also find that $\Re(S_{0+}/M_{1+}) = (7.1 \pm 0.8)\%$ at 
$W=1.232$ GeV is qualitatively consistent with most recent models and with
a recent measurement \cite{Elsner05} at $Q^2 = 0.2$ (GeV/$c$)$^2$ of
left-right cross section asymmetries at a pair of supplementary 
proton angles, 
but is inconsistent with a recent hypothesis \cite{Schmieden01} that a large 
negative value is needed to explain inconsistencies between SMR analyses 
at $Q^2 = 0.13$ (GeV/$c$)$^2$  using earlier data for forward versus 
backward $\theta_\pi$.
Truncation errors in the Legendre estimator for $\Re(S_{0+}/M_{1+})$
are quite severe \cite{Kelly05e}, even resulting in an incorrect sign at 
$Q^2 \sim 1$ (GeV/$c$)$^2$.
The analysis in Ref. \cite{Elsner05} relied on the MAID model instead
of the Legendre estimator but accurate, model-independent results require 
a phase-sensitive multipole analysis as performed here.  

Finally, we compared the fitted multipole amplitudes with calculations
based on a unitary isobar model (UIM).
The $\Re S_{0+}$ and $\Re E_{0+}$ multipoles strongly prefer pseudovector 
over pseudoscalar $\pi NN$ coupling and do not support the proposed mixing
between PV and PS coupling employed by the MAID model.
However, the UIM does not reproduce imaginary parts of $0+$ multipoles well,
with $\Im S_{0+}$ increasing more rapidly with $W$ than expected from
the pseudovector Born contribution.
Note that the ability of SAID to reproduce both real and imaginary
$S_{0+}$ data for $p\pi^0$ shows that failure of UIM to reproduce these
data does not necessarily mean inconsistency with unitarity;
a rigorous test of unitarity would require similar $n\pi^+$ data. 
The UIM reproduces $M_{1+}$ well, but the $W$ dependencies of $\Im E_{1+}$
and $\Im S_{1+}$ are only qualitatively consistent with the data.
If we select pure PV coupling, the data for $\Re M_{1-}$ are described
well by Born terms, suggesting that $_pA_{1/2}$ for the Roper is consistent
with zero.
Therefore, Roper excitation is dominantly longitudinal for
$Q^2 = 1.0$ (GeV/$c$)$^2$, where we find 
$_p S_{1/2} = (0.05 \pm 0.01)$ GeV$^{-1/2}$.
Although a larger PS admixture would permit appreciable transverse Roper
excitation, these findings tend to exclude the hybrid baryon model of the 
Roper resonance.
Clearly it would be of interest to extend these measurements to larger $W$.
 
In conclusion, recoil and/or target polarization data are essential to 
multipole analyses of meson electroproduction reactions,
providing access to the relative phase between resonant
and nonresonant contributions.
Although neutral pion electroproduction in the $\Delta$ region 
is the easiest example, this experiment demonstrates the
feasibility of the method and we hope that it will be
applied over wider kinematic ranges and to related reactions.
An advantage of this type of analysis is that it minimizes the
dependence upon models; however, it does not guarantee that
the fitted multipole amplitudes will depend smoothly on both
$W$ and $Q^2$.
Model-dependent analyses which adjust parameters of an effective
Lagrangian or unitary isobar model should produce kinematically smooth 
multipole amplitudes at the expense of possible bias.
Presumably, analyses of these types would also be less sensitive to 
variations of acceptance-averaged $W$ and $Q^2$ between bins of the 
angular variables, $(x,\phi)$.
Both types of analyses would benefit from more extensive coverage in $W$.
With sufficient kinematic coverage one hopes to obtain 
reliable transition form factors for overlapping resonances.

\begin{acknowledgments}
We thank Dr. L. Tiator for MAID and DMT multipole amplitudes,
Dr. T.-S. H. Lee for SL multipole amplitudes,  
Dr. R. Arndt for SAID subroutines and Dr. I. Aznauryan for discussions
of the unitary isobar model.
Finally, we thank Dr. T. Payerle for the first version of {\tt EPIPROD}.
This work was supported by DOE contract No. DE-AC05-84ER40150 Modification
No. M175 under which the Southeastern Universities Research Association (SURA)
operates the Thomas Jefferson National Accelerator Facility.  
We acknowledge additional grants from the U.S. DOE and NSF, the
Canadian NSERC, the Italian INFN, the French CNRS and CEA, and the
Swedish VR.
\end{acknowledgments}


\appendix
\section{Spin transport}
\label{appendix:spin}

The spin transport matrix consists of the sequence of transformations
\begin{equation}
\label{eq:spin}
F = R_\text{fpp} \, C \, R_\text{spectrometer} \, R_\text{hall} \, R_W \, T
= S \, T
\end{equation}
where 
\begin{itemize}
\item $R_W$ 
performs an active Wigner rotation of the polarization vector from the 
center of mass to the laboratory frame,
\item $R_\text{hall}$
performs a passive transformation from the reaction basis to the 
hall basis,
\item $R_\text{spectrometer}$
transforms from the hall basis to the spectrometer basis,
\item $C$ 
transports the spin through the magnetic spectrometer, and
\item $R_\text{fpp}$
transforms to the local FPP coordinate system.
\end{itemize}
The individual transformations are detailed below.

\subsection{Wigner rotation}
\label{section:wigner}

The polarization vector is transformed from the cm reaction basis
with $\hat{\ell}$ along $\vec{p}_{\rm cm}$ to the laboratory 
frame with $\hat{L}$ along $\vec{p}_{\rm lab}$
using a rotation of the form \cite{Giebink85}
\begin{equation}
\left( \begin{array}{c} P_S \\ P_N \\ P_L \end{array} \right)
= \left( 
\begin{array}{ccc} 
\cos{\theta_W} & 0 & \sin{\theta_W} \\
0 & 1 & 0  \\
-\sin{\theta_W} & 0 & \cos{\theta_W}
\end{array} \right)
\left( \begin{array}{c} P_t \\ P_n \\ P_\ell \end{array} \right)
\end{equation}
where 
\begin{equation}
\tan{\theta_W} = \frac{\beta \sin{\theta_{cm}}}
{\gamma_{cm} (\beta \cos{\theta_{cm}} + \beta_{cm})}
\end{equation}
Here 
$\beta_{cm} = p_{cm}/E_{cm}$ is the nucleon velocity in the cm and
$\beta=q/(m_p+\omega)$ is the velocity of the cm relative to the lab.
This matrix is identified as $R_W$ in Eq. (\ref{eq:spin}).

\subsection{Transformation to spectrometer frame}
The hall basis is defined with $\hat{z}$ along the beam line and
$\hat{y}$ vertically upward in the lab.
It is useful here to define the horizontal angle $\alpha$ in the
$xz$ plane measured counterclockwise from $\hat{z}$ and the vertical
angle $\beta$ to be positive above the horizontal plane.
Unit vectors $\hat{q}$ along the momentum transfer and $\hat{p}$ along
the nucleon momentum
plane then take the form
\begin{subequations}
\begin{eqnarray} 
\hat{q} &=& (\sin{\alpha_q}\cos{\beta_q},\sin{\beta_q},\cos{\alpha_q}\cos{\beta_q}) \\
\hat{p} &=& (\sin{\alpha_p}\cos{\beta_p},\sin{\beta_p},\cos{\alpha_p}\cos{\beta_p})
\end{eqnarray}
\end{subequations}
where both $\alpha_p$ and $\alpha_q$ are negative for our 
configuration with the proton spectrometer on the right side of the beam.
The laboratory reaction basis $(\hat{S},\hat{N},\hat{L})$ is now
defined by
\begin{subequations}
\begin{eqnarray} 
\hat{L} &=& \hat{p} \\
\hat{N} &\propto& \hat{q} \wedge \hat{p} \\
\hat{S} &=& \hat{N} \wedge \hat{L} 
\end{eqnarray}
\end{subequations}
where $\hat{L}$ is along the nucleon momentum, $\hat{N}$ is normal to the 
reaction plane, and $\hat{S}$ is sideways within that plane.
Therefore, the transformation from the laboratory reaction to
the hall frame employs
\begin{equation}
R_{\rm hall} = \left( \hat{S}, \hat{N}, \hat{L} \right) 
\end{equation}
obtained by using these unit vectors as the columns of a square matrix.

The {\tt COSY} \cite{COSY,Makino99} calculation of spin precession employs 
the {\tt TRANSPORT} \cite{TRANSPORT} 
coordinate system with $\hat{z}_s$ along the central axis of
the spectrometer, $\hat{x}_s$ downward, and $\hat{y}_s$ 
toward the left of the spectrometer midplane.
Thus, transformation from the hall to the spectrometer
frames is accomplished using
\begin{equation}
R_{\rm spectrometer} = \left( 
\begin{array}{ccc} 
0 & -1 & 0 \\
\cos{\alpha_0} & 0 & -\sin{\alpha_0}  \\
\sin{\alpha_0} & 0 & \cos{\alpha_0}
\end{array} \right)
\end{equation}
where the central angle of the spectrometer, $\alpha_0$, is 
also negative for our configuration.

\subsection{Spin transport}

The spin transport matrix $C_{i,j}$ relates components of spin
in the final spectrometer basis to those in the initial spectrometer
basis, where these bases differ by a rotation about the $\hat{y}_s$
axis through the bend angle $\Omega_0$ for the central ray.
Thus, the spin transport matrix $C^{(0)}_{i,j}$ for a pure dipole
would take the form
\begin{equation}
C^{(0)} = \left( 
\begin{array}{ccc} 
\cos{\chi_0} & 0 & -\sin{\chi_0}  \\
0 & 1 & 0 \\
\sin{\chi_0} & 0 & \cos{\chi_0}
\end{array} \right)
\end{equation}
where $\chi_0 = \gamma \kappa_p \Omega_0$,
$\kappa_p$ is the anomalous magnetic moment, and $\gamma=E/m$.
The central bend angle is $\Omega_0=45^\circ$ for HRS.

Elements of the spin-transport matrix  were computed by
the differential-algebra transport code {\tt COSY} using a 
magnetic model of HRS and were expanded in the form
\begin{equation}
C_{i,j} = \sum_{k,l,m,n,p} C_{ij}^{k \ell mnp} x^k \theta^\ell y^m
\phi^n \delta_K^p
\end{equation}
where $(x,\theta,y,\phi)$ are the reconstructed track variables
expressed in ${\tt TRANSPORT}$ form and $\delta_K=(K-K_0)/K_0$ 
is the kinetic energy displacement relative to the central value.
The expansions are carried to $5^\text{th}$ order.
Further details can be found in Ref. \cite{COSY,Pentchev03}.

\subsection{Transformation to FPP basis}

Tracks before and after scattering by the analyzer are reconstructed in
the form
\begin{subequations}
\begin{eqnarray} 
\hat{k}_i &=& (\sin{\alpha_i}\cos{\beta_i},\sin{\beta_i},\cos{\alpha_i}\cos{\beta_i}) \\
\hat{k}_f &=& (\sin{\alpha_f}\cos{\beta_f},\sin{\beta_f},\cos{\alpha_f}\cos{\beta_f})
\end{eqnarray}
\end{subequations}
where $\alpha$ and $\beta$ are cartesian angles.
The incident polarization is transformed to a basis aligned with $\hat{k}_i$ 
using
\begin{equation}
R_{\rm fpp} = \left( 
\begin{array}{ccc} 
\cos{\alpha_i} & 0 & -\sin{\alpha_i} \\
-\sin{\alpha_i}\sin{\beta_i} & \cos{\beta_i} &  -\cos{\alpha_i}\sin{\beta_i} \\
\sin{\alpha_i}\cos{\beta_i} & \sin{\beta_i} & \cos{\alpha_i}\cos{\beta_i}
\end{array} \right)
\end{equation}
Similarly, the polar and azimuthal scattering angles in the FPP are determined
using
\begin{equation}
\left( \begin{array}{c} 
\sin{\theta_{\rm fpp}}\cos{\phi_{\rm fpp}} \\
\sin{\theta_{\rm fpp}}\sin{\phi_{\rm fpp}} \\
\cos{\theta_{\rm fpp}}
\end{array} \right) = R_{\rm fpp} \hat{k}_f 
\end{equation}
such that 
\begin{equation}
\vec{F} \cdot \hat{n}_{\rm fpp} = 
F_2 \cos{\phi_{\rm fpp}} - F_1 \sin{\phi_{\rm fpp}}
\end{equation}
where 
$\hat{n}_{\rm fpp} = 
\hat{k}_i \wedge \hat{k}_f / | \hat{k}_i \wedge \hat{k}_f |$
is a unit vector normal to the FPP scattering plane.

\section{Maximization of likelihood}
\label{sec:maximization}

The likelihood function for polarization measurements takes the generic form
\begin{equation}
L = \prod_i \frac{1}{2\pi}(a_i + \lambda_i \cdot R)
\end{equation}
where the Latin index $i$ enumerates events satisfying the selection
criteria for a particular kinematical bin,
$R$ is an $n$-dimensional result vector where $R_\alpha$ represents
an acceptance-averaged quantity, identified by the Greek index $\alpha$,
that is assumed to be constant for all events in the bin,
$\lambda_{i\alpha}$ is an event-dependent coefficient that
weights the effect of parameter $R_\alpha$,
and $a_i$ includes the effects of instrumental asymmetry and
is assumed to be of order unity.
The scalar product denotes contraction with respect to $\alpha$.
The logarithm of the likelihood function is maximized with
respect to $R_\alpha$ when
\begin{equation}
\sum_i \frac{\lambda_{i\alpha}}{a_i + \lambda_i \cdot R} = 0
\end{equation}
Although this set of $n$ equations cannot be solved in closed form, 
an iterative solution can be obtained by combining successive substitution
with linearization of the summand.

If the asymmetry is small, linearization provides an approximate solution
in the form
\begin{equation}
R \approx \Lambda^{-1} \cdot V
\end{equation}
where
\begin{equation}
V_\alpha = \sum_i \frac{\lambda_{i\alpha}}{a_i}
\end{equation}
is a measurement vector and
\begin{equation}
\Lambda_{\alpha\beta} = 
\sum_i \frac{\lambda_{i\alpha}}{a_i} \frac{\lambda_{i\beta}}{a_i}
\end{equation}
is the design matrix for the experiment.
More generally, let 
\begin{equation}
\label{eq:Rseq}
R^{(k)} =  R^{(k-1)} + \Delta R^{(k)}
\end{equation}
represent a sequence of improved approximations to the result vector,
and let
\begin{subequations}
\begin{eqnarray}
V_\alpha^{(k)} &=& \sum_i \frac{\lambda_{i\alpha}}{a_i+\lambda_i \cdot R^{(k-1)}} \\
\Lambda_{\alpha\beta}^{(k)} &=& 
\sum_i \frac{\lambda_{i\alpha}\lambda_{i\beta}}{(a_i+\lambda_i \cdot R^{(k-1)})^2}
\end{eqnarray}
\end{subequations}
represent the measurement vector and design matrix for iteration $n$.
Thus, we obtain
\begin{equation}
\label{eq:deltaR}
\Delta R^{(k)} =  (\Lambda^{(k)})^{-1} \cdot V^{(k)}
\end{equation}
When the asymmetry is small, one normally begins with the
unbiased initial estimates $R^{(0)}_\alpha=0$ and expects
rapid convergence, but if the asymmetry is large it may be
advantageous to begin with model estimates for the
$R^{(0)}$ parameters because each iteration for a large
sample may require considerable computation time.

Let $\hat{R}$ represent the maximum-likelihood estimates of the model
parameters given by the convergence of Eqs. (\ref{eq:Rseq}-\ref{eq:deltaR})
and let $\Delta R = R - \hat{R}$ represent the deviation vector.
We assume that the likelihood function is described well by the Gaussian
approximation
\begin{equation}
\ln{L} \approx \log{L_0} - 
\frac{1}{2} \Delta R \cdot \Lambda \cdot \Delta R
\end{equation}
near its maximum.
Therefore, the covariance matrix is given by $\sigma = \hat{\Lambda}^{-1}$ 
where
\begin{equation}
\hat{\Lambda}_{\alpha\beta} = 
\sum_i \frac{\lambda_{i\alpha}\lambda_{i\beta}}{(a_i+\lambda_i \cdot \hat{R})^2}
\end{equation} 
and we estimate the parameter uncertainties 
\begin{equation}
\sigma_{R_\alpha} = \sigma_{\alpha\alpha}^{1/2}
\end{equation}
as standard deviations.

\section{Amplitudes}
\label{app:amplitudes}

The reaction amplitudes for any $A(e,e^\prime N)B$ process where $A$
has spin-$\frac{1}{2}$ and $B$ spin-$0$ that is governed by the
one-photon exchange mechanism can be expressed in terms of 
{\it helicity amplitudes} of the form
\begin{equation}
{\cal H}_{\lambda_f \lambda_i \lambda_\gamma}(W,Q^2,\theta,\phi) =\langle 
\lambda_f | {\cal F}_\mu \varepsilon^\mu | \lambda_i,\lambda_\gamma \rangle 
\end{equation}
where $\lambda_i$ and $\lambda_f$ are the initial and final helicities
of the nucleon, $\lambda_\gamma$ is the helicity of the virtual photon, 
${\cal F}^\mu$ is an appropriately normalized transition current operator,
and $\varepsilon^\mu$ is the virtual-photon polarization vector.
Since parity conservation \cite{Jacob59} requires
$|{\cal H}_{-\lambda_f -\lambda_i -\lambda_\gamma}| =
|{\cal H}_{ \lambda_f  \lambda_i  \lambda_\gamma}| $, 
it is sufficient to consider six independent amplitudes ${\cal H}_i$ 
for $(\lambda_f,\lambda_i,\lambda_\gamma)$ chosen as 
 $(-\frac{1}{2},-\frac{1}{2}, 1)$, 
 $(-\frac{1}{2}, \frac{1}{2}, 1)$, 
 $( \frac{1}{2},-\frac{1}{2}, 1)$, 
 $( \frac{1}{2}, \frac{1}{2}, 1)$, 
 $( \frac{1}{2}, \frac{1}{2}, 0)$, and 
 $( \frac{1}{2},-\frac{1}{2}, 0)$    
and numbered sequentially \cite{Jones65,Walker69}. 
The most general current operator for pseudoscalar meson production
consistent with parity conservation and other symmetries can be 
represented in term of CGLN amplitudes \cite{CGLN2,Dennery61} as
\begin{subequations}
\begin{eqnarray}
  i {\cal F}^0 &=& \frac{q}{\omega}\left( 
    {\cal F}_5^\prime \vec{\sigma}\cdot\hat{q}
  + {\cal F}_6^\prime \vec{\sigma}\cdot\hat{p} \right) \\
  i \vec{{\cal F}} &=& 
    {\cal F}_1 \vec{\sigma} 
- i {\cal F}_2 \vec{\sigma}\cdot\hat{p} \vec{\sigma}\times\hat{q}
+   {\cal F}_3 \hat{p} \vec{\sigma}\cdot\hat{q}
+   {\cal F}_4 \hat{p} \vec{\sigma}\cdot\hat{p}
+   {\cal F}_5 \hat{q} \vec{\sigma}\cdot\hat{q}
+   {\cal F}_6 \hat{q} \vec{\sigma}\cdot\hat{p}
\end{eqnarray}
\end{subequations}
where
\begin{subequations}
\begin{eqnarray}
{\cal F}_5^\prime &=& {\cal F}_5 + {\cal F}_3 \hat{p}\cdot\hat{q} + 
{\cal F}_1 \\
{\cal F}_6^\prime &=& {\cal F}_6 + {\cal F}_4 \hat{p}\cdot\hat{q} 
\end{eqnarray}
\end{subequations}
are combinations which simplify the multipole analysis.
Using phases for helicity states following the conventions
of Jacob and Wick \cite{Jacob59}, the helicity amplitudes are related
to CGLN coefficients by
\begin{subequations}
\label{eq:cgln-to-hamps}
\begin{eqnarray}
{\cal H}_1 &=& -e^{i\phi}\sqrt{2}\sin{\theta}\cos{(\frac{\theta}{2})}
({\cal F}_3 + {\cal F}_4) \\
{\cal H}_2 &=& \sqrt{2} \left( \cos{(\frac{\theta}{2})}
({\cal F}_2 - {\cal F}_1)
+ \sin{\theta}\sin{(\frac{\theta}{2})}({\cal F}_3 - {\cal F}_4) \right) \\
{\cal H}_3 &=& e^{2i\phi}\frac{1}{\sqrt{2}} 
\sin{\theta}\sin{(\frac{\theta}{2})}({\cal F}_3 - {\cal F}_4) \\
{\cal H}_4 &=& \sqrt{2} \left( \sin{(\frac{\theta}{2})}
({\cal F}_1 + {\cal F}_2)
+ \sin{\theta}\cos{(\frac{\theta}{2})}({\cal F}_3 + {\cal F}_4) \right) \\
{\cal H}_5 &=& \frac{Q}{\omega}\cos{(\frac{\theta}{2})}
({\cal F}_5^\prime + {\cal F}_6^\prime) \\
{\cal H}_6 &=& e^{i\phi}\frac{Q}{\omega}\sin{(\frac{\theta}{2})}
({\cal F}_5^\prime - {\cal F}_6^\prime) \\
\end{eqnarray}
\end{subequations}
where $\theta$ and $\phi$ refer to the meson.

CGLN amplitudes can be expanded in terms of {\it multipole amplitudes} 
as follows \cite{Dennery61}.
\begin{subequations}
\label{eq:mpamps-to-cgln}
\begin{eqnarray}
{\cal F}_1 &=& \sum_{\ell} 
\left(\ell M_{\ell +} + E_{\ell +}\right) P_{\ell+1}^\prime(x) +
\left( (\ell+1) M_{\ell -} + E_{\ell -}\right) P_{\ell-1}^\prime(x) \\
{\cal F}_2 &=& \sum_{\ell} 
\left( (\ell+1) M_{\ell +} + \ell M_{\ell -}\right) P_{\ell}^\prime(x) \\
{\cal F}_3 &=& \sum_{\ell} 
\left(E_{\ell +} - M_{\ell +}\right) P_{\ell+1}^{\prime\prime}(x) +
\left(  E_{\ell -} + M_{\ell -}\right) P_{\ell-1}^{\prime\prime}(x) \\
{\cal F}_4 &=& \sum_{\ell} 
\left(M_{\ell +} - E_{\ell +} - M_{\ell -} - E_{\ell -}\right) 
P_{\ell}^{\prime\prime}(x) \\
{\cal F}_5^\prime &=& \frac{\omega}{q} \sum_{\ell} 
(\ell+1)S_{\ell +}  P_{\ell+1}^{\prime}(x) -
\ell  S_{\ell -}  P_{\ell-1}^{\prime}(x) \\
{\cal F}_6^\prime &=& \frac{\omega}{q} \sum_{\ell} 
\left( \ell  S_{\ell -} -
(\ell+1)S_{\ell +} \right) P_{\ell}^{\prime}(x)
\end{eqnarray}
\end{subequations}
The multipole amplitudes can be projected from CGLN amplitudes using 
\cite{Jones65}
\begin{subequations}
\label{eq:cgln-to-mpamps}
\begin{eqnarray}
M_{\ell +} &=& \frac{1}{2(\ell+1)} \int_{-1}^{1} dx \;
\left[ P_{\ell} {\cal F}_1 - P_{\ell+1}{\cal F}_2 +
\frac{1}{2\ell+1}( P_{\ell+1} - P_{\ell-1} ){\cal F}_3 \right ] \\ 
M_{\ell -} &=& \frac{1}{2\ell} \int_{-1}^{1} dx \;
\left[ -P_{\ell} {\cal F}_1 + P_{\ell-1}{\cal F}_2 -
\frac{1}{2\ell+1}( P_{\ell+1} - P_{\ell-1} ){\cal F}_3 \right ] \\ 
E_{\ell +} &=& \frac{1}{2(\ell+1)} \int_{-1}^{1} dx \;
\left[ P_{\ell} {\cal F}_1 - P_{\ell+1}{\cal F}_2 -
\frac{\ell}{2\ell+1}( P_{\ell+1} - P_{\ell-1} ){\cal F}_3 +
\frac{\ell+1}{2\ell+3} (P_{\ell} - P_{\ell+2}){\cal F}_4 \right ] \\ 
E_{\ell -} &=& \frac{1}{2\ell} \int_{-1}^{1} dx \;
\left[ P_{\ell} {\cal F}_1 - P_{\ell-1}{\cal F}_2 -
\frac{\ell+1}{2\ell+1}( P_{\ell+1} - P_{\ell-1} ){\cal F}_3 +
\frac{\ell}{2\ell-1} (P_{\ell} - P_{\ell-2}){\cal F}_4\right ] \\ 
S_{\ell +} &=& \frac{q}{\omega} \frac{1}{2(\ell+1)} \int_{-1}^{1} dx \;
\left[ P_{\ell}{\cal F}_5^\prime + P_{\ell+1}{\cal F}_6^\prime \right] \\
S_{\ell -} &=& \frac{q}{\omega} \frac{1}{2\ell} \int_{-1}^{1} dx \;
\left[ P_{\ell}{\cal F}_5^\prime + P_{\ell-1}{\cal F}_6^\prime \right] 
\end{eqnarray}
\end{subequations}
where $x=\cos{\theta_\pi}$.

Often it is useful to express multipole amplitudes in terms of 
{\it partial-wave helicity amplitudes} \cite{Jones65} according to
\begin{subequations}
\label{eq:pwhamps-to-mpamps}
\begin{eqnarray}
M_{\ell +} &=& \left[ 2 A_{\ell +} - (\ell+2) B_{\ell +} \right] / 
               \left[ 2(\ell + 1) \right] \\
E_{\ell +} &=& \left[ 2 A_{\ell +} + \ell B_{\ell +} \right] / 
               \left[ 2(\ell + 1) \right] \\
S_{\ell +} &=&  C_{\ell +} / (\ell + 1)  \\
M_{\ell -} &=& \left[ 2 A_{\ell -} + (\ell-1) B_{\ell -} \right] / (2\ell) \\
E_{\ell -} &=& \left[ -2 A_{\ell -} + (\ell+1) B_{\ell -} \right] / (2\ell) \\
S_{\ell -} &=&  -C_{\ell -} / \ell   
\end{eqnarray}
\end{subequations}
where
\begin{subequations}
\label{eq:mpamps-to-pwhamps}
\begin{eqnarray}
A_{\ell +} &=& \frac{1}{2}\left[ \ell M_{\ell +} + (\ell+2) E_{\ell +} 
\right] \\
B_{\ell +} &=& E_{\ell +} - M_{\ell +} \\
C_{\ell +} &=& (\ell+1) S_{\ell +} \\
A_{\ell -} &=& \frac{1}{2}\left[ (\ell+1) M_{\ell -} - (\ell-1) E_{\ell -} 
\right] \\
B_{\ell -} &=& E_{\ell -} + M_{\ell -} \\
C_{\ell +} &=& - \ell S_{\ell -} 
\end{eqnarray}
\end{subequations}
differ from the usual Hebb-Walker convention \cite{Walker69} by using 
$S_{\ell \pm}$ in place of $L_{\ell \pm}$.

\section{Response functions}
\label{app:rsfns}

The response functions can be expressed in terms of helicity amplitudes
$H_i(W,Q^2,\theta) = {\cal H}_i(W,Q^2,\theta,0)$ for $\phi=0$ as follows.

\begin{subequations}
\label{eq:hamps_rsfns}
\begin{eqnarray}
             \; R_L   &=& \frac{q^2}{Q^2} \left( |H_5|^2 + |H_6|^2 \right) \\
\sin\theta   \; R_L^n &=& -2\frac{q^2}{Q^2} \Im H_5 H_6^* \\
             \; R_T   &=& \frac{1}{2} \left( |H_1|^2 + |H_2|^2 + |H_3|^2 + |H_4|^2 \right) \\
\sin\theta   \; R_T^N &=& \Im \left( H_1 H_3^* + H_2 H_4^* \right) \\
\sin\theta   \; R_{LT}      &=&  \frac{q}{\sqrt{2}Q} \Re \left( (H_1-H_4)H_5^* - (H_2+H_3)H_6^* \right) \\
             \; R_{LT}^n    &=& -\frac{q}{\sqrt{2}Q} \Im \left( (H_2+H_3)H_5^* + (H_1-H_4)H_6^* \right) \\
\sin\theta   \; R_{LT}^\ell &=&  \frac{q}{\sqrt{2}Q} \Im \left( (H_1+H_4)H_5^* - (H_2-H_3)H_6^* \right) \\
             \; R_{LT}^t    &=&  \frac{q}{\sqrt{2}Q} \Im \left( (H_2-H_3)H_5^* + (H_1+H_4)H_6^* \right) \\
\sin\theta   \; R_{LT}^\prime        &=& -\frac{q}{\sqrt{2}Q} \Im \left( (H_1-H_4)H_5^* - (H_2+H_3)H_6^* \right) \\
             \; R_{LT}^{\prime n}    &=& -\frac{q}{\sqrt{2}Q} \Re \left( (H_2+H_3)H_5^* + (H_1-H_4)H_6^* \right) \\
\sin\theta   \; R_{LT}^{\prime \ell} &=& -\frac{q}{\sqrt{2}Q} \Re \left( (H_1+H_4)H_5^* - (H_2-H_3)H_6^* \right) \\
             \; R_{LT}^{\prime t}    &=& -\frac{q}{\sqrt{2}Q} \Re \left( (H_2-H_3)H_5^* + (H_1+H_4)H_6^* \right) \\
\sin^2\theta \; R_{TT}      &=& \Re \left( H_2 H_3^* - H_1 H_4^* \right) \\
\sin\theta   \; R_{TT}^n    &=& \Im \left( H_1 H_2^* + H_3 H_4^* \right) \\
\sin^2\theta \; R_{TT}^\ell &=& \Im \left( H_2 H_3^* - H_1 H_4^* \right) \\
\sin\theta   \; R_{TT}^t    &=& \Im \left( H_3 H_4^* - H_1 H_2^* \right) \\
             \; R_{TT}^{\prime \ell} &=& \frac{1}{2} \left( |H_3|^2 + |H_4|^2 - |H_1|^2 - |H_2|^2 \right) \\ 
\sin\theta   \; R_{TT}^{\prime t}    &=& \Re \left( H_1 H_3^* + H_2 H_4^* \right) 
\end{eqnarray}
\end{subequations}

\section{Born baseline model}
\label{app:Born}

In this section we summarize the Born baseline model used for the
multipole analysis, including only the terms that contribute to the
$p\pi^0$ channel.
The electromagnetic vertices are represented by effective lagrangians of 
the form
\begin{subequations}
\begin{eqnarray}
{\cal{L}}_{\gamma NN} &=& -e {\bar N} \left[ F_1(Q^2) \gamma_\mu {\cal A}^\mu
 + F_2(Q^2) \frac{\sigma_{\mu \nu}}{2m_N} (\partial^\mu {\cal A}^\nu)
\right] N \\
{\cal{L}}_{\gamma \pi V} & = &e \frac{\lambda_{V}}{m_{\pi}}
           \varepsilon_{\mu \nu \alpha \beta} (\partial^{\mu} A^{\nu})
  \pi_3 \partial^{\alpha}(\delta_{i,3}\omega^{\beta} + \rho_i^{\beta})
  F_{\gamma \pi V}(Q^2) \\
\end{eqnarray}
\end{subequations}
where $N$ represents a nucleon field operator,
$A^{\mu}$ is the electromagnetic vector potential,  
$\bm{\pi}$ is the pion field as an isospin vector, 
$V \in \{\omega,\rho\}$ denotes a vector meson.
We used conventional dipole and Galster form factors for the nucleon and 
monopole form factors
\begin{equation}
F_{\gamma \pi V}(Q^2) = 
\left( 1 + \frac{Q^2}{m_{V}^2} \right)^{-1}
\end{equation}
for $\gamma\pi V$ vertices.
The $\gamma\pi V$ and $VNN$ parameters are listed in Table \ref{table:VNN}
and were taken from Ref. \cite{Drechsel99}.

We used pure pseudovector $\pi NN$ coupling 
\begin{equation}
{\cal{L}}_{\pi NN} = -\frac{g_{\pi NN}^{PV}}{2m_N}
{\bar N} \gamma_5 \gamma_\mu \bm{\tau}\cdot(\partial^\mu \bm{\pi}) N 
\end{equation}
with $g_{\pi NN}^{PV}=13.4$.
Drechsel {\it et al.}\ \cite{Drechsel99} proposed a more flexible $\pi NN$
model that interpolates between pseudovector coupling for small $p_\pi$ 
and pseudoscalar coupling for large $p_\pi$, but this variation only
affects real parts of $0+$ and $1-$ baseline multipoles and the fitted 
parameters simply compensate for variations of the pseudoscalar/pseudovector 
mixture anyway. 

Finally, the $VNN$ coupling is described by
\begin{equation}
{\cal{L}}_{V N N}  =  -{\bar{N}} \left[ \left( g_{V_1} \gamma_{\mu}
         + \frac{g_{V_2}}{2 m_N} \sigma_{\mu \nu} \partial^{\mu}\right)
 (\omega^{\nu} + \bm{\tau}\cdot\bm{\rho}^{\nu} ) \right] N F_{VNN}(t)
\end{equation}
where $\omega^{\nu}$ and $\bm{\rho}^{\nu}$ represent 
$\omega$ and $\rho$ fields.
A strong form factor,
\begin{equation}
F_{VNN}(t) = \frac{ \Lambda_{VNN}^2 - m_V^2 }{ \Lambda_{VNN}^2 - t }
\end{equation}
is applied to the $VNN$ vertex according to the prescription of
Brown {\it et al}.\ \cite{Brown86}. 

\begin{table}
\caption{Parameters for vector-meson vertices.
\label{table:VNN}}
\begin{ruledtabular}
\begin{tabular}{r r r r r r}
$ V $ & $ m_V $ [MeV]   & $ g_{V_1}$
& $ g_{V_2} $ & $ \Lambda_{VNN} $ [MeV]  & $ \lambda_V $ \\
\hline
$ \omega $ & $ 782.6 $ & $ 21 $ & $ -12 $ & $ 1200 $ & $ 0.314 $ \\
$ \rho $ & $ 769.0 $ & $ 2 $ & $ 13 $ & $ 1500 $ & $ 0.103 $ \\
\end{tabular}
\end{ruledtabular}
\end{table}


\end{document}